\pdfoutput=1
\newcommand*{\ATLASLATEXPATH}{}
\documentclass[atlasdraft=false, coverpage=false, texlive=2016, USenglish, cernpreprint=true]{\ATLASLATEXPATH atlasdoc}
 
\usepackage[subcaption]{\ATLASLATEXPATH atlaspackage}
\usepackage{\ATLASLATEXPATH atlasbiblatex}
 
\usepackage[BSM=true,jetetmiss=true,process=true]{\ATLASLATEXPATH atlasphysics}
 
\graphicspath{{logos/}{figures/}}
 
\usepackage{SUSY-2017-01-PAPER-defs}
 
\usepackage{float}
\floatstyle{plaintop}
\restylefloat{table}

\AtlasTitle{Search for chargino and neutralino production in final states with a Higgs boson and missing transverse momentum at $\sqrt{s} = 13$~TeV with the ATLAS detector}

\AtlasAbstract{
A search is conducted for the electroweak pair production of a chargino and a neutralino $pp \rightarrow \tilde\chi^\pm_1 \tilde\chi^0_2$, where the chargino decays into the lightest neutralino and a $W$ boson, $\tilde\chi^\pm_1 \rightarrow \tilde\chi^0_1 W^{\pm}$, while the neutralino decays into the lightest neutralino and a Standard Model-like 125 GeV Higgs boson, $\tilde\chi^0_2 \rightarrow \tilde\chi^0_1 h$. Fully hadronic, semileptonic, diphoton, and multilepton (electrons, muons) final states with missing transverse momentum are considered in this search. Higgs bosons in the final state are identified by either two jets originating from bottom quarks ($h \rightarrow b\bar{b}$), two photons ($h \rightarrow \gamma\gamma$), or leptons from the decay modes $h \rightarrow WW$, $h \rightarrow ZZ$ or $h \rightarrow \tau \tau$. The analysis is based on 36.1~fb$^{-1}$ of $\sqrt{s} = 13$~TeV proton--proton collision data recorded by the ATLAS detector at the Large Hadron Collider. Observations are consistent with the Standard Model expectations, and 95\% confidence-level limits of up to 680 GeV in $\tilde\chi^\pm_1/\tilde\chi^0_2$ mass are set in the context of a simplified supersymmetric model.
}

\AtlasRefCode{SUSY-2017-01}
 
\PreprintIdNumber{CERN-EP-2018-306}
 
\AtlasDate{30 July 2019}

\arXivId{1812.09432}
 
\HepDataRecord{86564}
 
\AtlasJournalRef{Phys.\ Rev.\ D. \textbf{100}, 012006}
\AtlasDOI{10.1103/PhysRevD.100.012006}
 
\hypersetup{pdftitle={ATLAS document},pdfauthor={The ATLAS Collaboration}}
 
\begin{document}
 
\maketitle
 
\tableofcontents

\section{Introduction}
\label{sec:intro}
 
Theoretical and experimental arguments suggest that the Standard Model (SM) is an effective
theory valid up to a certain energy scale. The observation by the ATLAS and CMS collaborations
of a particle consistent with the SM Higgs boson~\cite{HIGG-2012-27,CMS-HIG-12-028,HIGG-2014-14,HIGG-2015-07}
has brought renewed attention to the mechanism of electroweak symmetry breaking and the hierarchy problem~\cite{Sakai:1981gr,Dimopoulos:1981yj,Ibanez:1981yh,Dimopoulos:1981zb}: the Higgs boson mass is strongly sensitive to quantum corrections from physics at very high energy scales and
demands a high level of fine-tuning. Supersymmetry (SUSY)~\cite{Golfand:1971iw,Volkov:1973ix,Wess:1974tw,Wess:1974jb,Ferrara:1974pu,Salam:1974ig}
resolves the hierarchy problem by introducing for each known boson or fermion a new partner (superpartner) that shares the same mass and internal quantum numbers if supersymmetry is unbroken.
However, these superpartners have not been observed, so SUSY must be a broken symmetry and the mass scale of the supersymmetric particles is as yet undetermined.
The possibility of a supersymmetric dark matter (DM) candidate~\cite{Goldberg:1983nd,Ellis:1983ew} is
related closely to the conservation of \emph{R}-parity~\cite{Farrar:1978xj}. Under the \emph{R}-parity conservation hypothesis, the lightest supersymmetric particle (LSP) is stable.
If the LSP is weakly interacting, it may provide a viable DM candidate.
The nature of the LSP is defined by the mechanism that spontaneously breaks supersymmetry and the parameters of the chosen theoretical framework.
 
In the SUSY scenarios considered as benchmarks in this paper, the LSP is the lightest of the neutralinos (\nino) which, together with the charginos (\chinopm), represent the mass eigenstates formed from the mixture of the $\gamma, \Wboson, \Zboson$ and Higgs bosons'  superpartners (the higgsinos, winos and binos).
The neutralinos and charginos are collectively referred to as \textit{electroweakinos}.
Specifically, the electroweakino mass eigenstates are designated in order of increasing mass as $\tilde{\chi}_{i}^{\pm}$ $(i=1,2)$ (charginos) and $\tilde{\chi}_{j}^{0}$ $(j=1,2,3,4)$ (neutralinos).
In the models considered in this paper, the compositions of the lightest chargino (\chinoonepm) and next-to-lightest neutralino (\ninotwo) are wino-like and the two particles are nearly mass degenerate, while the lightest neutralino (\ninoone) is assumed to be bino-like.
 
Naturalness considerations~\cite{Barbieri:1987fn,deCarlos:1993yy} suggest that the
lightest of the charginos and neutralinos have masses near the electroweak scale.
Their direct production may be the dominant mechanism at the Large Hadron Collider
(LHC) if the superpartners of the gluon and quarks are heavier than a few TeV.
In SUSY models where the masses of the heaviest (pseudoscalar, charged) MSSM Higgs boson and
the superpartners of the leptons have masses larger than those of the lightest chargino and next-to-lightest
neutralino, the former might decay into the \ninoone and a \Wboson boson ($\chinoonepm \rightarrow W \ninoone$),
while the latter could decay into the \ninoone and the lightest MSSM Higgs boson ($h$, SM-like), or \Zboson boson
($\ninotwo \rightarrow h / Z \ninoone$)~\cite{Fayet:1976et,Fayet:1977yc,Farrar:1978xj}.
The decay via the Higgs boson is dominant for many choices of the parameters as long as the mass-splitting between the two
lightest neutralinos is larger than the Higgs boson mass and the higgsinos are heavier than the
winos. SUSY models of this kind, where sleptons are not too heavy although with masses above that of $\chinoonepm$ and $\ninotwo$,
could provide a possible explanation for the discrepancy between measurements of the muon's anomalous magnetic moment $g-2$
and SM predictions~\cite{Das:2014kwa,Ajaib:2015yma,Chakraborti:2015mra,Endo:2017zrj}.
 
\begin{figure}[h!]
 
\begin{subfigure}[b]{0.24\textwidth}
\centering\includegraphics[width=\textwidth]{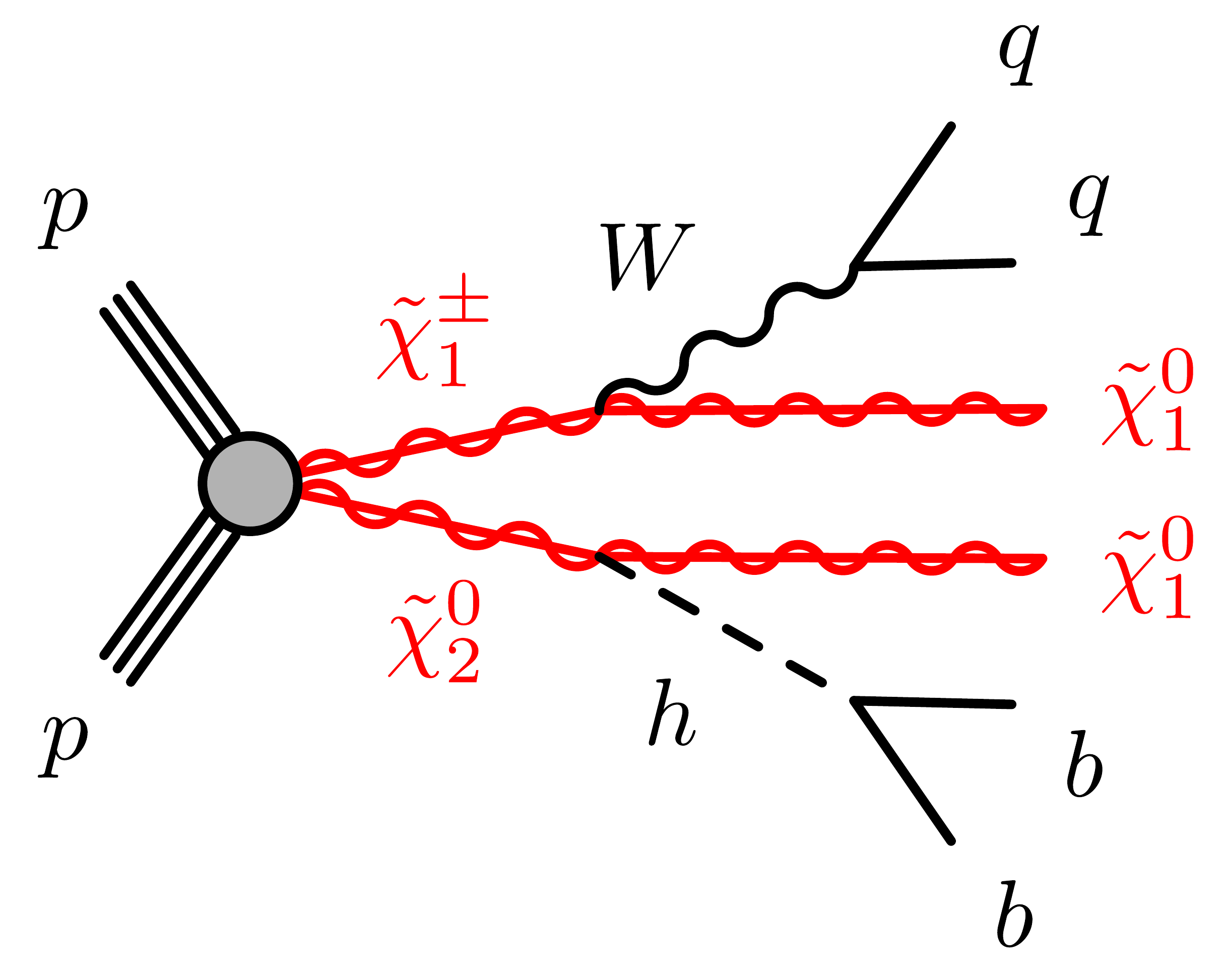}
\caption{\hadronic\label{fig:diagram-had}}
\end{subfigure}
\begin{subfigure}[b]{0.24\textwidth}
\centering\includegraphics[width=\textwidth]{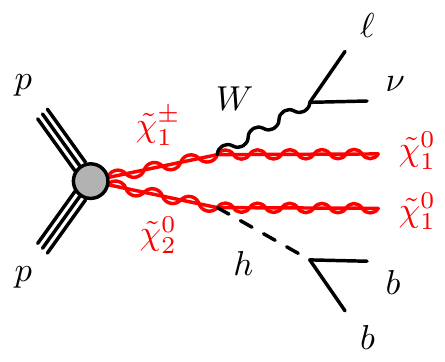}
\caption{\oneLbb \label{fig:diagram-1lbb}}
\end{subfigure}
\begin{subfigure}[b]{0.24\textwidth}
\centering\includegraphics[width=\textwidth]{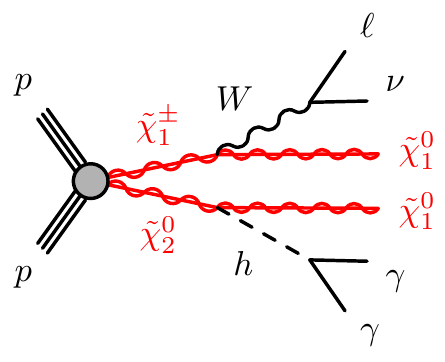}
\caption{\oneLgamgam \label{fig:diagram-photon}}
\end{subfigure}
\begin{subfigure}[b]{0.24\textwidth}
\centering\includegraphics[width=\textwidth]{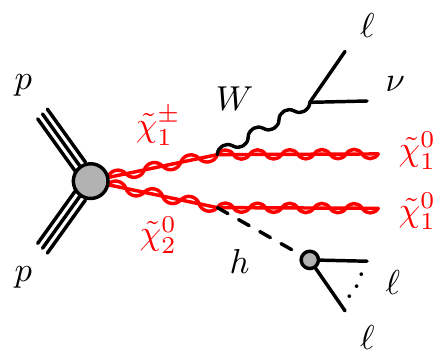}
\caption{\twoL,\threeL \label{fig:diagram-lep}}
\end{subfigure}
 
\caption{
Diagrams illustrating the signal scenarios considered for the pair production of chargino and next-to-lightest neutralino targeted by the \subref{fig:diagram-had} hadronic (\hadronic) and \subref{fig:diagram-1lbb} \oneLbb, \subref{fig:diagram-photon} \oneLgamgam, \subref{fig:diagram-lep} \twoL, \threeL leptonic channel selections. In \subref{fig:diagram-had} and \subref{fig:diagram-1lbb} the Higgs boson decays into two $b$-quarks. In \subref{fig:diagram-photon}, the diphoton channel is shown with $h \rightarrow \gamma \gamma$. In \subref{fig:diagram-lep}, the visible multilepton  final state of the Higgs boson is shown. Leptons are either electrons or muons.
}
\label{fig:diagrams}
\end{figure}

This paper presents a search in proton-proton collision produced at the LHC at a center-of-mass energy $\sqrt{s} = 13$~TeV for the direct pair production of mass-degenerate charginos and next-to-lightest neutralinos that promptly decay as $\chinoonepm \rightarrow W \ninoone$ and $\ninotwo \rightarrow h \ninoone$. The search targets hadronic and leptonic decays of both the $W$ and Higgs bosons.
Three Higgs decay modes are considered: decays into a pair of $b$-quarks, a pair of photons, or a pair of \Wboson or \Zboson bosons or $\tau$-leptons, where at least one of the $\Wboson/\Zboson/\tau$ decays leptonically.
Four signatures are considered, illustrated in Figure~\ref{fig:diagrams}.
All final states contain missing transverse momentum ($\vec{p}_{\mathrm{T}}^{\mathrm{miss}}$,
with magnitude $\etmiss$) from neutralinos, and in some cases neutrinos.
Events are characterized by the various decay modes of the $W$ and Higgs bosons.  The signatures considered have:
jets, with two of them originating from the fragmentation of $b$-quarks, called \emph{b-}jets, and either no leptons (\hadronic,  \Fig{\ref{fig:diagram-had}}),
or exactly one lepton ($\ell= e, \mu$)
(\oneLbb, \Fig{\ref{fig:diagram-1lbb}});
two photons and one lepton (\oneLgamgam, \Fig{\ref{fig:diagram-photon}});
only leptons (\Fig{\ref{fig:diagram-lep}}) such that the final state contains either two leptons with the same electric charge, \twoL, or three leptons, \threeL .
 
A simplified SUSY model~\cite{Alwall:2008ag,Alves:2011wf} is considered for the optimization of the search and the interpretation of results.
The $\chinoonepm \rightarrow W \ninoone$ and $\ninotwo \rightarrow h \ninoone$ decays are assumed to have 100\% branching ratio.
The Higgs boson mass is set to 125 GeV
and its branching ratios are assumed to be the same as in the SM.
The Higgs boson candidate can be fully reconstructed with \hadronic, \oneLbb and \oneLgamgam signatures, while \twoL and \threeL final states are sensitive to decays $h \rightarrow WW$, $h \rightarrow ZZ$ and $h \rightarrow \tau \tau$.
Previous searches for charginos and neutralinos at the LHC targeting decays via the Higgs boson into leptonic final states have been reported by the ATLAS~\cite{SUSY-2014-05} and CMS~\cite{Sirunyan:2018ubx} collaborations; a search in the hadronic channel is also reported in this paper.
 
\section{ATLAS detector}
\label{sec:detector}
 
The ATLAS detector~\cite{PERF-2007-01} is a multipurpose particle detector with a forward--backward symmetric cylindrical
geometry and nearly 4$\pi$ coverage in solid angle.\footnote{ATLAS uses a right-handed coordinate system with its origin at the nominal
interaction point in the center of the detector. The positive $x$-axis is defined by the direction from the interaction point to the center of the LHC ring, with the positive $y$-axis pointing upwards, while the beam direction defines the $z$-axis. Cylindrical coordinates
$(r,\phi)$ are used in the transverse plane, $\phi$ being the azimuthal angle around the $z$-axis. The pseudorapidity $\eta$ is
defined in terms of the polar angle $\theta$ by $\eta=-\ln\tan(\theta/2)$. Rapidity is defined as $y$ = $0.5 \ln[(E + p_z)/(E-p_z)]$ where $E$
denotes the energy and $p_z$ is the component of the momentum along the beam direction. The angular distance $\DeltaR$ is defined as $\sqrt{(\Delta y)^2 + (\Delta \phi)^2}$.}
The inner tracking detector consists of pixel and microstrip silicon detectors
covering the pseudorapidity region $|\eta|<2.5$, surrounded by a transition radiation tracker
which enhances electron identification in the region $|\eta|<2.0$.
A new inner pixel layer, the insertable B-layer~\cite{ATLAS-TDR-19, Abbott:2018ikt}, was added at a mean radius of 3.3~cm during the period between Run~1 and Run~2 of the LHC.
The inner detector is surrounded by a thin superconducting solenoid providing an axial 2~T magnetic field and by a fine-granularity lead/liquid-argon
(LAr) electromagnetic calorimeter covering $|\eta|<3.2$. A steel/scintillator-tile calorimeter provides hadronic coverage in the central pseudorapidity
range ($|\eta|<1.7$).
The endcap and forward regions ($1.5<|\eta|<4.9$) of the hadronic calorimeter are made of LAr active layers with either copper or tungsten
as the absorber material.
A muon spectrometer with an air-core toroid magnet system surrounds the calorimeters.
Three layers of high-precision tracking chambers provide coverage in the range $|\eta|<2.7$, while dedicated fast chambers allow triggering
in the region $|\eta|<2.4$. The ATLAS trigger system consists of a hardware-based level-1 trigger followed by a software-based high-level trigger~\cite{TRIG-2016-01}.
 
 
\section{Data and Monte Carlo simulation}
\label{sec:datamc}
 
The data used in this analysis were collected in \pp collisions at the LHC with a center-of-mass energy of 13~\TeV\ and a 25~ns proton bunch crossing interval during 2015 and 2016. The full dataset corresponds to an integrated luminosity of 36.1~\ifb~after requiring that all detector subsystems were operational during data recording.
The uncertainty in the combined 2015+2016 integrated luminosity is 2.1\%. It is derived, following a methodology similar to that detailed in Ref.~\cite{DAPR-2013-01}, and using the LUCID-2 detector for the baseline luminosity measurements~\cite{LUCID2}, from calibration of the luminosity scale using $x$--$y$ beam-separation scans.
Each event includes on average 13.7 and 24.9 inelastic \pp collisions in the same bunch crossing (pileup) in the 2015 and 2016 datasets, respectively. In the \hadronic and \oneLbb channels, events are required to pass $\etmiss$\ triggers with period-dependent thresholds. These triggers are fully efficient for events with $\etmiss>200~\GeV$ reconstructed offline. Data for the \oneLgamgam signature were collected with a diphoton trigger which selects events with at least two photons, with transverse momentum thresholds on the highest- and second-highest \pt photons of 35~GeV and 25~GeV, respectively. A combined set of dilepton and single-lepton triggers was used for event selection in the \twoL and \threeL channels.

Monte Carlo (MC) samples of simulated events are used to model the signal and to aid in the estimation of SM background processes, with the exception of multijet processes, which are estimated from data. All simulated samples were produced using the ATLAS simulation infrastructure~\cite{SOFT-2010-01} and GEANT4~\cite{Agostinelli:2002hh}, or a faster simulation based on a parameterization  of the calorimeter response and GEANT4 for the other detector systems. The simulated events were reconstructed with the same algorithm as that used for data.

SUSY signal samples were generated with \madgraph~v2.2.3~\cite{Alwall:2014hca} (v2.3.3 for \hadronic) at leading order (LO) and interfaced to \pythia~v8.186~\cite{Sjostrand:2007gs} (v8.212 for \hadronic) with the A14~\cite{ATL-PHYS-PUB-2014-021} set of tuned parameters (tune) for the modeling of the parton showering (PS), hadronization and underlying event. The matrix element (ME) calculation was performed at tree level and includes the emission of up to two additional partons. The ME--PS matching was done using the CKKW-L~\cite{Lonnblad:2012ix} prescription, with a matching scale set to one quarter of the chargino and next-to-lightest neutralino mass. The NNPDF23LO~\cite{Ball:2012cx} parton distribution function (PDF) set was used. The cross-sections used to evaluate the signal yields are calculated to next-to-leading-order (NLO) accuracy in the strong coupling constant, adding the resummation of soft gluon emission at next-to-leading-logarithm accuracy (NLO+NLL)~\cite{Beenakker:1997ut,Beenakker:2010nq,Beenakker:2011fu}. The nominal cross-section and its uncertainty are taken as the midpoint and half-width of an envelope of cross-section predictions using different PDF sets and factorization and renormalization scales, as described in Ref.~\cite{Borschensky:2014cia}.

Background samples were simulated using different MC event generators depending on the process.
All background processes are normalized to the best available theoretical calculation of their respective cross-sections. The event generators, the accuracy of theoretical cross-sections, the underlying-event parameter tunes, and the PDF sets used in simulating the SM background processes are summarized in \Tab{\ref{tab-MCsamples}}.
For all samples, except those generated using \sherpa~\cite{Gleisberg:2008ta}, the \evtgen~v1.2.0~\cite{Lange:2001uf} program was used to simulate the properties of the bottom- and charm-hadron decays. Several samples produced without detector simulation are employed to estimate systematic uncertainties associated with the specific configuration of the MC generators used for the nominal SM background samples. They include variations of the renormalization and factorization scales, the CKKW-L matching scale, as well as different PDF sets and fragmentation/hadronization models. Details of the MC modeling uncertainties are discussed in~\Sect{\ref{sec:systematics}}.
 
\begin{table}[b!]
\centering
{\footnotesize
\begin{tabular}{ccccc}
\toprule
Process &    Generator                   & Tune & PDF set  &  Cross-section  \\
& + fragmentation/hadronization  &      &          &  order \\
\midrule
$W/\Zjets$ & \SHERPA-2.2.1~\cite{Gleisberg:2008ta} & Default & {\textsc NNPDF3.0NNLO}~\cite{Ball:2012cx} & NNLO \\
\\
{$\ttbar$} & \POWHEGBOX v2~\cite{Alioli:2010xd,Frixione:2007vw}
& {\textsc PERUGIA2012}~\cite{perugia} & {\textsc CT10}~\cite{Lai:2010vv} & NNLO \\
& + \PYTHIA-6.428~\cite{pythia6} & & & +NNLL \\
\\
{Single top} & \POWHEGBOX v1 or v2
& {\textsc PERUGIA2012} & {\textsc CT10} & NNLO  \\
& + \PYTHIA-6.428 & & & +NNLL \\
\\
{Diboson}  & & & & \\
{$WW$, $WZ$, $ZZ$}  & \SHERPA-2.2.1 & Default & {\textsc NNPDF3.0NNLO} & NLO \\
\\
{$\ttbar+X$} & & & & \\
 
{$\ttbar W/Z$} & \MADGRAPH-2.2.2~\cite{Alwall:2014hca}
& {\textsc A14}~\cite{ATL-PHYS-PUB-2014-021} & {\textsc NNPDF2.3} & NLO \\
4 top quarks & + \PYTHIA-8.186~\cite{Sjostrand:2007gs} & & & \\
\\
{$\ttbar h$} & \MGMCatNLO-2.2.1
& {\textsc UEEE5}~\cite{ueee5} & {\textsc CT10} & NLO \\
& + \herwigpp-2.7.1 & & & \\
\\
{$Wh$, $Zh$} & \PYTHIA-8.186~\cite{Lange:2001uf} & A14 & NNPDF2.3 & NLO \\
\\
\bottomrule
\end{tabular}
}
\caption{List of generators used for the different
processes. Information is given about the underlying-event
tunes, the PDF sets and the perturbative QCD highest-order accuracy (LO, NLO, next-to-next-to-leading order, NNLO, and next-to-next-to-leading-log, NNLL)
used for the normalization of the different samples.}
\label{tab-MCsamples}
\end{table}
 
 
\section{Event reconstruction and object definitions}
\label{sec:obj}
 
Common event-quality criteria and object reconstruction definitions are applied for all analysis channels, including standard data-quality requirements to select events taken during optimal detector operation. In addition, each analysis channel applies selection criteria that are specific to the objects and kinematics of interest in those final states, which are described in \Sect{\ref{sec:analysis}}.
 
Events are required to have at least one primary vertex, defined as the vertex associated with at least two tracks with $\pt>0.4$~\GeV~and with the highest sum of squared transverse momenta of associated tracks~\cite{ATL-PHYS-PUB-2015-026}. Quality criteria are imposed to reject events that contain at least one jet arising from non-collision sources or detector noise~\cite{ATLAS-CONF-2015-029}.
 
Electron candidates are reconstructed from energy clusters in the electromagnetic calorimeter and inner-detector tracks. They are required to satisfy the loose likelihood identification criteria, have B-layer hits (the \textit{loose} requirement), and be isolated~\cite{PERF-2017-03,ATLAS-CONF-2016-024}. These identification criteria are based on several properties of the electron candidates, including calorimeter-based shower shapes, inner-detector track hits and impact parameters, and comparisons of calorimeter cluster energy to track momentum. Corrections for energy contributions due to pileup are included.  For all but the \lgg channel, electrons are also required to have $\pt > 20~\gev$ and $|\eta|<2.47$; for the \lgg channel they are required to have $\pt > 15~\gev$ and $|\eta|<2.37$. These electrons are used in the overlap removal procedure that is described below, and to apply lepton selections and vetoes in the various analysis channels, in some cases with additional selections applied.
 
Photon candidates are reconstructed from energy clusters in the electromagnetic calorimeter~\cite{PERF-2013-04} in the region $|\eta| < 2.37$, after removing the transition region between barrel and endcap calorimeters, $1.37<|\eta|<1.52$. Photons are classified as unconverted photons if they do not have tracks from a conversion vertex matched to the cluster, and as converted if they do~\cite{PERF-2017-02}. Identification criteria are applied to separate photon candidates from \pizero or other neutral hadrons decaying into two photons~\cite{PERF-2013-04}. Strict identification requirements based on calorimeter shower shapes are used to identify the so-called \textit{tight} photons, which are used in the \oneLgamgam analysis channel. In this case, photons are required to satisfy an isolation criterion based on the sum of the calorimeter energy in a cone of $\DeltaR = 0.4$ centered on the direction of the candidate photon, minus the energy of the photon candidate itself and energy expected from pileup interactions. The resulting isolation transverse energy is required to be less than $2.45~\GeV + 0.022\times \ET^{\gamma}$, where $\ET^{\gamma}$ is the candidate photon's transverse energy.
Photons are calibrated using comparisons of data with MC simulation~\cite{PERF-2017-03} and required to have $\ET > 25~\gev$.
For both the electrons and photons, additional criteria are applied to remove poor quality or fake electromagnetic clusters resulting from instrumental problems.
 
Muon candidates are reconstructed from matching tracks in  the inner detector and muon spectrometer. They are required to meet \textit{medium} quality and identification criteria and to be isolated, as described in Ref.~\cite{PERF-2015-10}, and to have $\pt > 20~\gev$ ($\pt > 10~\gev$ for the \lgg analysis) and $|\eta|<$~2.5. These muons are used in the overlap removal procedure and to apply lepton selections and vetoes in the various analysis channels, in some cases with additional selections applied.
Events containing muons from calorimeter punch-through or poorly measured tracks are rejected if any muon has a large relative $q/p$ error, or $\sigma(q/p) / |q/p| > 0.2$, where $q$ is the charge of the track and $p$ is the momentum. Cosmic-ray muons are rejected after the muon--jet overlap removal by requiring the transverse and longitudinal impact parameters to be $|d_{0}| < 0.25$~mm and $|z_{0} \sin \theta|< 0.5$~mm, respectively.
 
Jets are reconstructed from three-dimensional topological energy clusters~\cite{PERF-2014-07} in the calorimeter using the \antikt jet algorithm~\cite{Cacciari:2008gp} with a radius parameter of 0.4. Each topological cluster is calibrated to the electromagnetic scale prior to jet reconstruction. The reconstructed jets are then calibrated to the energy scale of stable final state particles\footnote{Stable particles in the MC simulation event record are those that have a lifetime $\tau$ such that $c\tau > 10$~mm. 
Jets of this kind are referred to as particle jets.} in the MC simulation by a jet energy scale (JES) correction derived from \sqsthirt data and simulations~\cite{PERF-2016-04}. Further selections are applied to reject jets within $|\eta|< 2.4$  that originate from pileup interactions by means of a multivariate algorithm using information about the tracks matched to each jet~\cite{PERF-2014-03,PERF-2016-04}. Candidate jets are required to have $\pt > 20~\GeV$ and $|\eta|< 2.8$.
 
A jet is tagged as a $b$-jet by means of a multivariate algorithm called MV2c10 using information about the impact parameters of inner-detector tracks matched to the jet, the presence of displaced secondary vertices, and the reconstructed flight paths of $b$- and $c$-hadrons inside the jet~\cite{PERF-2012-04,ATL-PHYS-PUB-2016-012,PERF-2016-05}. Jets tagged as $b$-jets must have $|\eta| < 2.5$.  Several operating points are available, corresponding to various efficiencies obtained in $\ttbar$ simulated events. The 77\% efficiency point was found to be optimal for most SUSY models considered in this paper and is used in the analysis. This configuration corresponds to a background rejection of 6 for $c$-jets, 22 for $\tau$-leptons and 134 for light-quark and gluon jets \cite{PERF-2012-04,ATL-PHYS-PUB-2016-012,PERF-2016-05}, estimated using $\ttbar$ simulated events.
 
The \etmiss in the event is defined as the magnitude of the negative vector sum of the \pt of all selected and calibrated physics objects in the event, with an extra term added to account for soft energy in the event that is not associated with any of the selected objects. This soft term is calculated from inner-detector tracks matched to the primary vertex to make it more resilient to pileup contamination~\cite{PERF-2016-07}.
 
Overlaps between reconstructed objects are accounted for and removed in a prioritized sequence. If a reconstructed muon shares an inner-detector track with an electron, the electron is removed.
Jets within $\DeltaR = 0.2$ of an electron are removed. Electrons that are reconstructed within $\DeltaR = 0.4$ of any surviving jet are then removed, except in the case of the \hadronic channel, where $\DeltaR = \min(0.4, 0.04 +10~\GeV/\pt^{e})$, thereby allowing a high-\pt electron
to be slightly closer to a jet than $\DeltaR = 0.4$. If a
jet is reconstructed within $\DeltaR = 0.2$ of a muon and the jet has fewer than three associated tracks or the muon energy constitutes a large fraction ($> 50$~\%) of the jet energy, then the jet is removed. Muons reconstructed within a cone of size $\DeltaR = \min(0.4,0.04+10~\GeV/\pt^{\mu})$ around the axis of any surviving jet are
removed. If an electron (muon) and a photon are found within $\Delta R = 0.4$, the object is interpreted as
electron (muon) and the overlapping photon is removed from the event.
Finally, if a jet and a photon are found within $\Delta R < 0.2$, the object is interpreted as photon and the overlapping jet is removed from the event; otherwise, if $\Delta R < 0.4$,  the object is interpreted as a jet and the overlapping photon is discarded.

 
 
\section{Kinematic requirements and event variables}
\label{sec:variables}
 
Different analysis channels' signal regions are optimized to target different mass hierarchies of the particles involved.
The event selection criteria are defined on the basis of kinematic requirements for the objects described in the previous
section and event variables are presented below. In the following, jets are ordered according to decreasing \pt, and \pt
thresholds depend on the analysis channel.
 
\begin{itemize}
 
\item $\njet$ is the number of jets with $|\eta| < 2.8$ and $\pt$ above an analysis-dependent \pt threshold.
 
\item $\nbjet$ is the number of $b$-jets with $|\eta| < 2.5$ with $\pt$ above an analysis-dependent \pt threshold.
 
\item \DEtaLL is the pseudorapidity difference between the two leading leptons in the case of multilepton channels.
 
\item The minimum azimuthal angle $\dphimin$ between the \ptmiss and the $\vec{p}_{\mathrm{T}}$ of each of the four leading jets in the event is useful for rejecting events with mismeasured jet energies leading to \etmiss in the event, and is defined as:
 
\begin{equation}
\label{eq:dphimin}
\dphimin = \textrm{min}_{i\leq 4} \Delta\phi
\left(  \ptmiss,
\vec{p}_{\mathrm{T},i}^{\mathrm{jet}}
\right) \notag
\end{equation}
\noindent where $\textrm{min}_{i\leq 4}$ selects the jet the minimizes $\Delta\phi$.
 
\item The effective mass \meff is defined as the scalar sum of the $\pt$ of jets, leptons and $\met$, which aids in establishing the mass scale of the processes being probed, and is defined as:
 
\begin{equation}
\label{eq:meff}
\meff = \sum_{i}^{\njet} \pTjeti  + \sum_{j}^{\nlep} p_{\mathrm{T}, j}^{\ell} + \met. \notag
\end{equation}
 
\item \mbb is the invariant mass of the two leading $b$-jets in the event, and serves as a selection criterion for jet pairs to be considered as Higgs boson candidates.
 
\item \mnonbb corresponds to the invariant mass of the two highest-$\pt$ jets in the event not identified as $b$-jets. This observable,
used in the \hadronic~channel, serves as a selection criterion for jet pairs to be considered as $W$ boson candidates.

\item \mgg is the invariant mass of the two leading photons in the event, and serves as a selection criterion for photon pairs to be considered as Higgs boson candidates.

\item \mljj is the invariant mass of the jet
(when requiring exactly one jet), or the two leading jets system (when requiring two or more jets), and the closest lepton. The angular distance \DeltaR is used as the distance measure between the lepton and the jet.
 
\item \mlll is the invariant mass of the three selected leptons.
 
\item \mtw is the transverse mass formed by the \met and the leading lepton in the event. It is defined as:
 
\begin{equation}
\label{eq:mtw}
\mtw  = \sqrt{2 p_{\mathrm{T}}^{\ell} \MET \left(1 - \cos \Delta\phi(\ell,\ptmiss) \right) } \notag
\end{equation}
and is used to reduce the $W$+jets and \ttbar~backgrounds.
 
\item \mTbmin is the minimum transverse mass formed by $\met$ and up to two of the highest-\pt $b$-jets in the event, defined as:
 
\begin{equation}
\label{eq:mtb}
\mTbmin =  \mathrm{min}_{i\leq 2}
\left ( \sqrt{ 2\pt^{b\textrm{-jet}_i}
\met \left( 1-\cos \Delta\phi(\ptmiss,
\vec{p}_{\mathrm{T}}^{b\textrm{-jet}_i}) \right) }
\right ). \notag
\end{equation}
\noindent where $\textrm{min}_{i\leq 2}$ selects the $b$-jet the minimizes the transverse mass.

\item The lepton--\MET--$\gamma$ transverse mass \mTWgami is calculated with respect to the $i^{\mathrm{th}}$ photon $\gamma_{i}$, ordered in terms of decreasing \et, the \met, and the identified lepton $\ell$. It is defined as:
 
\begin{equation}
\label{eq:mtwgamma}
\begin{split}
\left( \mTWgami \right)^{2} =
& \ 2 \et^{\gamma_i} \MET       \left(1 - \cos \Delta\phi (\gamma_i,\ptmiss)  \right) \\
& + 2 \pt^{\ell}     \MET       \left(1 - \cos \Delta\phi (\ell    ,\ptmiss)  \right) \\
& + 2 \et^{\gamma_i} \pt^{\ell} \left(1 - \cos \Delta\phi (\gamma_i,\ell)  \right). \notag
\end{split}
\end{equation}
 
\item \mctbb is the contransverse mass variable~\cite{Tovey:2008ui, Polesello:2009rn} and is defined for the \bbbar system as:
 
\begin{equation}
\mctbb =  \sqrt{2 \pt^{b_1} \pt^{b_2}  \left ( 1+\cos\Delta\phi_{bb} \right )}, \notag
\label{eq:mctbb}
\end{equation}
 
where $\pt^{b_1}$ and $\pt^{b_2}$ are transverse momenta of the two leading $b$-jets and $\Delta\phi_{bb}$ is the azimuthal angle between them.
It is one of the main discriminating variables in selections targeting Higgs bosons decaying into $b$-quarks and is effective in suppressing the background from top-quark pair production.

\item $\mttwo$ is referred to as the stransverse mass and is closely related to \mt. It is used to bound the masses of particles produced in pairs and each decaying into one particle that is detected and another particle that contributes to the missing transverse momentum~\cite{Lester:1999tx,Barr:2003rg}.  In the case of a dilepton final state, it is defined by:
 
\begin{equation}
\mttwo = \min_{q_\mathrm{T}}{\left[ \max \left(\mt (\vec{p}_{\mathrm{T}}^{\, \ell, 1},\vec{q}_{\mathrm{T}}),\mt (\vec{p}_{\mathrm{T}}^{\, \ell,  2}, \ptmiss - \vec{q}_{\mathrm{T}}) \right) \right]}, \notag
\end{equation}
 
where $\vec{q}_{\mathrm{T}}$ is the transverse vector that minimizes the larger of the two transverse masses \mt, and $\vec{p}_{\mathrm{T}}^{\,  \ell, 1}$ and $\vec{p}_{\mathrm{T}}^{\,  \ell, 2}$ are the leading and subleading transverse momenta of the two leptons in the pair.
 
\item The \lgg variable \dphiWH is the azimuthal angle between the \Wboson boson and Higgs boson candidates. The \Wboson boson is defined by the sum of the lepton $\vec{p}_{\mathrm{T}}^{\ell}$ and \ptmiss vectors, and the Higgs boson by the sum of the transverse momentum vectors of the two photons.
 
\end{itemize}
 
\section{Analysis strategy}
\label{sec:analysis}
 
 
The hadronic and leptonic decay modes of the \Wboson and Higgs bosons are divided into four independent and mutually exclusive analysis channels according to key features of the visible final state: hadronic decays of both the \Wboson and $h$ (\hadronic, \Sect{\ref{sec:0Lbb}}); hadronic $h$ decays with leptonic \Wboson decays (\oneLbb, \Sect{\ref{sec:1Lbb}}); diphoton $h$ decays with leptonic \Wboson decays (\lgg, \Sect{\ref{sec:1Lgamgam}}); multilepton $h$ decays via $\Wboson,\Zboson,\tau$ and leptonic \Wboson decays (\twoL and \threeL, \Sect{\ref{sec:SS3L}}). Event selections and background estimation methods specific to each analysis channel are described here, as well as the signal, control, and validation region definitions (SR, CR, and VR, respectively).
 
The expected SM backgrounds are determined separately for each SR, and independently for each channel, with a profile likelihood fit~\cite{Baak:2014wma}, referred to as a background-only fit. The background-only fit uses the observed event yield in the associated CRs as a constraint to adjust the normalization of the dominant background processes assuming that no signal is present. The CRs are designed to be enriched in specific background contributions relevant to the analysis, while minimizing the signal contamination, and they are orthogonal to the SRs. The inputs to the background-only fit for each SR include the number of events observed in the associated CR and the number of events predicted by simulation in each region for all background processes. They are both described by Poisson statistics. The systematic uncertainties, described in \Sect{\ref{sec:systematics}}, are included in the fit as nuisance parameters. They are constrained by Gaussian distributions with widths corresponding to the sizes of the uncertainties and are treated as correlated, when appropriate, between the various regions. The product of the various probability density functions forms the likelihood, which the fit maximizes by adjusting the background normalization and the nuisance parameters. Finally, the reliability of the MC extrapolation of the SM background estimates outside of the control regions is evaluated in validation regions orthogonal to CRs and SRs.
 
\subsection{Fully hadronic signature (\hadronic)}
\label{sec:0Lbb}
 
\subsubsection{Event selection}
\label{sec:0Lbb:eventsel}
 
The fully hadronic analysis channel exploits the large branching ratios for both \Wqqbar and \hbb. Missing transverse momentum triggers are used for the trigger selection for the analysis, with an offline requirement of $\met>200$~GeV.  Stringent event selections based on the masses of both the \Wboson and Higgs boson candidates, the presence of exactly two \bjets, and the kinematic relationships of the final-state jets and \met, are required in order to reduce the significant backgrounds from \ttbar, \Zjets, \Wjets and single-top \Wt production. Events are characterized by having four or five jets with $\pt > 30$~GeV, exactly two of which are identified as $b$-jets, and large \meff, \mctbb, and \mTbmin. Two signal regions are defined, specifically targeting either high (HM) or low (LM) \ninotwo and \chinoonepm masses (\SRHMhad and \SRLMhad, respectively). The selections used are shown in \Tab{\ref{tab:SRs-hadronic}}. The \meff and \mTbmin selections are particularly effective in reducing the \ttbar contributions, which is the dominant background for both signal regions. The \Zjets  and single-top contributions are also significant, whereas the contribution from multijet production is found to be negligible and is not included.  Control regions are used to constrain the normalizations of the \ttbar, \Zjets, and \Wt backgrounds with the data, while other processes are estimated using simulation. The \bbbar invariant mass is required to be consistent with the Higgs boson mass, $105~<\mbb~<~135~\GeV$, for all signal regions. All CRs and VRs select sidebands in the \mbb spectrum in order to remain orthogonal to the two SRs. These are further described in \Sect{\ref{sec:0Lbb:bckgest}}.
 
\begin{table}[!ht]
\centering
\begin{tabular}{l | c c}
\toprule
Variable         & \SRHMhad       & \SRLMhad        \\
\midrule
\nlep            & $=0$           & $=0$            \\
\njet ($\pt>30~\GeV$) & $\in[4,~5]$    & $\in[4,~5]$     \\
\nbjet           & $=2$           & $=2$            \\
\dphimin         & $>0.4$         & $>0.4$          \\
$\met~[\gev]$    & $>250$         & $>200$          \\
$\meff~[\gev]$   & $>900$         & $>700$          \\
$\mbb~[\gev]$    & $\in[105,135]$ & $\in[105,135]$  \\
$\mnonbb~[\gev]$ & $\in[75,90]$   & $\in[75,90]$    \\
$\mct~[\gev]$    & $>140$         & $>190$          \\
$\mTbmin~[\gev]$ & $>160$         & $>180$          \\
\bottomrule
\end{tabular}
\caption{\label{tab:SRs-hadronic} Signal region definitions for the fully hadronic \hadronic analysis channel.}
\end{table}
 
\subsubsection{Background estimation}
\label{sec:0Lbb:bckgest}
 
The background contributions to \SRHMhad and \SRLMhad are estimated using fits to the data for \ttbar, \Zjets, and single-top production in specially designed control regions.
 
The three control regions used for estimating the \ttbar (CRHad-TT), \Zjets (CRHad-Zj), and \Wt (CRHad-ST) contributions are further divided into high-mass (HM) and low-mass (LM) categories in order to follow the design of the SRs. These control regions are defined primarily by inverting the selections on \mbb, \mctbb, \mTbmin, and by requiring the presence of a lepton in some cases. The \ttbar background is estimated using $\mctbb, \mTbmin < 140~\GeV$ and $\mbb>~135~\GeV$ selections, while retaining the other SR requirements. This approach isolates the \ttbar contribution while suppressing single-top and \Zjets events, yielding a sample estimated to be 94\% pure in \ttbar events with negligible signal contamination. Background events from \Wt are estimated by requiring exactly one lepton and $\mctbb>200~\GeV$, $\mTbmin>180~\GeV$, and $\mbb>195~\GeV$, and relaxing the \meff requirement for HM to $\meff>700~\GeV$. The \Zjets contribution is isolated using an opposite-sign, same-flavor, high-\pt $2\ell$ requirement with $\ptllead>140~\GeV$ and $75 < \mll < 105~\GeV$, which reduces the \ttbar contribution to this control region. These leptons are then treated as invisible when calculating the \etmiss. \Fig{\ref{fig:had-analysis-cr}} shows the distribution of two key observables: the \met in the \ttbar high-mass control region (\Fig{\ref{fig:had-CRtt-highmass}}) and the \mbb distribution in the \Zjets low-mass control region (\Fig{\ref{fig:had-CRZ-lowmass}}). The yields estimated with the background-only fit are reported in \Tab{\ref{tab:had-yields-CRs}}. The normalization factors are found to be $0.88\pm0.10$ ($0.85\pm0.04$), $1.47\pm0.32$ ($1.22\pm0.35$), and $0.54\pm0.25$ ($0.57\pm0.22$) for \ttbar, \Zjets, and \Wt in the high-mass (low-mass) signal region, respectively. The errors include statistical and systematic uncertainties. No diboson MC simulation events are found to contribute to the CRHad-ST regions.
 
To validate the background prediction, three sets of validation regions are defined so as to be similar, but orthogonal, to the SRs. The \ttbar VRs for each SR (VRHad-TT, for HM or LM) reverse the \mct selections, requiring $\mctbb < 140~(190)~\GeV$ for HM (LM), select the sideband $\mbb>135~\gev$ (orthogonal to the SRs), but retain the SR selection on \mTbmin. In order to validate the \Wt and \Zjets estimates, VRs are defined using sideband regions in the \mbb  and \mnonbb spectra, either by vetoing the SR range in both of these variables, $\mbb\notin[105,~135]~\gev$ and $\mnonbb\notin[75,~90]~\gev$ (VRHad-SB for HM and LM), or by selecting the $\mbb>135~\gev$ sideband and imposing a $\Wboson$ mass requirement on the non-$b$-tagged dijet invariant mass, $75<\mnonbb<90~\gev$ (VRHad-bbhigh, for HM or LM).
 
The number of events predicted by the background-only fit is compared with the data in the VRs in the upper panel of \Fig{\ref{fig:had-pulls-VR}}.  The pull, defined by the difference between the observed number of events ($n_{\textrm{obs}}$) and the predicted background yield ($n_{\textrm{pred}}$) divided by the total uncertainty ($\sigma_{\textrm{tot}}$), is shown for each region in the lower panel. No evidence of significant background mismodeling is observed in the VRs.
 
\begin{figure}[htbp]
\centering
\begin{subfigure}[b]{0.5\textwidth}
\centering\includegraphics[width=\textwidth]{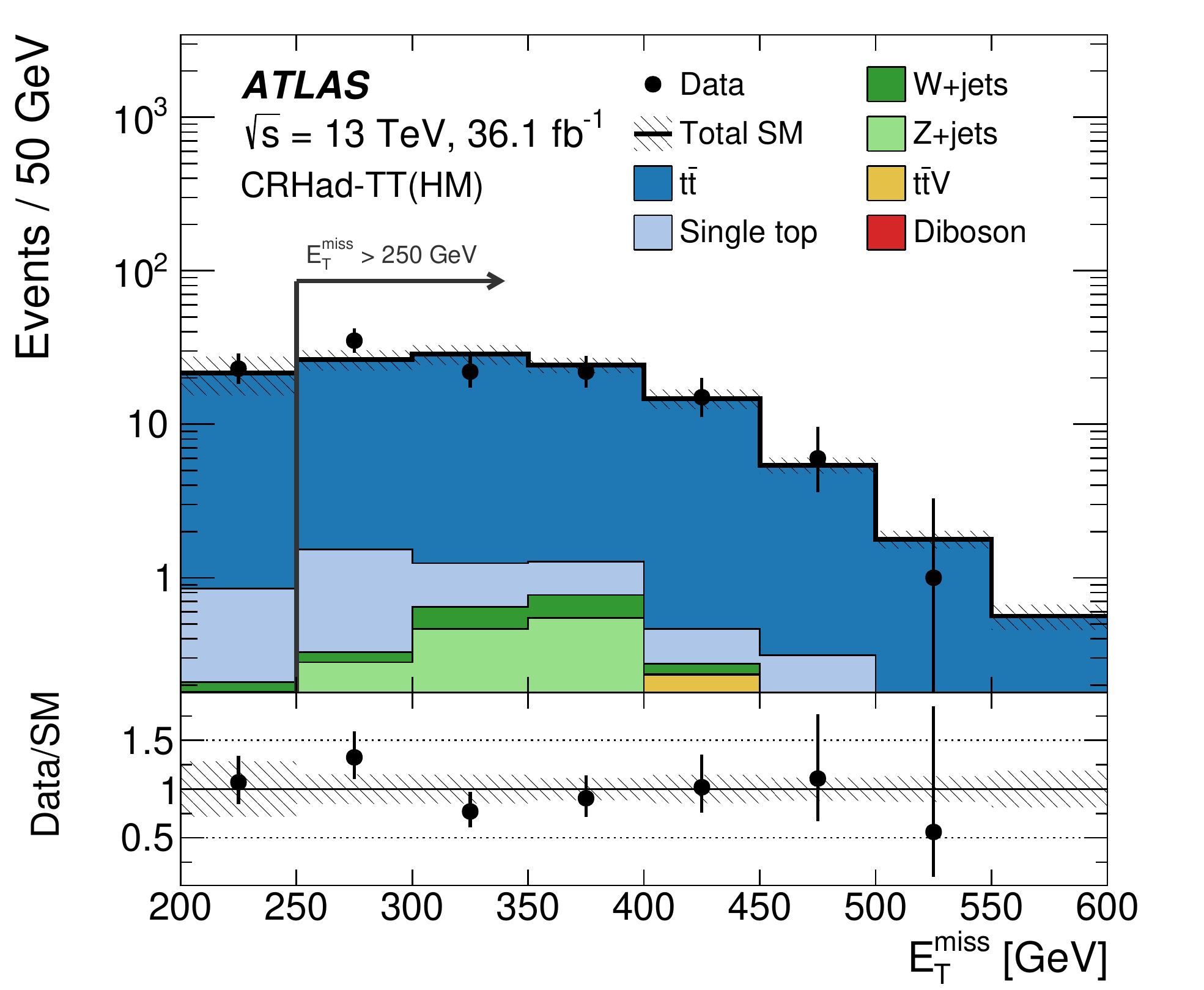}
\caption{\ttbar high-mass control region}
\label{fig:had-CRtt-highmass}
\end{subfigure}
\begin{subfigure}[b]{0.5\textwidth}
\centering\includegraphics[width=\textwidth]{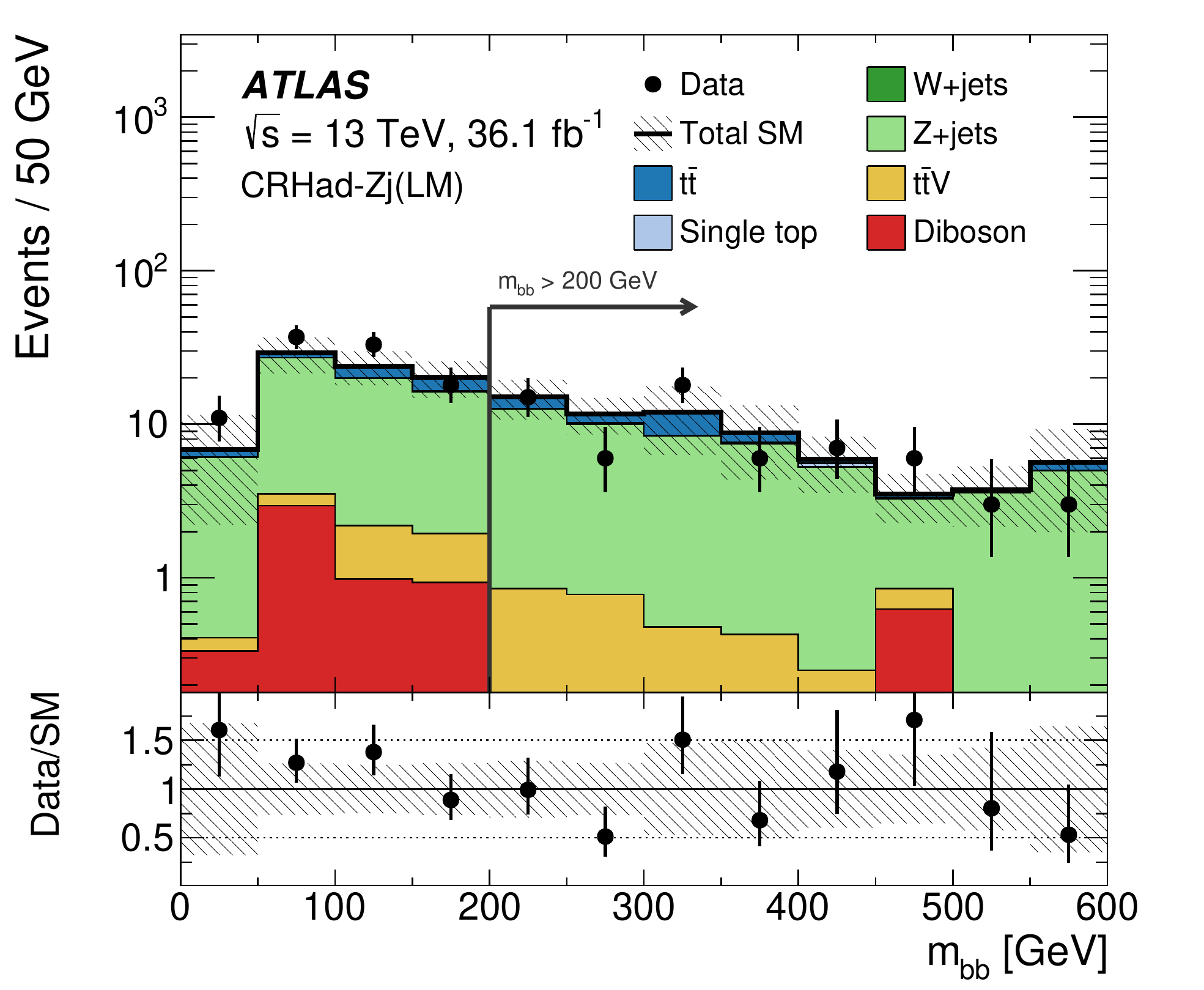}
\caption{\Zjets low-mass control region}
\label{fig:had-CRZ-lowmass}
\end{subfigure}
\caption{Comparisons of data with SM predictions in \ttbar and \Zjets control regions for representative kinematic distributions: \subref{fig:had-CRtt-highmass} \etmiss for the \ttbar high-mass control region and \subref{fig:had-CRZ-lowmass} \mbb for the \Zjets low-mass control region. Predictions from simulation are shown after the background-only fit. The arrow indicates the selection on that variable used to define the corresponding CRs. The uncertainty bands include statistical and systematic uncertainties.}
\label{fig:had-analysis-cr}
\end{figure}
 
\begin{table}[!ht]
\begin{center}\renewcommand\arraystretch{1.6}
\setlength{\tabcolsep}{0.3pc}
\scalebox{0.7}{
\begin{tabular}{l r r r r r r}
\toprule
\textbf{CR channels} & CRHad-TT(HM) & CRHad-ST(HM) & CRHad-Zj(HM) & CRHad-TT(LM) & CRHad-ST(LM) & CRHad-Zj(LM) \\
\midrule
Observed events & $102$ & $17$ & $39$ & $695$ & $23$ & $78$ \\
\midrule
Fitted bkg events & $102\pm10$ & $17\pm4$ & $39\pm6$ & $695\pm26$ & $23\pm5$ & $78\pm9$ \\
\midrule
\ttbar & $97\pm11$ & $3.7\pm2.0$ & $2.9\pm2.4$ & $659\pm34$ & $4.7\pm2.3$ & $10^{+12}_{-10}$ \\
Single top & $2.7^{+3.5}_{-2.7}$ & $10\pm5$ & $0.8^{+0.9}_{-0.8}$ & $19\pm19$ & $15\pm6$ & $1.0\pm0.9$ \\
\Wjets & $0.5^{+0.6}_{-0.5}$ & $2.2\pm1.1$ & $0.0059\pm0.0025$ & $3.9\pm3.1$ & $2.8\pm1.2$ & $0.0059\pm0.0026$ \\
\Zjets & $1.1\pm0.6$ & $0.08\pm0.07$ & $32\pm7$ & $9.5\pm3.2$ & $0.09\pm0.04$ & $63\pm17$ \\
\ttV & $0.63\pm0.14$ & $0.62\pm0.16$ & $2.0\pm0.4$ & $3.1\pm0.5$ & $0.80\pm0.17$ & $3.7\pm0.6$ \\
Diboson & $0.08^{+0.14}_{-0.08}$ & $<0.07$ & $0.8\pm0.8$ & $1.16\pm0.34$ & $<0.07$ & $0.8\pm0.5$ \\
\bottomrule
\end{tabular}
}
\end{center}
\caption{\label{tab:had-yields-CRs}Fit results in the control regions for the \hadronic channel. The results are obtained from the control regions using the background-only fit. The errors shown are the statistical plus systematic uncertainties. Uncertainties in the fitted yields are symmetric by construction, where the negative error is truncated when reaching zero event yield.}
\end{table}
 
 
\begin{figure}[!ht]
\centering
\includegraphics[width=0.90\textwidth]{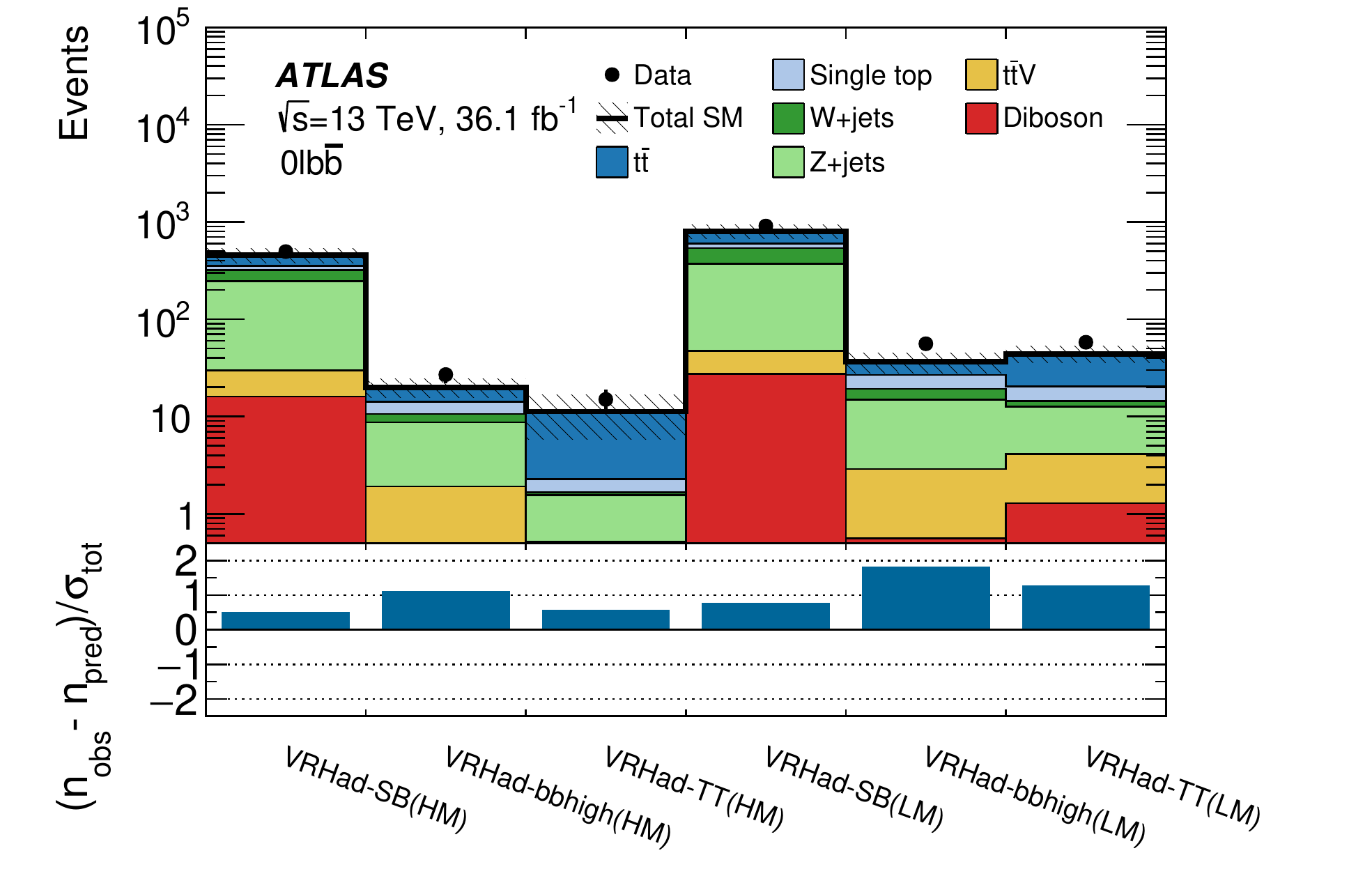}
\caption{ Comparison of the predicted backgrounds with the observed numbers of events in the VRs associated with the \hadronic channel.
The normalization of the backgrounds is obtained from the fit to the CRs. The upper panel shows the observed number of events and the predicted background yield.
All uncertainties are included in the uncertainty band.
The lower panel shows the pulls in each VR. }
\label{fig:had-pulls-VR}
\end{figure}
 
 
\subsection{Single-lepton plus di-$b$-jet signature (\oneLbb)}
\label{sec:1Lbb}
 
\subsubsection{Event selection}
\label{sec:1Lbb:eventsel}
 
The events considered in the one-lepton plus two-$b$-jets channel are also recorded with the \etmiss
trigger, with an offline requirement of $\met>200~\GeV$.
Events with exactly one electron or muon are selected if they also contain two or three jets with $\pt>25$~GeV,
two of which must be $b$-tagged.
Discriminating variables are used to separate the signal from backgrounds, and include
\etmiss, \mtw, the invariant mass of the two $b$-jets and their
contransverse mass.
The dominant SM background contributions in the \oneLbb channel are
\ttbar, \Wjets, and single-top (\Wt channel) production.
Three sets of signal regions  are defined and optimized to target different
LSP  and next-to-lightest neutralino or chargino mass hierarchies. The three regions,
labeled as \SRlbbLow, \SRlbbMedium, and \SRlbbHigh, are summarized in \Tab{\ref{tab:1LbbRegions}}.
\SRlbbLow provides sensitivity to signal models with a mass-splitting between LSP and next-to-lightest
neutralino similar to the Higgs boson mass, while \SRlbbMedium and -High target
mass-splittings between 150 and 250 GeV and above 250 GeV, respectively.
The \mctbb distribution of the \ttbar background has
an upper endpoint approximately equal to the top-quark mass, and thus
this background is efficiently suppressed by
requiring $\mctbb > 160~\GeV$ in all regions. The \Wjets background is reduced by selecting
events with  $\mtw > 100~\GeV$. The three SRs require $100< \mtw <140~\GeV$, $140< \mtw <200~\GeV$,
and $\mtw > 200~\GeV$ for \SRlbbLow, -Medium and -High, respectively.
Finally, the \bbbar invariant mass is required to be $105~<\mbb~<~135~\GeV$,
consistent with the Higgs boson mass, for all regions.
 
\begin{table}[!htpb]
\centering
\begin{tabular}{ l | c c c }
\toprule
Variable               & \SRlbbLow &   \SRlbbMedium &  \SRlbbHigh \\
\midrule
\nlep                  & \multicolumn{3}{c}{$=1$} \\
$\ptl$ [\GeV]          & \multicolumn{3}{c}{$>27$}  \\
\Njet ($\pt>25~\GeV$)  & \multicolumn{3}{c}{$=2$ or 3} \\
\nbjet                 & \multicolumn{3}{c}{$=2$} \\
\etmiss\ [\GeV]        & \multicolumn{3}{c}{$>200$} \\
$\mctbb$ [\GeV]        & \multicolumn{3}{c}{$>160$} \\
$\mtw$ [\GeV]          & $\in[100,140]$ & $\in[140,200]$ & $>200$ \\
$\mbb$ [\GeV]          & \multicolumn{3}{c}{$\in[105,135]$} \\ 
\hline
\end{tabular}
\caption{Summary of the event selection for signal regions of the \oneLbb channel.}
\label{tab:1LbbRegions}
\end{table}

\subsubsection{Background estimation}
\label{sec:1Lbb:bckgest}
 
The contributions from the \ttbar, \Wt, and \Wjets background sources are estimated from MC simulation, but with yields that are normalized to data in dedicated CRs. The contribution from multijet production, where the lepton is misidentified as a jet or originates from a heavy-flavor hadron decay or photon conversion, is found to be negligible and neglected in the following. The remaining sources of background (single-top $t$- and $s$-channels, \Zjets, diboson, \Zh, and \Wh production), including their total yields, are estimated from simulation.
 
Three sets of CRs, CR1Lbb-TT, CR1Lbb-ST and CR1Lbb-Wj, are designed to estimate the \ttbar, \Wt, and \Wjets background processes, respectively.
The acceptance for \ttbar events is increased in CR1Lbb-TT by requiring $\mctbb<160~\GeV$ and inverting the selection on \mbb. Three \ttbar~CRs are defined as a function of \mtw\ mirroring the Low, Medium and High SR selections. Contributions from \Wjets events are estimated using a common CR1Lbb-Wj for all SRs, where events are required to have  $40<\mtw<100~\GeV$ and \mbb $< 80$ GeV. CR1Lbb-ST is designed to be orthogonal to the three CR1Lbb-TTs and CR1Lbb-Wj by requiring events to have $\mctbb>160~\GeV$, $\mbb>195~\GeV$ and $\mtw>100~\GeV$.
 
\begin{figure}[htbp]
\centering
 
\begin{subfigure}[b]{0.45\textwidth}
\centering\includegraphics[width=\textwidth]{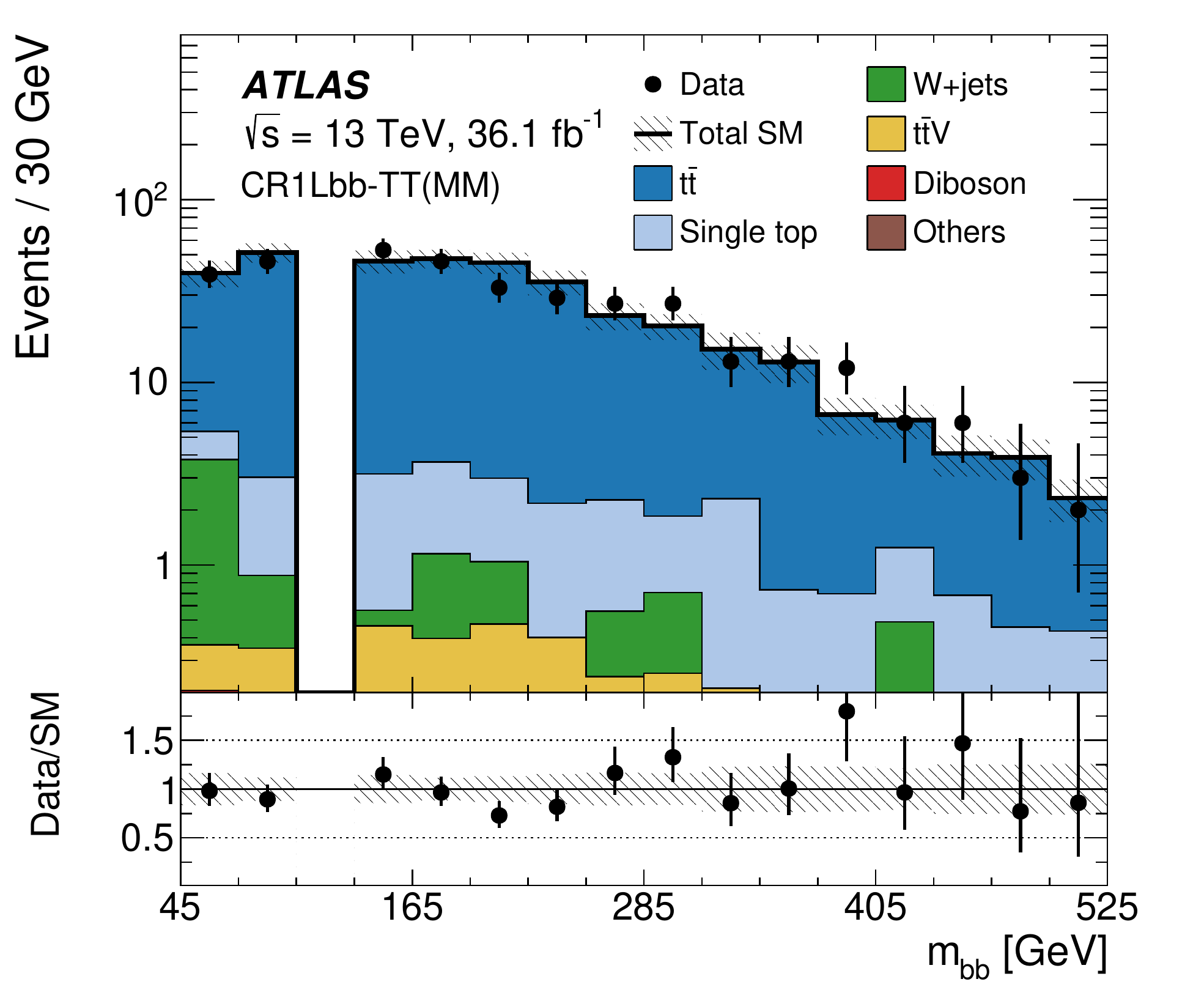}
\caption{\mbb for \ttbar medium-mass (MM) control region}
\label{fig:onelbb-CRtt-lowmass}
\end{subfigure}
\begin{subfigure}[b]{0.45\textwidth}
\centering\includegraphics[width=\textwidth]{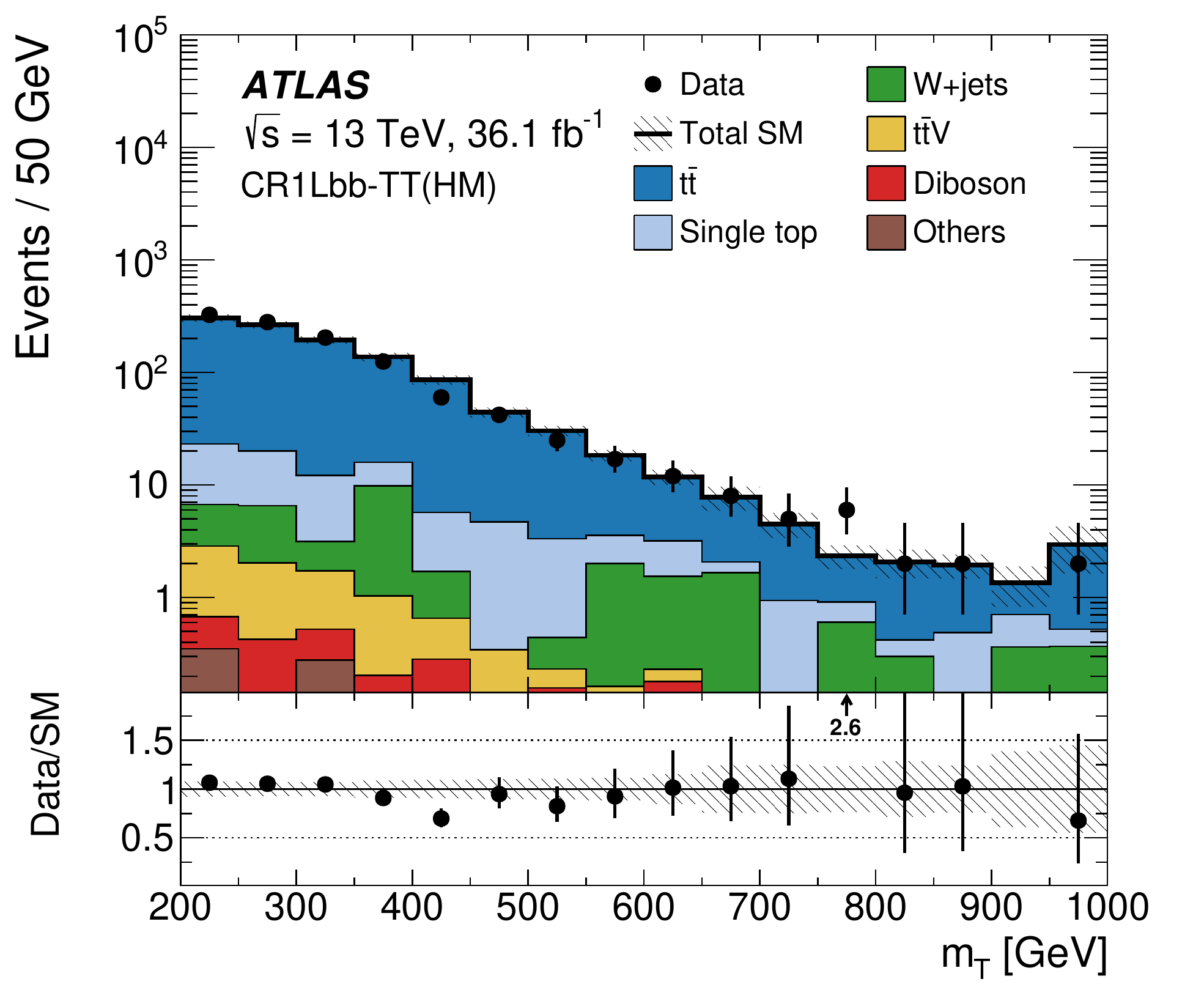}
\caption{\mtw for \ttbar high-mass (HM) control region}
\label{fig:onelbb-CRtt-highmass}
\end{subfigure}
 
\begin{subfigure}[b]{0.45\textwidth}
\centering\includegraphics[width=\textwidth]{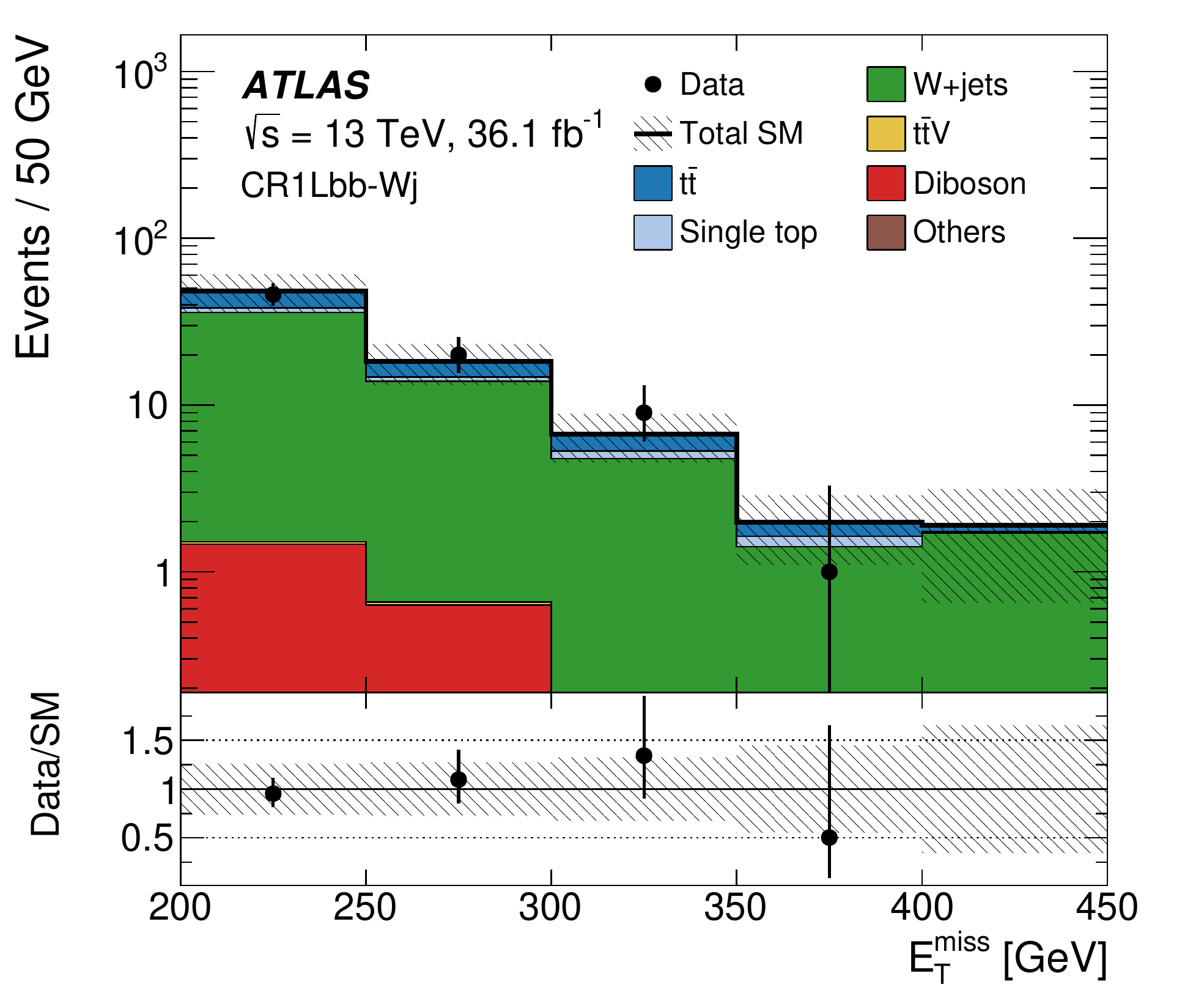}
\caption{\etmiss for \Wjets control region}
\label{fig:onelbb-CRW}
\end{subfigure}
\begin{subfigure}[b]{0.45\textwidth}
\centering\includegraphics[width=\textwidth]{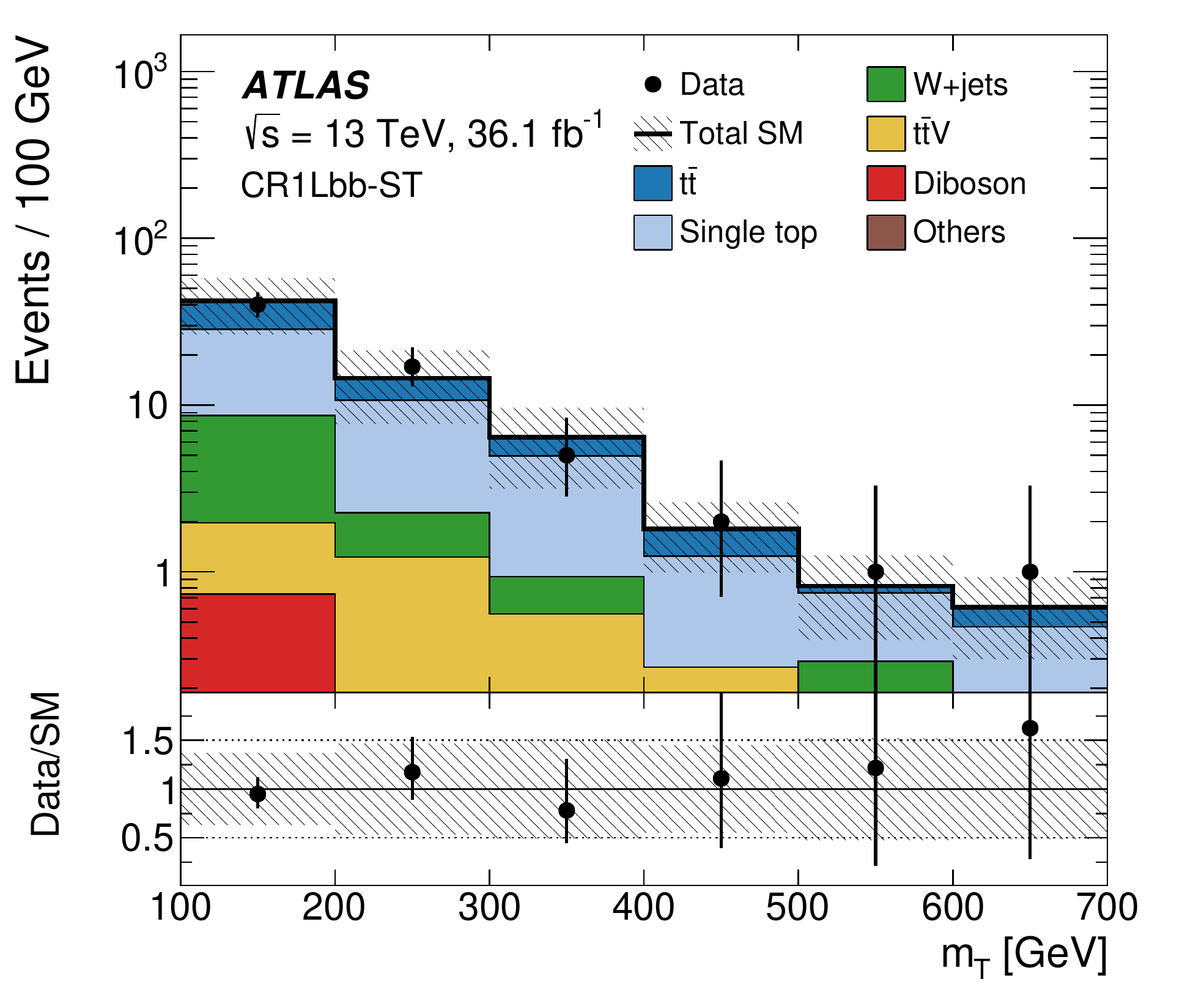}
\caption{\mtw for single-top control region}
\label{fig:onelbb-singletop}
\end{subfigure}
 
\caption{Comparison of data with SM predictions in \ttbar, \Wjets, and single-top control regions for representative kinematic
distributions: \subref{fig:onelbb-CRtt-lowmass} \mbb for CR1Lbb-TT medium, \subref{fig:onelbb-CRtt-highmass} \mtw for CR1Lbb-TT high, \subref{fig:onelbb-CRW} \met for CR1Lbb-Wj and \subref{fig:onelbb-singletop} \mtw for CR1Lbb-ST.
Predictions from MC simulation are shown after the background-only fit. The uncertainty bands include statistical and systematic uncertainties.}
 
\label{fig:onelbb-analysis-cr}
\end{figure}
 
\begin{table}[bp]
\begin{center}\renewcommand\arraystretch{1.6}
\setlength{\tabcolsep}{0.0pc}
\scalebox{0.7}{
\begin{tabular*}{\textwidth}{@{\extracolsep{\fill}}lrrrrr}
\noalign{\smallskip}\hline\noalign{\smallskip}
\textbf{CR channels}           & CR1Lbb-TT(LM)            & CR1Lbb-TT(MM)            & CR1Lbb-TT(HM)            & CR1Lbb-Wj            & CR1Lbb-ST              \\[0.05cm] \hline
 
Observed events          & $192$              & $359$              & $1115$              & $72$              & $65$                    \\
\noalign{\smallskip}\hline\noalign{\smallskip}
Fitted bkg events         & $192 \pm 14$          & $359 \pm 19$          & $1115 \pm 34$          & $72 \pm 9$          & $65 \pm 8$              \\
\noalign{\smallskip}\hline\noalign{\smallskip}
\ttbar\         & $147 \pm 33$          & $325 \pm 32$          & $1020 \pm 90$          & $15 \pm 14$          & $20_{-20}^{+23}$              \\
Single top         & $28 \pm 25$          & $22_{-22}^{+24}$          & $60_{-60}^{+70}$          & $4_{-4}^{+6}$          & $33 \pm 25$              \\
$W$+jets         & $16 \pm 7$          & $7.3 \pm 2.7$          & $25 \pm 11$          & $51 \pm 17$          & $8 \pm 4$              \\
\ttV          & $1.16 \pm 0.20$          & $2.8 \pm 0.4$          & $6.9 \pm 1.1$          & $0.079 \pm 0.022$          & $3.2 \pm 0.6$              \\
Diboson         & $0.57 \pm 0.24$          & $0.92 \pm 0.29$          & $1.3 \pm 0.4$          & $2.1 \pm 1.1$          & $0.84 \pm 0.28$              \\
Others            & $0.125 \pm 0.032$          & $0.20 \pm 0.06$          & $1.9 \pm 0.5$          & $0.24 \pm 0.17$          & $0.10 \pm 0.04$              \\
\noalign{\smallskip}\hline\noalign{\smallskip}
\end{tabular*}
}
\end{center}
\caption{
Fit results in the control regions for the \oneLbb channel. The results are obtained from the control regions using the background-only fit.
The category ``Others'' includes contributions from \Wh production and \Zjets.  The errors shown
are the statistical plus systematic uncertainties. Uncertainties in the fitted yields are symmetric by construction, where the negative error
is truncated when reaching zero event yield. }
\label{tab:1Lbb-yields-CRs}
\end{table}
 
 
The yields estimated with the background-only fit are reported in \Tab{\ref{tab:1Lbb-yields-CRs}}.
The normalization factors are found to be between $0.89^{+0.21}_{-0.20}$ and $1.15\pm0.13$ for the three SRs' \ttbar\ estimates, $1.1^{+0.7}_{-1.1}$ for \Wt and $1.4 \pm 0.5$
for \Wjets, where the errors include statistical and systematic uncertainties.
\Fig{\ref{fig:onelbb-analysis-cr}} shows representative comparisons of data with MC simulation for \mbb, \mtw and \etmiss distributions
in \ttbar, \Wjets and single-top control regions. The data agree well with the SM predictions in all distributions.
 
To validate the background predictions, two sets of VRs are defined similarly but orthogonal to the SRs. VR1Lbb-onpeak regions are defined similarly to the three CR1Lbb-TT regions but requiring $105<\mbb<135~\GeV$. VR1Lbb-offpeak requires $\mctbb>160~\GeV$, \mbb\ below 95 GeV or in the range 145--195 GeV and $\etmiss>180~\GeV$. The yields and pulls in each VR are shown in \Fig{\ref{fig:onelbb-analysis-pulls-VR}} after the background-only fit. The data event yields are found to be consistent with background expectations.
 
\begin{figure}[!ht]
\centering
\includegraphics[width=0.90\textwidth]{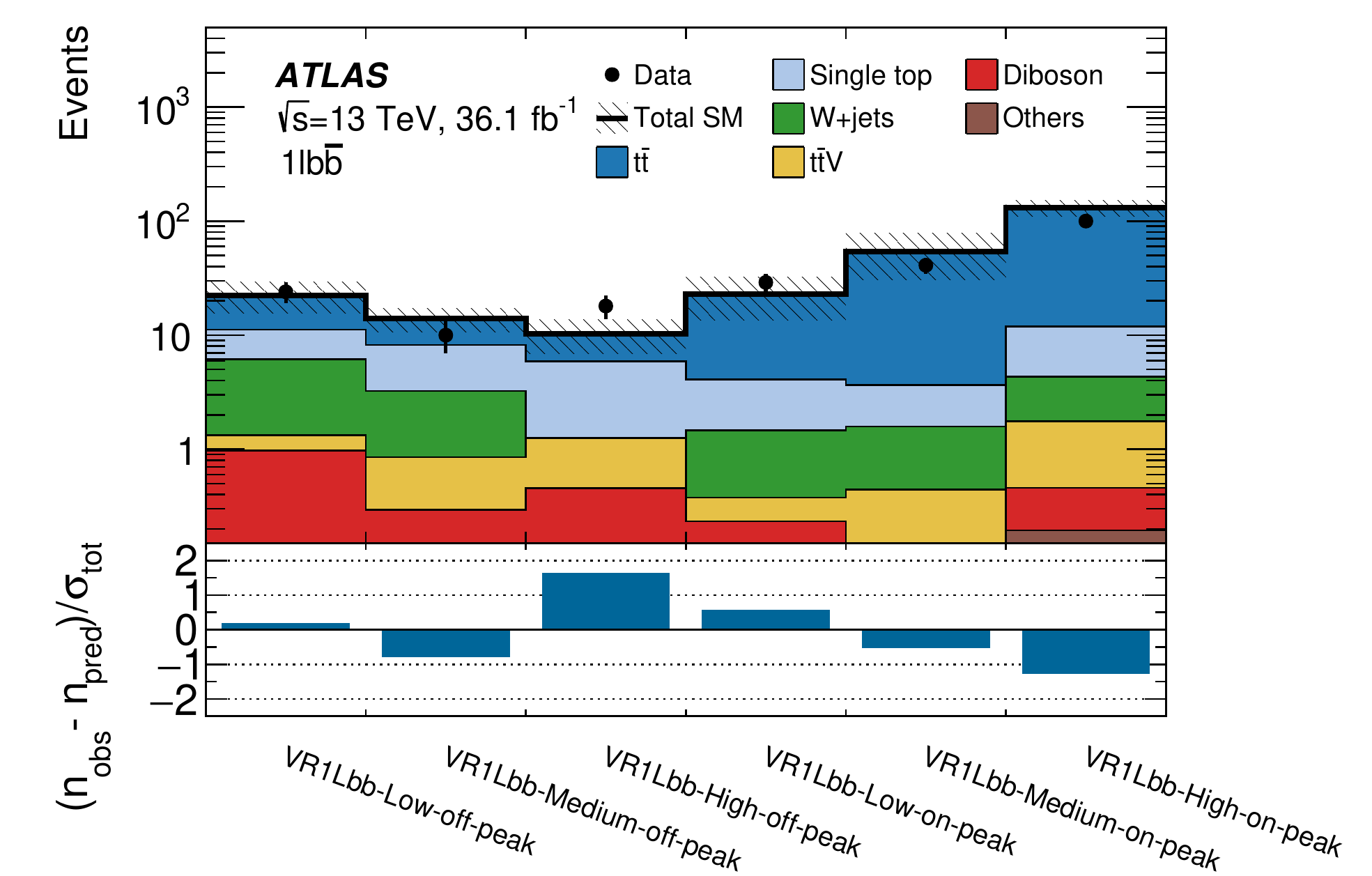}
\caption{ Comparison of the predicted backgrounds with the observed numbers of events in the validation regions associated with the \oneLbb channel.
The normalization of the backgrounds is obtained from the fit to the CRs. The upper panel shows the observed number of events and the predicted background yield.
All uncertainties are included in the uncertainty band.
The lower panel shows the pulls in each VR. }
\label{fig:onelbb-analysis-pulls-VR}
\end{figure}

 
\subsection{Single-lepton plus diphoton signature (\lgg)}
\label{sec:1Lgamgam}
 
\subsubsection{Event selection}
\label{sec:1Lgamgam:eventsel}
 
Events used in the single-lepton plus diphoton (\lgg) channel were
recorded with a diphoton trigger using a trigger-level requirement of
$\ET > 35$ GeV and $\ET > 25$ GeV for the leading and subleading photons, respectively.
For these events, the selection requires exactly one lepton
($e$ or $\mu$) with $\pt > 25~\GeV$ and exactly two photons.
To achieve full trigger efficiency, the leading and subleading photons are required to have a minimum \ET of 40~GeV and 31~GeV, respectively.
The diphoton invariant mass \mgg, which is measured in the
region of the Higgs boson mass with a resolution of approximately 1.7~GeV, is required to
lie within the mass window $120<\mgg<130$~GeV.
This effectively rejects SM backgrounds without a Higgs boson in the final state, referred to as \textit{non-peaking} backgrounds.
These backgrounds, which include SM diphoton and $V\gamma\gamma$ ($V = W, Z$) production,
vary slowly across the selected mass window and can be reliably estimated
from sidebands above and below the narrow mass window assuming a flat distribution. Observables
such as \met, \mtw, \mTWgamone, \mTWgamtwo, \dphiWH and the number of $b$-jets
provide additional discrimination between signal and both the peaking backgrounds
(containing a Higgs boson decaying into two photons) and the non-peaking backgrounds.
 
The dominant peaking background arises from \Wh production, which can be
difficult to distinguish from the signal, which itself includes both
a \Wboson and a Higgs boson. After applying a series of selection criteria
optimized to separate signal from both the peaking and non-peaking backgrounds
(see \Tab{\ref{tab:lyy_sel}}), the resulting inclusive SR is subdivided
into a region largely depleted of \Wh backgrounds (\SRlyyA) and a SR with
a significant contribution from \Wh production (\SRlyyB).
 
\begin{table}[!htpb]
\centering
\begin{tabular}{ l | c c c }
\toprule
Variable               & \SRlyyA     &    &  \SRlyyB \\
\midrule
\ngamma                & & $=2$             & \\
\ptg [\GeV]            & & $> (40,31)$      & \\
\nlep                  & & $=1$             & \\
\ptl [\GeV]            & & $> 25$           & \\
\etmiss\ [\GeV]        & & $>40$            & \\
\dphiWH                & & $> 2.25$         & \\
\mgg [\GeV]            & & $\in[120,130]$   & \\
\nbjet  ($\pt>30~\GeV$) & & $=0$             & \\
\mTWgamone [\GeV]      & & $\ge 150$        & \\
\mTWgamtwo [\GeV]      & $>140$      &      & $\in[80,140]$ \\
\mtw [\GeV]            & $>110$      &      & $<110$      \\
\bottomrule
\end{tabular}
\caption{Summary of the event selection for the two regions of the \lgg channel, \SRlyyA and \SRlyyB. }
\label{tab:lyy_sel}
\end{table}
 
\subsubsection{Background estimation}
\label{sec:1Lgamgam:bckgest}
 
Non-peaking backgrounds are estimated separately for \SRlyyA and \SRlyyB by measuring the event yields, per 10~GeV in \mgg, in the lower and upper sidebands within $105 < \mgg < 120~\GeV$ and $130 < \mgg < 160~\GeV$, respectively.
The yields are determined by fitting a constant function to the observed events in sidebands. Results obtained by fitting a linear function are found to be consistent. The observation of 1 (15) event(s) in the sidebands around \SRlyyA (\SRlyyB) leads to an expectation of $0.22 \pm 0.22$ ($3.3 \pm 0.9$) non-peaking background events, with the uncertainty dominated by the number of events in the sideband data sample.
 
Peaking backgrounds are estimated from MC simulations of the production of the Higgs boson through gluon--gluon and vector-boson fusion, and of Higgs boson production in association with a \Wboson or \Zboson boson. Production of a Higgs boson in association with a \ttbar pair is also taken into account, although this contribution is suppressed by the requirement that the events contain no $b$-jets.
A value of $(2.28 \pm 0.08) \times 10^{-3}$ is assumed for the $h \rightarrow \gamma \gamma$
branching ratio~\cite{deFlorian:2016spz}.
Production of \Wh events, with a subsequent decay of the Higgs boson into two photons, is expected to account for approximately 90\% of the peaking background in the two SRs.
Altogether, a total of $0.14 \pm 0.02$ ($2.01 \pm 0.30$) events are expected
to arise from peaking backgrounds in \SRlyyA (\SRlyyB).

 
\subsection{Same-sign dilepton and three-lepton signatures (\twoL, \threeL)}
\label{sec:SS3L}
 
 
Two- or three-lepton (multilepton) signatures arise when the $\Wboson$ boson produced in conjunction with the Higgs boson decays semileptonically and the Higgs boson itself decays into one of $\Wboson\Wboson$, $\Zboson\Zboson$ or $\tau\tau$, and these in turn yield at least one other lepton in the final state.
Final-state neutrinos and lightest neutralinos all contribute to sizable \met in multileptonic signal events.
Two sets of signal regions, kinematically orthogonal, are defined according to the presence of either exactly two leptons with same-sign electric charge (\twoL analysis), or exactly three leptons satisfying various requirements on lepton-flavor and electric-charge combinations (\threeL analysis).
The \twoL and \threeL analyses share the same trigger.
Events must pass a trigger selection that combines single- and two-lepton triggers into a logical OR, where trigger thresholds on lepton $\pt$ between 8 and 140 GeV are applied
in conjunction with trigger-specific lepton identification criteria. Selected leptons have offline
requirements of $\pT>25~\GeV$ to ensure that trigger efficiencies
are maximal and uniform in the relevant phase space.
For both analyses, events with additional leptons are removed, and a $b$-jet veto is applied such that there are zero $b$-jets with $\pt>20~\GeV$. Non-$b$-tagged jets are not vetoed, and are required in some signal regions to account for hadronic decays of intermediate particles (e.g. $W$ bosons), or for the presence of initial-state radiation.
Jets in both the \twoL and \threeL signal regions are required to have $\pt > 20~\GeV$ and pass
the quality and kinematic selections described in \Sect{\ref{sec:obj}}.
The signal region optimization and background estimations are developed independently for \twoL and \threeL events.
 
Two primary sources of background are distinguished in these analyses. The first category is the reducible background, which includes events containing at least one fake or non-prompt (FNP) lepton (referred to as fake background) and, for the \twoL analysis only, events containing electrons with misidentified charge (referred to as charge-flip background). This background primarily arises from the production of top-quark pairs.  The FNP lepton typically originates from heavy-flavor hadron decays in events containing top quarks, or \Wboson or \Zboson bosons. Those are suppressed for the \twoL and \threeL analyses by vetoing \btagged jets, while hadrons misidentified as leptons, electrons from photon conversions, and leptons from pion or kaon decays in flight remain as other possible sources. Data-driven methods are used for the estimation of this reducible background in the signal and validation regions. The second background category is the irreducible background from events with two same-sign prompt leptons or at least three prompt leptons. It is estimated using simulation samples and is dominated by diboson ($WZ$ and $ZZ$) production. Dedicated validation regions with enhanced contributions from these processes, and small signal contamination, are defined to verify the background predictions from the simulation.
 
Details of the estimates of both the reducible and irreducible backgrounds for each analysis are given in the following subsections.
 
\subsubsection{\twoL event selection and background estimation}
\label{sec:SS3L:SS:eventsel}
 
Two signal regions are defined for the \twoL analysis channel, requiring either exactly one jet (\SRSSJone) or two to three (\SRSSJtwothr) jets.
In both regions, events must satisfy $\met>100~\GeV$, while region-specific requirements are applied on the transverse mass \mt, the effective mass \meff, the stransverse mass \mttwo, and the kinematic variable \mljj, which in signal events provides an estimate of the visible mass of the Higgs boson candidate. The \twoL signal region selections are summarized in \Tab{\ref{tab:SS-SRDef}}.
 
\begin{table}[!htpb]
\centering
\begin{tabular}{l | c  c }
\toprule
Variable                &  \SRSSJone  & \SRSSJtwothr   \\
\midrule
\DEtaLL                 & $< 1.5$     &  -             \\
\njet  ($\pt>20~\GeV$) & $=1$        & $=2$ or 3      \\
\nbjet                  & $=0$        & $=0$           \\
\met  [\GeV]            & $> 100$     & $> 100$        \\
\mt   [\GeV]            & $> 140$     & $> 120$        \\
\meff [\GeV]            & $> 260$     & $> 240$        \\
$m_{\ell j(j)}$ [\GeV]  & $<180$      & $<130$         \\
\mttwo [\GeV]           & $>80$       & $>70$          \\
\bottomrule
\end{tabular}
\caption{Summary of the event selections for the \twoL signal regions.}
\label{tab:SS-SRDef}
\end{table}
 
The reducible FNP background is estimated using the matrix method~\cite{TOPQ-2010-01,ATLAS-CONF-2014-058}. The matrix method uses both relaxed and more stringent lepton identification criteria in order to isolate the contributions from FNP leptons in a given data sample. The two sets of identification criteria that are used are referred to as \textit{tight} and \textit{loose}. The matrix method relates the number of events containing prompt or FNP leptons to the number of observed events with tight or loose-but-not-tight leptons using the probability, $\mathcal{O}(10^{-1}$--$10^{-2})$, for loose prompt or FNP leptons to satisfy the tight criteria.
Inputs to the method are the probability for loose prompt leptons to satisfy the tight selection criteria, estimated using $\Zboson \rightarrow \ell \ell$ events, and  the probability for loose FNP leptons to satisfy the tight selection criteria, determined from data in SS control regions enriched in non-prompt
leptons.
Final yields for FNP backgrounds are validated in VRs.
\Fig{\ref{fig:SS-VR-fakes}} shows the \met distribution in the VR for the \twoL channel in the case of electrons (VRSS-ee) and good agreement is found between data and predictions.
 
Charge misidentification is only relevant for electrons and the contribution of charge-flip events to the SRs and VRs is estimated using the data.
The electron charge-flip probability is extracted in a $Z \rightarrow ee$ data sample using a likelihood fit which takes as input the numbers of same-sign and opposite-sign electron pairs observed in a 80--100~GeV electron-pair mass window. It is treated as a free parameter of the fit and it is found to be between $2\times 10 ^{-4}$ and $10^{-3}$ depending on the \pt and $\eta$ of the electron.
Sources of SM irreducible background arise from $WZ$ and $ZZ$ events and are evaluated using simulation.
 
The background estimates are validated in dedicated VRs defined for each signal region and referred to as VRSS-j1 and VRSS-j23. In VRSS-j1, events are required to pass all selections as in \SRSSJone but for \met, required to be between 70~GeV and 100~GeV, and $\mljj > 130~\GeV$. No selections are applied on \meff and \mttwo, while \mt is required to be above 140~GeV. VRSS-j23 is equivalent to \SRSSJtwothr, with \mt required to be between 65~GeV and 120~GeV and \mljj above 130~GeV. The total numbers of events observed in data and predicted by the background estimation for the \twoL VRs are shown in \Fig{\ref{fig:SS-and-3L-analysis-pulls-VR}}, together with the pull estimates.
 
\begin{figure}[htbp]
\centering
\begin{subfigure}[b]{0.45\textwidth}
\includegraphics[width=\textwidth]{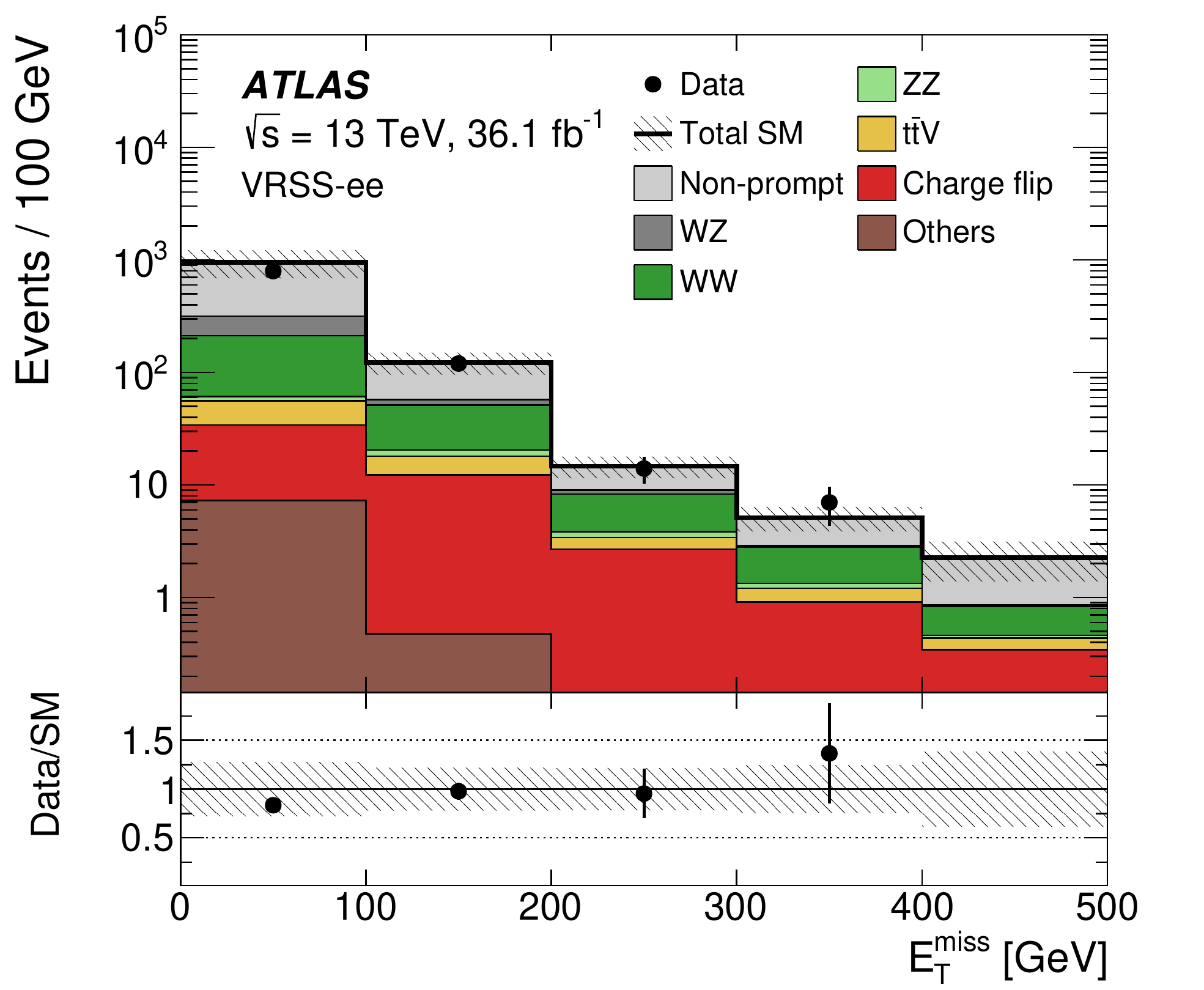}
\caption{\met in \twoL VRSS-ee validation region}
\label{fig:SS-VR-fakes}
\end{subfigure}
\begin{subfigure}[b]{0.45\textwidth}
\centering\includegraphics[width=\textwidth]{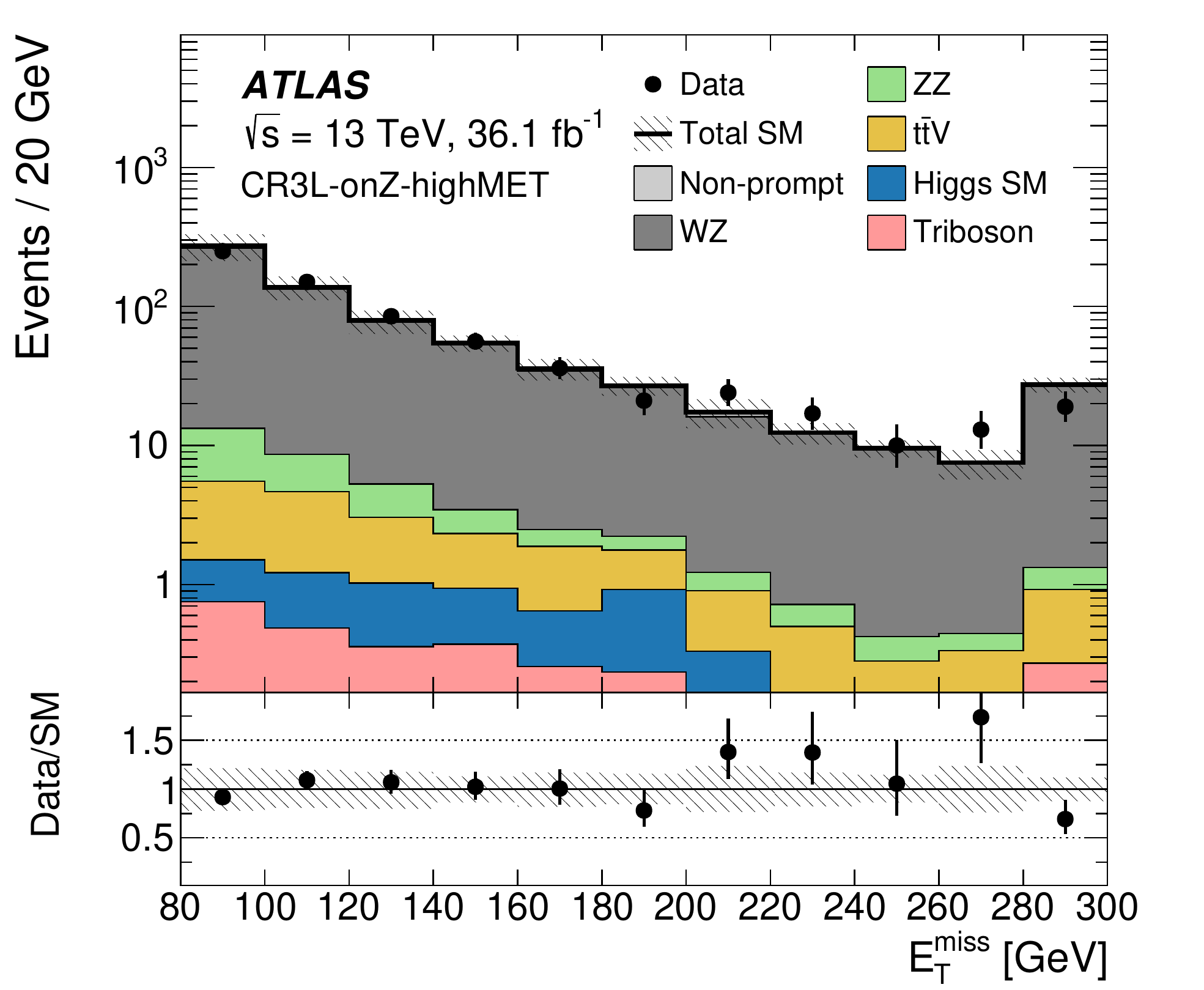}
\caption{\met in \threeL \Zboson-boson control}
\label{fig:3L-CR}
\end{subfigure}
\caption{\subref{fig:SS-VR-fakes} \met distribution in the electron-type VRSS-ee for \twoL channel (used to estimate FNP leptons background) and \subref{fig:3L-CR} \met distribution in the on-shell \Zboson-boson CR for \threeL (used to estimate the $WZ$ normalization).}
\label{fig:multi-analysis-cr}
\end{figure}
 
\begin{figure}[!ht]
\centering
\includegraphics[width=0.90\textwidth]{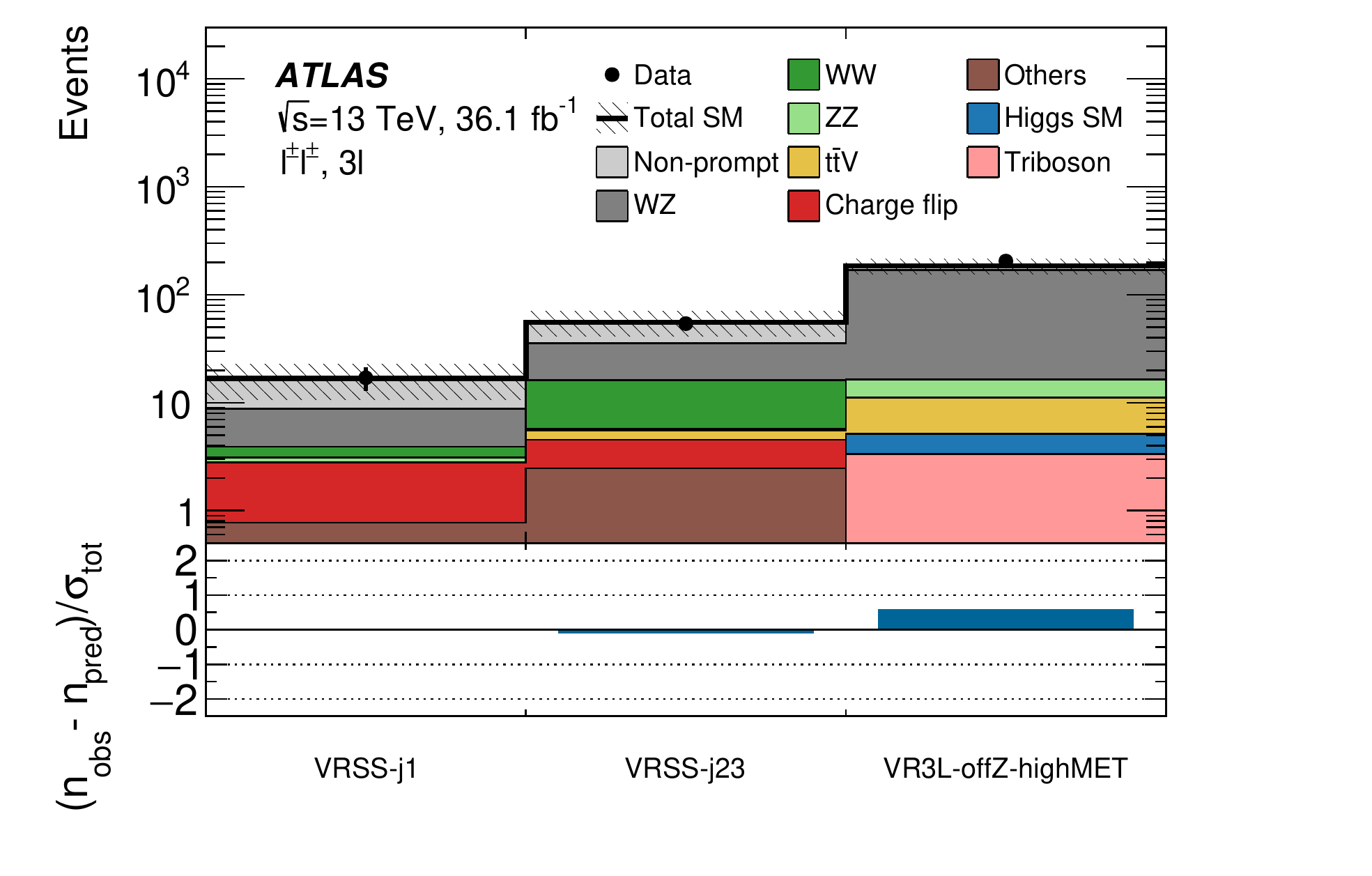}
\caption{Results of the likelihood fit extrapolated to the VRs associated with both the \twoL and \threeL channels. The normalization of the backgrounds is obtained from the fit to the CRs. The upper panel shows the observed number of events and the predicted background yield. All uncertainties are included in the uncertainty band. The lower panel shows the pulls in each VR.}
\label{fig:SS-and-3L-analysis-pulls-VR}
\end{figure}
 
\subsubsection{\threeL event selection}
\label{sec:SS3L:3L:eventsel}
 
Events in the \threeL signal regions are categorized according to flavor and charge-sign combinations of the leptons in the event. Appropriate selection criteria tailored to each region are applied in order to reject lepton-rich background processes while at the same time maximizing signal significance. The event selections applied in the \threeL signal regions are summarized in Tables~\ref{tab:3L-DFOS-SRcuts} and~{\ref{tab:3L-SFOS-SRcuts}.
In \textit{different-flavor opposite-sign} (DFOS) signal regions, two of the leptons are required to have the \textit{same flavor and same-sign} (SFSS) electric charge (the \textit{SFSS lepton pair}), while the third lepton (the \textit{DFOS lepton}) must have different flavor and opposite charge to the other two leptons. The DFOS lepton and the SFSS lepton closest to it in $\Delta\phi$ (the \textit{near lepton}) are taken to originate from the Higgs boson decay. The \DeltaR between these two leptons is called \DROSnear, and their invariant mass, which in signal events gives an estimate of the Higgs boson visible mass, is called \mDFOSnear. The azimuthal angle between the two SFSS leptons is called \dphiSS.
In \textit{same-flavor opposite-sign} (SFOS) signal regions, there must be at least one pair of leptons of the same flavor and with opposite-sign charge (the \textit{SFOS lepton pair}). When only one SFOS lepton pair exists, the invariant mass \mSFOSmin must be greater than $20~\GeV$ and lie outside the 81.2--101.2~GeV interval, to suppress low-mass resonances and \Zboson-rich backgrounds.
If two SFOS pairs exist, the chosen SFOS pair has a lower $m_{\mathrm{T}}^{\ell,\mathrm{min}}$ for the third highest \pt lepton, and the invariant mass requirement is applied to this pair.
The variable $m_{\mathrm{T}}^{\ell,\mathrm{min}}$, defined in analogy with \mTbmin, is also used to identify the unique transverse mass value obtained from the lepton not in the SFOS pair in events for which only one such pair exists. Both the DFOS and SFOS events are further separated into orthogonal signal regions, depending on whether at least one light jet of $\pt>20~\GeV$ is present in the event or not. Region-dependent requirements are placed on \met, as well as on other kinematic variables.
 
\begin{table}[!htpb]
\centering
\begin{tabular}{l | c c c}
\toprule
Variable           & \SRThrLDFOSJ & \SRThrLDFOSJa & \SRThrLDFOSJb\\
\midrule
\Njet  ($\pt>20~\GeV$) & $= 0$        & $>0$          & $>0$             \\
\nbjet             & $= 0$        & $=0$          & $=0$             \\
\met [\GeV]        & $>60$        & $\in[30,100]$ & $>100$           \\
\mDFOSnear [\GeV]  & $<90$        & $<60$         & $<70$            \\
\DROSnear          & -            & $<1.4$        & $<1.4$           \\
\dphiSS            & -            & -             & $<2.8$           \\
\bottomrule
\end{tabular}
\caption{Summary of the event selection for DFOS \threeL signal regions.}
\label{tab:3L-DFOS-SRcuts}
\end{table}
 
\begin{table}[!htbp]
\centering
\begin{tabular}{l | c c c }
\toprule
Variable         & \SRThrLSFOSJa  & \SRThrLSFOSJb   & \SRThrLSFOSJ      \\
\midrule
\Njet  ($\pt>20~\GeV$) & $= 0$          & $= 0$           & $>0$   \\
\nbjet           & $= 0$          & $= 0$           & $=0$   \\
\met [\GeV]      & $\in[80,120]$  & $>120$          & $>110$ \\
\mtmin [\GeV]    & $>110$         & $>110$          & $>110$ \\
$\mSFOSmin$      & $>20~\GeV, \notin[81.2,101.2]$ &  $>20~\GeV, \notin[81.2,101.2]$ &  $>20~\GeV, \notin[81.2,101.2]$\\
\bottomrule
\end{tabular}
\caption{Summary of the event selection for SFOS \threeL signal regions.
}
\label{tab:3L-SFOS-SRcuts}
\end{table}
 
The reducible FNP lepton background in the \threeL channel is dominated by \ttbar and \Zjets processes, and it is estimated using the same approach as for the \twoL analysis. The irreducible background is dominated by $WZ$ production and is estimated using a dedicated control region. The normalization of the $WZ$ background is constrained in this region to reduce systematic uncertainties due to the MC modeling and experimental sources. The $WZ$ CR uses a three-lepton selection in which a SFOS pair has an invariant mass in the $Z$ peak region, $81.2 < m_{\ell\ell} < 101.2$ GeV, the \met is above 80 GeV, and a \btag veto is applied. The estimate from the background-only fit leads to a normalization factor of 1.11$\pm$0.13 for the $WZ$ background and the  \met distribution in the CR is shown in \Fig{\ref{fig:3L-CR}}. Its validity is cross-checked by comparing the SM estimates with data from a VR (referred to as VR3L-offZ-highMET) where events are required to have \met above 120 GeV and $\mSFOSmin$ outside of the \Zboson peak region.
 
The total number of events observed in data and predicted by the background estimation for the \threeL VR are shown in \Fig{\ref{fig:SS-and-3L-analysis-pulls-VR}}, together with the pull estimates.

 
\section{Systematic uncertainties}
\label{sec:systematics}
 
 
Several sources of experimental and theoretical systematic uncertainties in the signal and background estimates are considered in these analyses.
Their impact is reduced through the normalization of the dominant backgrounds in the control regions defined with
kinematic selections resembling those of the corresponding signal region.
Experimental and theoretical uncertainties are included as nuisance parameters with Gaussian constraints
in the likelihood fits, taking into account correlations between different regions.
Uncertainties due to the numbers of events in the CRs are also included in the fit for each region.
 
Theory uncertainties for \ttbar processes are dominant for the \hadronic and \oneLbb analysis channels, ranging from 15\% to 20\% for the \oneLbb channel to nearly 50\% for the low-mass signal region (\SRLMhad) of the \hadronic analysis. Generator uncertainties are assessed by comparing \POWPYTHIA6 with \SHERPA 2.2.1, and the parton shower models are tested by comparing \POWPYTHIA6 with \POWHERWIG{}++. Scale variations are evaluated by varying the \hdamp parameter between $\mtop$ and $2\times\mtop$, and the renormalization and factorization scales up and down by a factor of two. Systematic uncertainties in the contributions from single-top production also account for the impact of interference terms between  single-resonant and double-resonant top-quark production.  Statistical uncertainties are included via the control regions in the data by which the processes are normalized and the size of the simulation samples used for evaluating theoretical systematic uncertainties.
Relaxed selections are used to reduce the statistical uncertainty of theory estimates of top-quark contributions. In particular, the \mCT selection is loosened for both \hadronic and \oneLbb, as are the \mTbmin and \meff selections for the \hadronic channel.
The \Zjets\ and \Wjets\ modeling uncertainties are estimated using the nominal \SHERPA 2.2.1 samples by considering different merging (CKKW-L) and resummation scales, PDF variations from the NNPDF30NNLO replicas, as well as the envelope of changes resulting from seven-point scale variations of the renormalization and factorization scales.
The various components are added in quadrature.
 
Theory uncertainties in both the \Wh production cross-section and the modeling of the \Wh final state also contribute to the uncertainty of the peaking backgrounds in the \oneLgamgam analysis.
They are estimated by varying the nominal PDF error sets, the QCD factorization scale, the parameters associated with the underlying event and parton shower, and the
NLO electroweak correction factors associated with the simulation of the \Wh process. These variations  lead to a fractional uncertainty of $\pm 5.5$\% in the expected contribution of \Wh production to the \lgg SRs.
 
Theory uncertainties related to the estimation of the $\Wboson\Zboson$ background are among the most significant for the multilepton analysis channels (\twoL and \threeL). The effects of PDF choice and the scale of the strong coupling constant, $\alphas$, on the $\Wboson\Zboson$ background are assessed using the same procedure as described above for scale variations in top-quark production processes: by varying the relevant parameters and measuring the impact on the quantities of interest.
 
The dominant detector-related systematic effects differ depending on the analysis channel.
Experimental uncertainties related to the jet energy resolution are significant in the case of \oneLbb, accounting for nearly 20\% of the total systematic uncertainty on the background estimation in the \SRlbbMedium region. Uncertainties related to the jet energy scale contribute to approximately a 30\% systematic uncertainty in the \SRHMhad region.
Uncertainties of the $b$-tagging efficiency and mistagging rates are subdominant for \oneLbb and \hadronic\ channels, and are estimated by varying the $\eta$-, \pt- and flavor-dependent scale factors applied to each jet in the simulation within a range that reflects the systematic uncertainty of the measured tagging efficiency and mistagging rates.
The effects of experimental uncertainty in the \oneLgamgam channel are dominated by uncertainties in the photon, lepton and jet energy scale and resolution.
The uncertainty on the contribution from non-peaking background is dominated by the effect of the limited number of events in the \mgg sidebands. An additional contribution from the uncertainty in the shape of the non-peaking background \mgg distribution was found to be negligible.
The \twoL/\threeL channels are dominated in several signal regions by experimental systematic uncertainties related to the estimation of background contributions due to FNP leptons. These systematic uncertainties are evaluated with various studies including $Z \to \ell\ell$ efficiency comparisons, variations of kinematic selections, modifications to the definition of the control regions, and alternative trigger selections.
For the \twoL channel, these are the dominant uncertainties and have similar contributions from each source.

The dominant systematic uncertainties in the various signal regions are summarized in \Tab{\ref{tab:systematics-all}}.
 
\begin{table}
\begin{center}
\setlength{\tabcolsep}{0.0pc}
 
\footnotesize{
 
\begin{tabular*}{\textwidth}{@{\extracolsep{\fill}}lcc}
\toprule
\textbf{\hadronic channel} & &              \\
\midrule
\toprule
Uncertainty of region                                  & \SRHMhad     & \SRLMhad    \\
\midrule
Total background expectation                           & $2.5$       & $8$      \\
\midrule
Total background uncertainty                           & $\pm 1.3$   & $\pm 4$  \\
\midrule
Systematic, experimental                               & $\pm 0.9$   & $\pm 1.2$  \\
Systematic, theoretical                                & $\pm 0.7$   & $\pm 3$  \\
Statistical, MC samples                                & $\pm 0.5$   & $\pm 0.8$  \\
Statistical, $\mu_{\mathrm{TT,ST,Zj}}$ scale-factors   & $\pm 0.25$   & $\pm 0.5$   \\
\bottomrule
\end{tabular*}
 
\begin{tabular*}{\textwidth}{@{\extracolsep{\fill}}lccc}
\toprule
\textbf{\oneLbb channel}   & &           \\
\midrule
\toprule
Uncertainty of region                                 & \SRlbbLow   & \SRlbbMedium  & \SRlbbHigh \\
\midrule
Total background expectation                           & $5.7$      & $2.8$         & $4.6$      \\
\midrule
Total background uncertainty                           & $\pm 2.3$  & $\pm 1.0$     & $\pm 1.2$  \\
\midrule
Systematic, experimental                               & $\pm 1.3$  & $\pm 0.7$    & $\pm 0.6$ \\
Systematic, theoretical                                & $\pm 2.2$  & $\pm 0.9$    & $\pm 0.7$ \\
Statistical, MC samples                                & $\pm 1.1$  & $\pm 0.5$    & $\pm 0.6$ \\
Statistical, $\mu_{\mathrm{TT,ST,Wj}}$ scale-factors   & $\pm 0.8$ & $\pm 0.6$    & $\pm 1.3$  \\
\bottomrule
\end{tabular*}
 
\begin{tabular*}{\textwidth}{@{\extracolsep{\fill}}lcc}
\toprule
\textbf{\lgg channel} & &              \\
\midrule
\toprule
Uncertainty of region                                  & \SRlyyA              & \SRlyyB       \\
\midrule
Total background expectation                           & $0.36$               & $5.3$        \\
\midrule
Total background uncertainty                           & $\pm 0.22$           & $\pm 1.0$    \\
\midrule
Systematic, experimental                               & $\pm 0.018$           & $\pm 0.27$    \\
Systematic, theoretical                                & $\pm 0.008$           & $\pm 0.11$    \\
Statistical, MC samples                                & $\pm 0.006$           & $\pm 0.024$    \\
Statistical, non-peaking                               & $\pm 0.22$           & $\pm 0.9$    \\
\bottomrule
\end{tabular*}
 
\begin{tabular*}{\textwidth}{@{\extracolsep{\fill}}lcc}
\toprule
\textbf{  \twoL channel} & &              \\
\midrule
\toprule
Uncertainty of region                                  & \SRSSJone    & \SRSSJtwothr   \\
\midrule
Total background expectation                           & $6.7$        & $5.3$          \\
\midrule
Total background uncertainty                           & $\pm 2.2$    & $\pm 1.6$      \\
\midrule
Systematic, experimental                               & $\pm 2.1$    & $\pm 1.6$.     \\
Systematic, theoretical                                & $\pm 0.21$   & $\pm 0.28$     \\
Statistical, MC samples                                & $\pm 0.4$   & $\pm 0.34$      \\
\bottomrule
\end{tabular*}
 
\begin{tabular*}{\textwidth}{@{\extracolsep{\fill}}lcccc}
\toprule
\textbf{  \threeL channel} & &   & &           \\
\midrule
\toprule
Uncertainty of region           & \SRThrLDFOSJ   & \SRThrLDFOSJa(b)           & \SRThrLSFOSJa(b)            & \SRThrLSFOSJ \\
\midrule
Total background expectation    &  $2.05$        &  $8$($1.7$)            & $3.8$($2.37$)              & $11.4$     \\
\noalign{\smallskip}\hline\noalign{\smallskip}
 
Total background uncertainty    & $\pm 0.98$     & $\pm 4$ ($\pm 0.7$)    &  $\pm 1.7$ ($\pm 0.96$)    & $\pm 2.6$  \\ \midrule
Systematic, experimental        & $\pm 0.8 $    & $\pm 4\ $ ($\pm 0.5 $) & $\pm 1.7$ ($\pm 0.8 $)    & $\pm 2.0$  \\
Systematic, theoretical         & $\pm 0.11$     & $\pm 0.25$ ($\pm 0.16$)    & $\pm 0.15$ ($\pm 0.22$)     & $\pm 1.5$  \\
Statistical, MC samples         & $\pm 0.6$     & $\pm 1.2$ ($\pm 0.4$)    & $\pm 0.6$ ($\pm 0.4$)     & $\pm 0.9$  \\
Statistical, $\mu_{\mathrm{WZ}}$ scale-factors   & $\pm 0.022$ & $\pm 0.12(\pm 0.06)$ & $\pm 0.30(\pm 0.24)$ & $\pm 0.9$  \\
\bottomrule
\end{tabular*}
}
\end{center}
\caption[Breakdown of uncertainty on background estimates]{
Dominant systematic uncertainties in the background estimates in the various signal regions, expressed in terms of number of events.
Individual uncertainties can be correlated, and do not necessarily add up quadratically to
the total background uncertainty.
For the \threeL channel, numbers in parentheses indicate the results for the (b) signal region in each case.
\label{tab:systematics-all}}
\end{table}
 
\section{Results}
\label{sec:results}
 
No significant differences between the observed and expected yields are found in the search regions for each of the analysis channels considered.
The results are translated into upper limits on contributions from physics processes beyond the SM (BSM) for each signal region and are used to set exclusion limits at the 95\% confidence level (CL) on the common mass of the charginos and next-to-lightest neutralinos for various values of the LSP mass in the simplified model considered in the analysis.

\Tab{\ref{tab:yields-0Lbb}} provides the event yields and SM expectation for the \hadronic analysis channel in the two signal regions (\SRHMhad, \SRLMhad) after the background-only fit. The errors shown are the statistical plus systematic uncertainties. \Tab{\ref{tab:yields-SR-1Lbb}} reports the observed number of events in the three SRs for the \oneLbb signature compared to the SM expectations. Good agreement is found between data and SM predictions for both \hadronic signal regions and two of the three \oneLbb signal regions; \SRlbbMedium exhibits a mild excess.
For the \oneLgamgam channel, the expected SM backgrounds, broken down by source, are summarized along
with their estimated uncertainties in \Tab{\ref{tab:lyy_sr_content}}.
A mild excess of observed events relative to expected SM backgrounds is seen in each of the two signal regions, corresponding to $p_0$-values of 0.027 and 0.087 for \SRlyyA and \SRlyyB, respectively. Finally, Tables~\ref{tab:yields-2LSS}, \ref{tab:SR3L-SFOS} and~\ref{tab:SR3L-DFOS} report the observed and predicted SM backgrounds for the various multilepton signal regions.
 
\begin{table}[!ht]
\centering
\begin{tabular*}{0.8\textwidth}{ @{\extracolsep{\fill}} l r r }
\hline \hline
\textbf{SR channels} & \SRHMhad & \SRLMhad \\
\hline
Observed events      & $1$                    & $7$                    \\
\hline
Fitted bkg events    & $2.5 \pm 1.3$        & $8 \pm 4$        \\
\hline
\ttbar               & $1.1 \pm 0.9$        & $4 \pm 4$        \\
Single top (\Wt)     & $0.15_{-0.15}^{+0.16}$ & $0.44 \pm 0.33$        \\
\Wjets               & $0.1_{-0.1}^{+0.3}$ & $1.0 \pm 0.7$        \\
\Zjets               & $1.0 \pm 0.7$        & $1.7 \pm 1.0$        \\
\ttV                 & $0.09 \pm 0.03$        & $0.40 \pm 0.08$        \\
Diboson              & $<0.01$        & $0.3_{-0.3}^{+0.4}$ \\[0.2cm]
\hline \hline
\end{tabular*}
\caption{\label{tab:yields-0Lbb} Event yields and SM expectation for the \hadronic channel after the background-only fit for the \SRHMhad and \SRLMhad regions. The errors shown are the statistical plus systematic uncertainties. Uncertainties in the fitted yields are symmetric by construction, where the negative error is truncated when reaching zero event yield.}
\end{table}
\begin{table}
\begin{center}
\setlength{\tabcolsep}{0.0pc}
{
\begin{tabular*}{0.8\textwidth}{@{\extracolsep{\fill}}lrrr}
\noalign{\smallskip}\hline \hline\noalign{\smallskip}
\textbf{SR channels}           & \SRlbbLow            & \SRlbbMedium            & \SRlbbHigh              \\ \hline
Observed events          & $6$              & $7$              & $5$                    \\
\noalign{\smallskip}\hline\noalign{\smallskip}
Fitted bkg events         & $5.7 \pm 2.3$          & $2.8 \pm 1.0$          & $4.6 \pm 1.2$              \\
\noalign{\smallskip}\hline\noalign{\smallskip}
 
\ttbar         & $3.4 \pm 2.9$          & $1.4 \pm 1.0$          & $1.1 \pm 0.6$              \\
Single top (\Wt)         & $1.4_{-1.4}^{+1.4}$          & $0.8_{-0.8}^{+0.9}$          & $1.2 \pm 1.1$              \\
\Wjets         & $0.6 \pm 0.4$          & $0.20 \pm 0.11$          & $1.6 \pm 0.6$              \\
\ttV            & $0.10 \pm 0.04$          & $0.32 \pm 0.09$          & $0.54 \pm 0.14$              \\
Diboson         & $0.12_{-0.12}^{+0.15}$          & $0.05 \pm 0.03$          & $0.08 \pm 0.02$              \\
Others            & $0.10 \pm 0.05$ & $0.03 \pm 0.01$          & $0.04 \pm 0.02$              \\
\noalign{\smallskip}\hline\hline\noalign{\smallskip}
\end{tabular*}
}
\end{center}
 
\caption{Event yields and SM expectation after the background-only fit
in the \oneLbb channel for the \SRlbbLow, \SRlbbMedium, and \SRlbbHigh regions.
The category ``Others'' includes contributions from three- and four-top production and SM Higgs processes. The errors shown are the statistical plus systematic uncertainties. Uncertainties in the fitted yields are symmetric by construction,
where the negative error is truncated when reaching zero event yield.}
\label{tab:yields-SR-1Lbb}
\end{table}

\begin{table}[!htpb]
\centering
 
\begin{tabular*}{0.8\textwidth}{ @{\extracolsep{\fill}} l r r }
\hline \hline
\textbf{SR channels}        & \SRlyyA          & \SRlyyB \\
\hline
Observed events               &       2          &     9           \\
\hline
Total bkg events              &  $0.37 \pm 0.22$ & $5.3 \pm 1.0$ \\
\hline
\Wh background                &  $0.09 \pm 0.01$ & $1.8 \pm 0.3$ \\
Other peaking backgrounds     &  $0.04 \pm 0.01$ & $0.19 \pm 0.02$ \\
Non-peaking background        &  $0.22 \pm 0.22$ & $3.3 \pm 0.9$ \\
 
\hline \hline
\end{tabular*}
\caption{Expected numbers of peaking and non-peaking SM background
events in the \oneLgamgam channel for \SRlyyA and \SRlyyB.
Non-peaking-background uncertainty is dominated by the statistical uncertainty
in the sideband fits.
The peaking background uncertainties include both theoretical (production rate)
and experimental (detector effect) contributions, as described in the text.
The uncertainties in the \Wh and \textit{Other peaking} backgrounds are
taken to be fully correlated.
Also shown are the observed numbers of events in \SRlyyA and \SRlyyB.
}
\label{tab:lyy_sr_content}
\end{table}
\begin{table}
\begin{center}
\setlength{\tabcolsep}{0.0pc}
{
\begin{tabular*}{0.8\textwidth}{@{\extracolsep{\fill}}lrr}
\noalign{\smallskip}\hline\hline\noalign{\smallskip}
\textbf{SR channels}                & \SRSSJone                & \SRSSJtwothr        \\
\noalign{\smallskip}\hline\noalign{\smallskip}
Observed events                     & $2$                      & $8$              \\
\noalign{\smallskip}\hline\noalign{\smallskip}
Fitted bkg events                   & $6.7 \pm 2.2$          & $5.3 \pm 1.6$          \\
\noalign{\smallskip}\hline\noalign{\smallskip}
FNP events                  & $3.3 \pm 2.1$          & $1.8 \pm 1.5$          \\
$WZ$                        & $2.2 \pm 0.5$          & $1.9 \pm 0.6$        \\
Rare                        & $0.44 \pm 0.13$          & $0.73 \pm 0.17$        \\
\ttV                        & $0.12 \pm 0.05$          & $0.14 \pm 0.05$        \\
$WW$                        & $0.17 \pm 0.03$          & $0.51 \pm 0.07$        \\
$ZZ$                        & $0.06 \pm 0.03$          & $0.07 \pm 0.04$        \\
Charge-flip events          & $0.47 \pm 0.07$          & $0.27 \pm 0.03$        \\
\noalign{\smallskip}\hline\hline\noalign{\smallskip}
\end{tabular*}
}
\end{center}
\caption{Event yields and SM expectation for the \twoL~signal regions \SRSSJone and  \SRSSJtwothr after the background-only fit. The category `Rare'' includes contributions from triboson, three- and four-top production and SM Higgs processes.
The errors shown are the statistical plus systematic uncertainties.}
\label{tab:yields-2LSS}
\end{table}
\begin{table}
\begin{center}
\setlength{\tabcolsep}{0.0pc}
{
\begin{tabular*}{0.8\textwidth}{@{\extracolsep{\fill}}lrrr}
\noalign{\smallskip}\hline\hline\noalign{\smallskip}
\textbf{SR channels}                    & \SRThrLSFOSJa            & \SRThrLSFOSJb            & \SRThrLSFOSJ              \\
\noalign{\smallskip}\hline\noalign{\smallskip}
Observed events           & $0$              & $3$              & $11$                    \\
\noalign{\smallskip}\hline\noalign{\smallskip}
Fitted bkg events              & $3.8 \pm 1.7$          & $2.4 \pm 1.0$          & $11.5 \pm 2.6$              \\
\noalign{\smallskip}\hline\noalign{\smallskip}
$WZ$              & $2.5 \pm 1.2$          & $2.0 \pm 0.9$          & $7.4 \pm 2.3$              \\
$ZZ$                 & $0.10 \pm 0.04$          & $0.07 \pm 0.02$          & $ 0.29 \pm 0.09 $           \\
\ttV             & $0.09 \pm 0.03$          & $0.02 \pm 0.01$          & $1.9 \pm 0.5$              \\
Tribosons                & $0.57 \pm 0.29$          & $0.16 \pm 0.08$          & $1.4 \pm 0.4$              \\
Higgs SM               & $0.24_{-0.24}^{+0.25}$          &  $0.07 \pm 0.07$         & $0.07 \pm 0.04$              \\
FNP events                & $0.27_{-0.27}^{+0.31}$          & $0.11_{-0.11}^{+0.20}$          & $0.4_{-0.4}^{+0.5}$              \\
\noalign{\smallskip}\hline\hline\noalign{\smallskip}
\end{tabular*}
}
\end{center}
\caption{Event yields and SM expectation after the background-only fit
in the \threeL channel for the \SRThrLSFOSJa, \SRThrLSFOSJb and \SRThrLSFOSJ regions.
The category ``Higgs'' includes contributions from $\ttbar$+Higgs boson production.
The errors shown are the statistical plus systematic uncertainties.
Uncertainties in the fitted yields are symmetric by construction,
where the negative error is truncated when reaching zero event yield.}
\label{tab:SR3L-SFOS}
\end{table}
 
\begin{table}
\begin{center}
\setlength{\tabcolsep}{0.0pc}
{
\begin{tabular*}{0.8\textwidth}{@{\extracolsep{\fill}}lrrrr}
\noalign{\smallskip}\hline\hline\noalign{\smallskip}
\textbf{SR channels}                  & \SRThrLDFOSJ            & \SRThrLDFOSJa            & \SRThrLDFOSJb              \\
\noalign{\smallskip}\hline\noalign{\smallskip}
Observed events                 & $0$              & $7$              & $1$                    \\
\noalign{\smallskip}\hline\noalign{\smallskip}
Fitted bkg events                & $2.1 \pm 1.0$          & $8.3 \pm 3.8$          & $1.7 \pm 0.7$              \\
\noalign{\smallskip}\hline\noalign{\smallskip}
$WZ$              & $0.18 \pm 0.13$          & $1.01 \pm 0.27$          & $0.54 \pm 0.16$              \\
$ZZ$              &  $0.0017\pm 0.0012$          & $0.06 \pm 0.02$          & $0.03 \pm 0.01$              \\
\ttV                & $0.0013\pm 0.0013$          & $0.79 \pm 0.29$          & $0.43 \pm 0.16$              \\
Tribosons              & $0.52 \pm 0.28$          & $0.66 \pm 0.22$          & $0.23 \pm 0.08$              \\
Higgs SM            & $0.39 \pm 0.15$          & $0.1_{-0.1}^{+0.5}$          & $0.05 \pm 0.04$              \\
FNP                  & $1.0 \pm 0.9$          & $5.6 \pm 3.8$          & $0.4_{-0.4}^{+0.6}$              \\
\noalign{\smallskip}\hline\hline\noalign{\smallskip}
\end{tabular*}
}
\end{center}
\caption{Event yields and SM expectation after the background-only fit
in the \threeL channel for the \SRThrLDFOSJ, \SRThrLDFOSJa and \SRThrLDFOSJb regions.
The category ``Higgs'' includes contributions from $\ttbar$+Higgs boson production.
The errors shown are the statistical plus systematic uncertainties.
Uncertainties in the fitted yields are symmetric by construction,
where the negative error is truncated when reaching zero event yield.}
\label{tab:SR3L-DFOS}
\end{table}
 
\Tab{\ref{tab:upper}} summarizes the observed ($S^{95}_\mathrm{obs}$) and expected ($S^{95}_\mathrm{exp}$) 95\% CL upper limits on the number of signal events and on the observed visible cross-section, $\sigma^{}_\textrm{vis}$, for all channels and signal regions.
Upper limits on contributions from new physics processes are estimated using the so-called model-independent fit.
The CL$_\mathrm{s}$~method~\cite{Junk:1999kv,Read:2002hq} is used to derive the confidence level
of the exclusion for a particular signal model; signal models with a CL$_\mathrm{s}$~value below 0.05 are excluded at 95\% CL.
When normalized to the integrated luminosity of the data sample, results can be interpreted as corresponding to observed upper limits on $\sigma^{}_\textrm{vis}$, defined as the product of the production cross-section, the acceptance and the selection efficiency of a BSM signal.
The $p_0$-values, which represent the probability of the SM background alone to fluctuate to the observed number of events or higher, are also provided.

\begin{table}[!ht]
\begin{center}\renewcommand\arraystretch{1.3}
\begin{tabular*}{\textwidth}{@{\extracolsep{\fill}}lcccc}
\noalign{\smallskip}\hline\noalign{\smallskip}
\textbf{}       & $\sigma^{}_\textrm{vis}$ [fb] & $S^{95}_\mathrm{obs}$ & $S^{95}_\mathrm{exp}$ & $p_0$-value \\ \hline
 
\SRLMhad        & $0.26$ &  $9.4$ & $ { 9.5 }^{ +3.3 }_{ -1.9 }$ & $ 0.50$ \\
\SRHMhad        & $0.10$ &  $3.6$ & $ { 4.3 }^{ +1.6 }_{ -1.0 }$ & $ 0.50$ \\ \hline
 
\SRlbbLow       & $0.23$ & $8.3$  &  $8.0^{ +3.3 }_{ -2.2 }$     & $0.46$ \\
\SRlbbMedium    & $0.28$ & $10.0$ & $ { 5.6 }^{ +2.9 }_{ -1.7 }$ 
& $0.04$ \\
\SRlbbHigh      & $0.18$ & $6.4$  &  $6.1^{ +3.1 }_{ -1.9 }$     & $0.44$ \\ \hline
 
\SRlyyA         & $0.15$ &  $5.5$ &  $3.2 ^{+0.9}_{-0.1}$        & $0.03$  \\
\SRlyyB         & $0.28$ & $10.1$ &  $6.4 ^{+2.6}_{-1.6}$        & $0.09$  \\ \hline
 
\SRSSJone       & $0.12$ &  $4.2$ & $ { 6.1 }^{ +2.7 }_{ -1.5 }$ & $ 0.50$ \\
\SRSSJtwothr    & $0.27$ &  $9.9$ & $ { 6.6 }^{ +3.4 }_{ -1.1 }$ & $ 0.17$ \\ \hline
 
\SRThrLSFOSJa   & $0.08$ &  $3.0$ & $ { 4.4 }^{ +1.9 }_{ -1.3 }$ & $ 0.47$ \\
\SRThrLSFOSJb   & $0.16$ &  $5.9$ & $ { 5.0 }^{ +2.0 }_{ -1.2 }$ & $ 0.35$ \\
\SRThrLSFOSJ    & $0.26$ &  $9.2$ & $ { 9.4 }^{ +3.8 }_{ -2.5 }$ & $ 0.50$ \\

\SRThrLDFOSJ    & $0.08$ &  $3.0$ & $ { 3.8 }^{ +1.4 }_{ -0.9 }$ & $ 0.43$ \\
\SRThrLDFOSJa   & $0.25$ &  $9.0$ & $ { 9.2 }^{ +3.3 }_{ -2.0 }$ & $ 0.50$ \\
\SRThrLDFOSJb    & $0.10$ &  $3.7$ & $ { 4.0 }^{ +1.6 }_{ -0.5 }$ & $ 0.50$ \\
\hline
 
\end{tabular*}
\end{center}
\caption{ From left to right, the observed 95\% CL upper limits on the visible cross-sections $\sigma^{}_\textrm{vis}$,
the observed ($S^{95}_\mathrm{obs}$) and expected ($S^{95}_\mathrm{exp}$) 95\% CL upper limits on the number of
signal events with $\pm1 \sigma$ excursions of the expectation, and the discovery $p$-value ($p_0$),
truncated at 0.5. }
\label{tab:upper}
\end{table}
 
For the \hadronic analysis channel, \Fig{\ref{fig:had-results-sr}} shows the distributions of \met and \mbb in the \SRHMhad and \SRLMhad SRs, respectively.
\Fig{\ref{fig:SR1lbb}} shows the data distributions of \mctbb~and \met for the \oneLbb analysis in the \SRlbbHigh and \SRlbbMedium SRs compared to the SM expectations.
\Fig{\ref{fig:lgg_myy}} shows the \mgg distribution, separately for \SRlyyA and \SRlyyB, before the final selection applied to \mgg.
Observed and predicted distributions of \mljj (\SRSSJone) and \mttwo (\SRSSJtwothr) for the \twoL signature are shown in \Fig{\ref{fig:twoL-results}}.
The data agree well with the SM expectations in all distributions and for all channels, and no significant deviations are observed.

\begin{figure}[htbp]
\centering
\begin{subfigure}[b]{0.45\textwidth}
\centering\includegraphics[width=\textwidth]{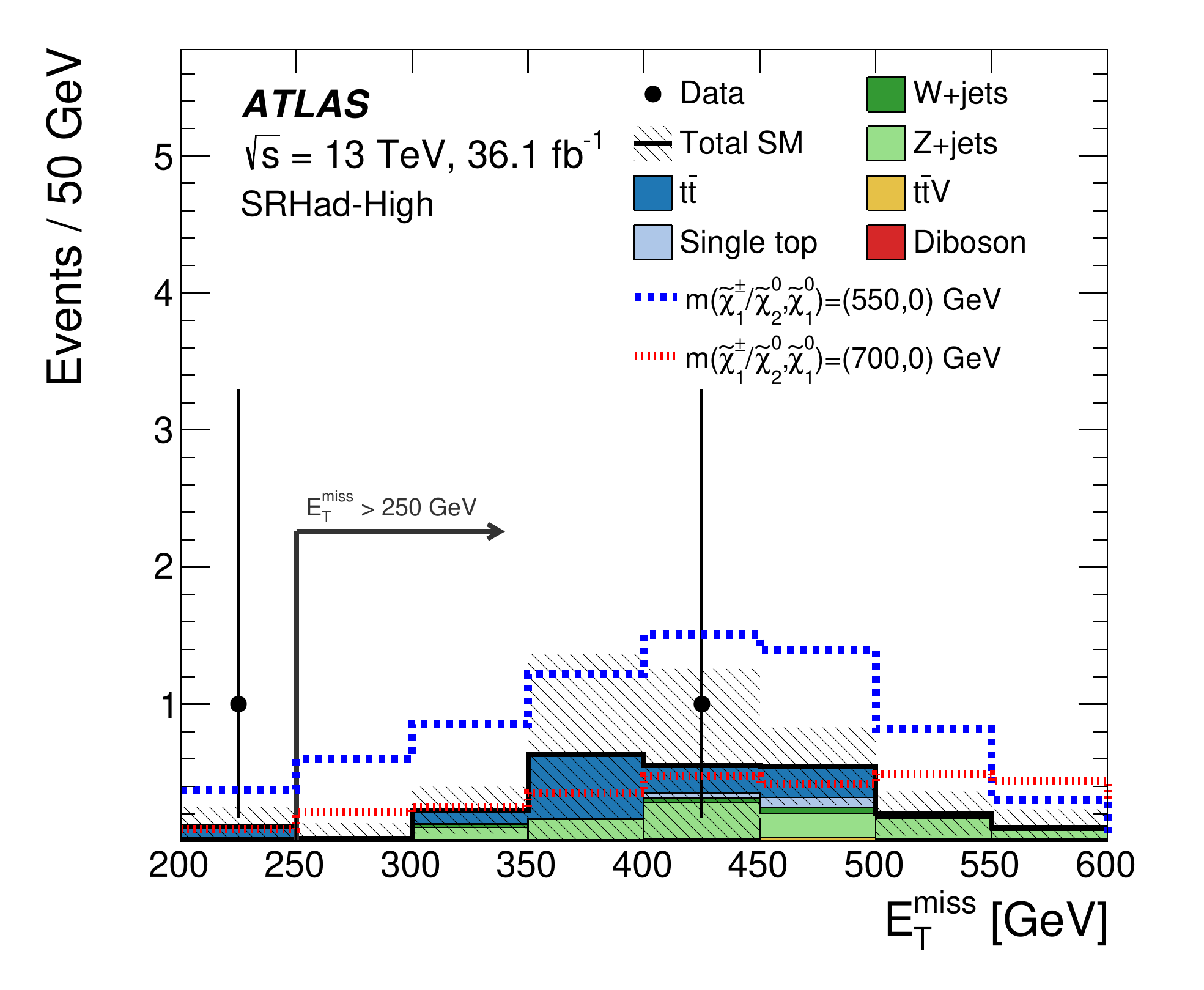}
\caption{\met in \SRHMhad}
\label{fig:had-SR-highmass-met}
\end{subfigure}
\begin{subfigure}[b]{0.45\textwidth}
\centering\includegraphics[width=\textwidth]{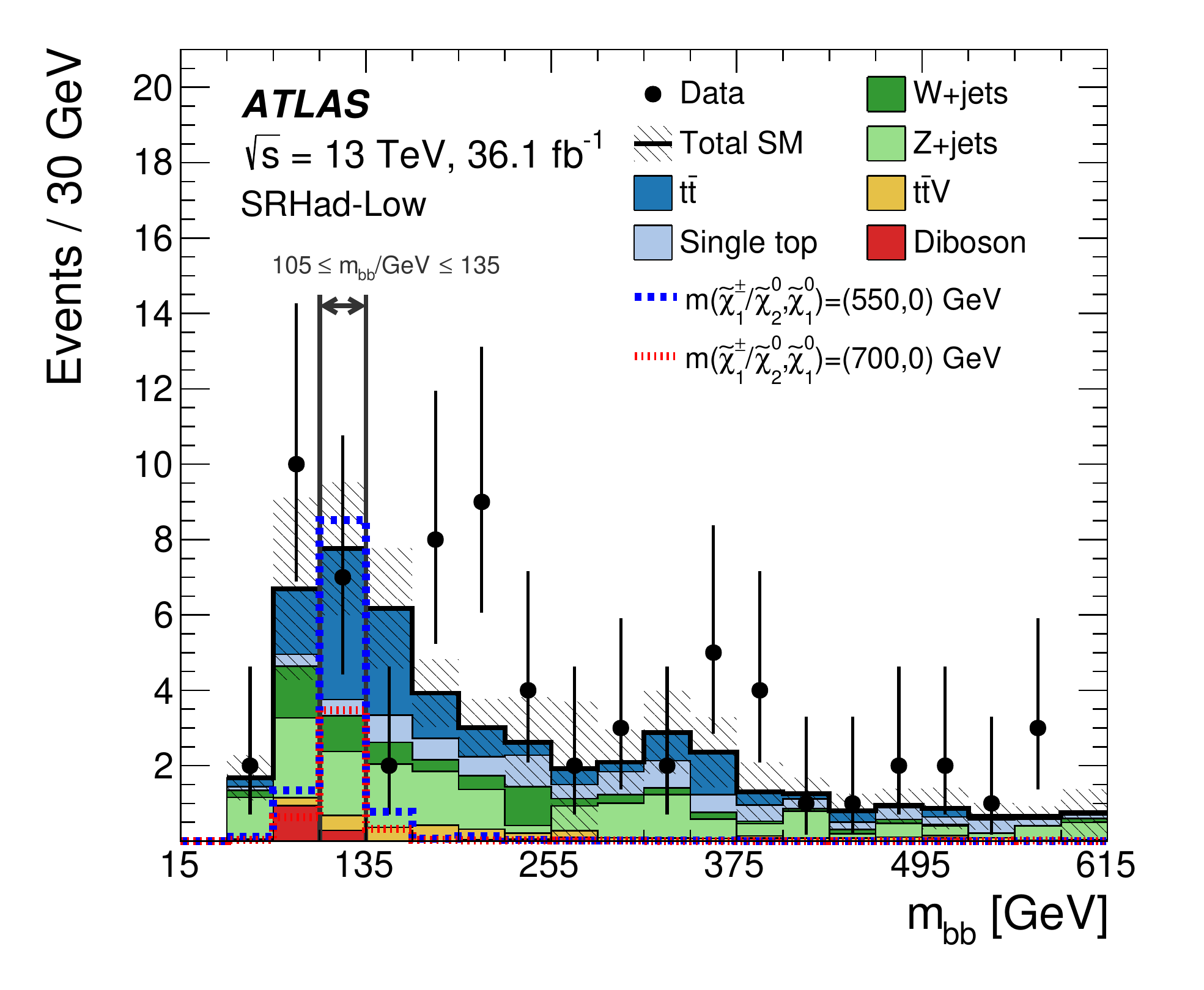}
\caption{\mbb in \SRLMhad}
\label{fig:had-SR-lowmass-mbb}
\end{subfigure}
\caption{Data and SM predictions in SRs for the \hadronic analysis for \subref{fig:had-SR-highmass-met} \met in \SRHMhad and \subref{fig:had-SR-lowmass-mbb} \mbb in \SRLMhad. All SRs selections but the one on the quantity shown are applied. All uncertainties are included in the uncertainty band. Two example SUSY models are superimposed for illustrative purposes. }
\label{fig:had-results-sr}
\end{figure}
 
\begin{figure}[!ht]
\centering
\begin{subfigure}{0.45\textwidth}
\centering\includegraphics[width=\textwidth]{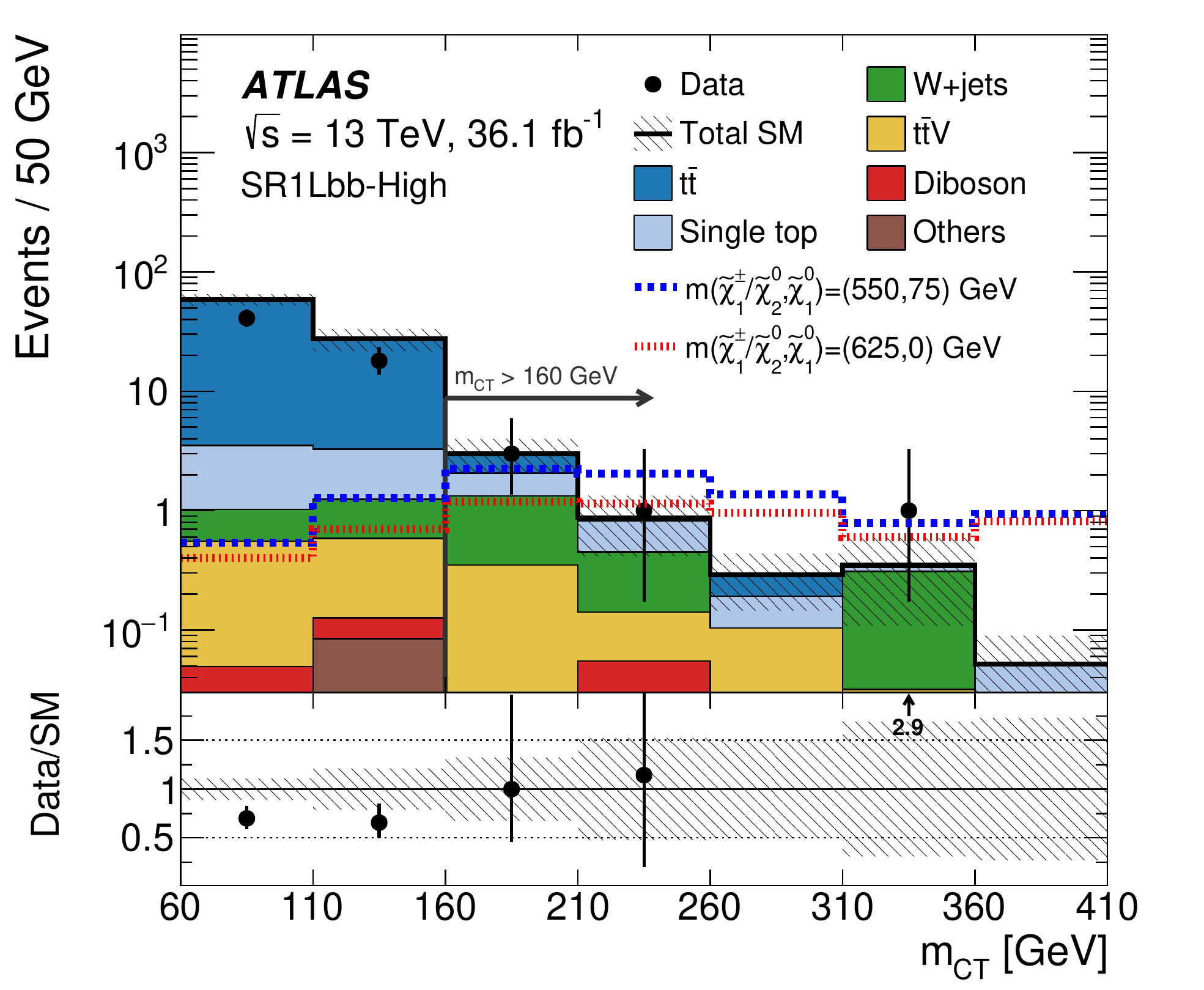}
\caption{\mCT in \SRlbbHigh}
\label{fig:SR1lbb-SRHM-mCT}
\end{subfigure}
\begin{subfigure}{0.45\textwidth}
\centering\includegraphics[width=\textwidth]{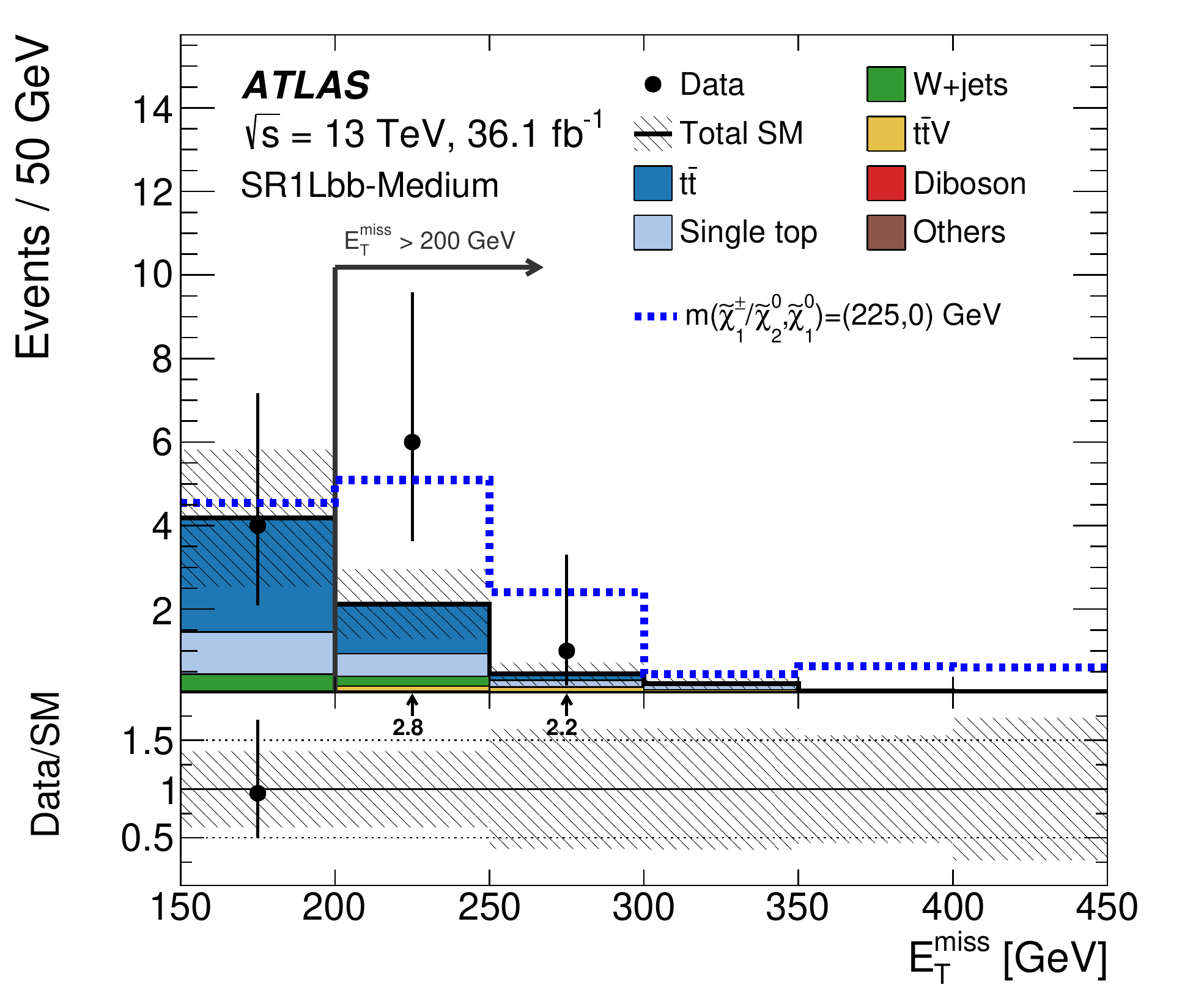}
\caption{\met in \SRlbbMedium}
\label{fig:SR1lbb-SRLM-met}
\end{subfigure}
 
\caption{Data and SM predictions in SRs for the \oneLbb analysis for \subref{fig:SR1lbb-SRHM-mCT} \mCT in \SRlbbHigh and \subref{fig:SR1lbb-SRLM-met} \met in \SRlbbMedium. All SRs selections but the one on the quantity shown are applied. All uncertainties are included in the uncertainty band. Example SUSY models are superimposed for illustrative purposes. }
\label{fig:SR1lbb}
\end{figure}
 
\begin{figure}[htbp]
\centering
\begin{subfigure}[b]{0.5\textwidth}
\centering\includegraphics[width=\textwidth]{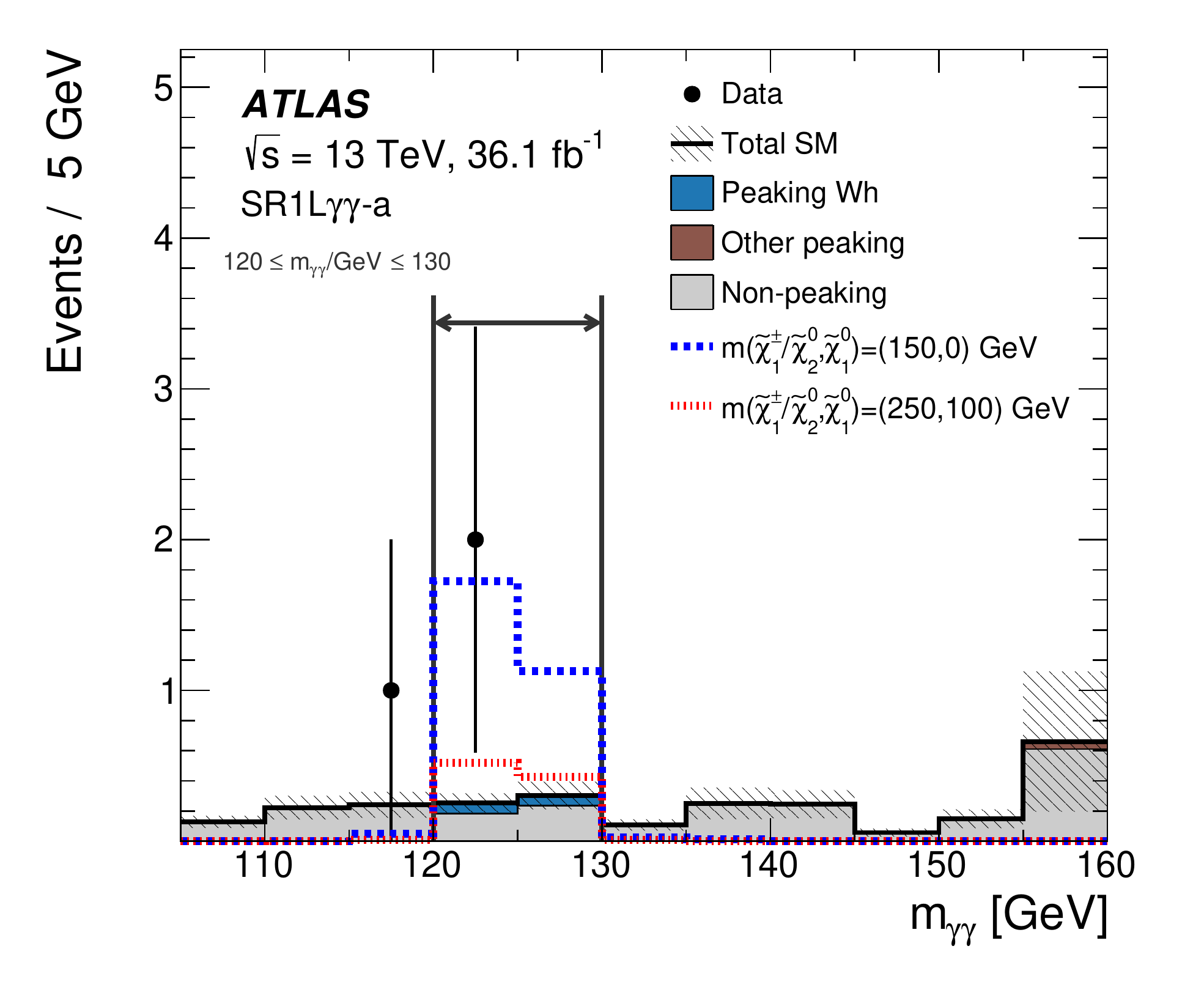}
\caption{\mgg in \SRlyyA}
\label{fig:lgg_myy_SRa}
\end{subfigure}
\begin{subfigure}[b]{0.5\textwidth}
\centering\includegraphics[width=\textwidth]{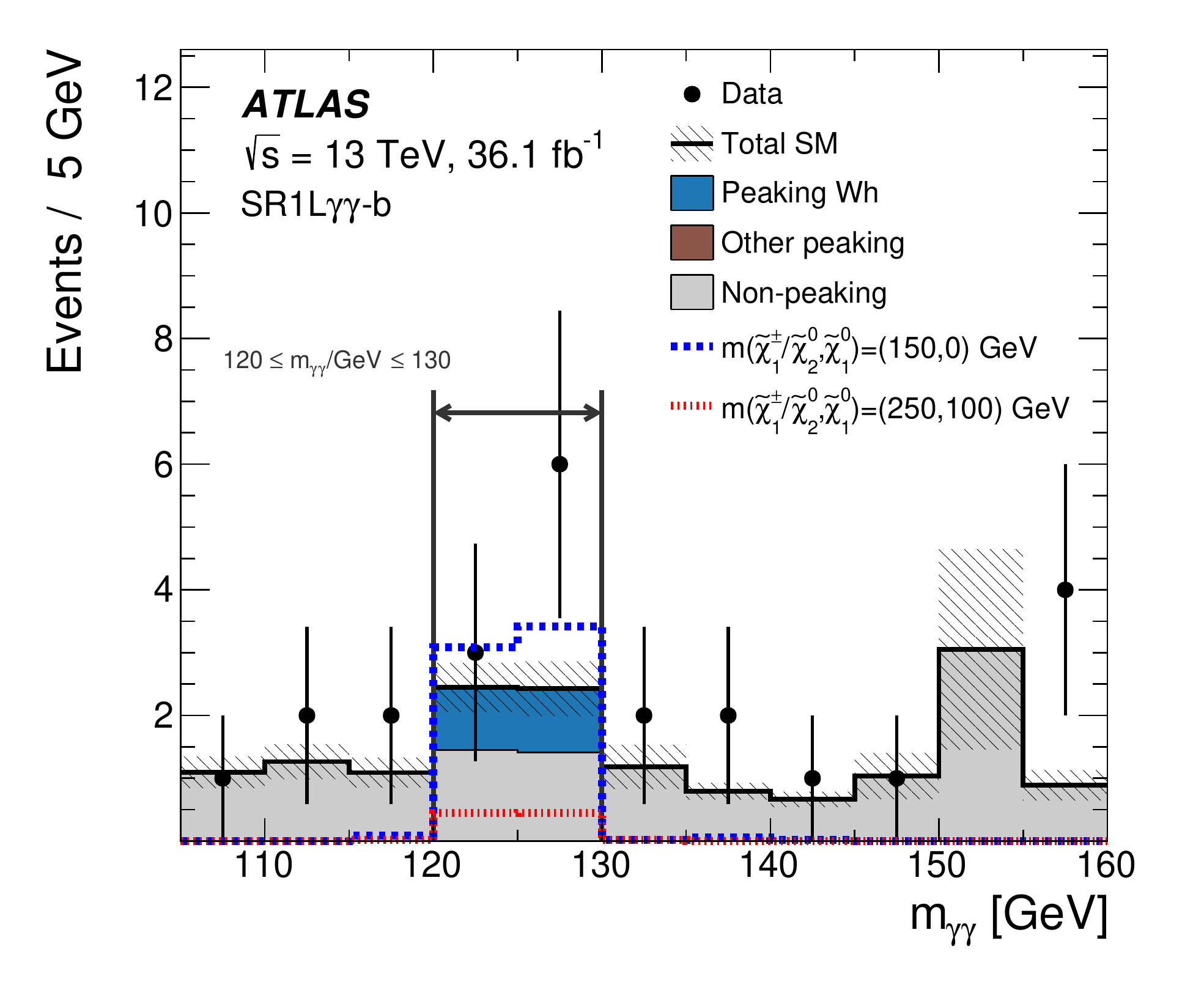}
\caption{\mgg in \SRlyyB}
\label{fig:lgg_myy_SRb}
\end{subfigure}
\caption{
Distributions of \mgg before the final requirement on \mgg
in \subref{fig:lgg_myy_SRa} \SRlyyA and \subref{fig:lgg_myy_SRb}
\SRlyyB. The expected contributions from both the peaking and non-peaking backgrounds are
shown as stacked colored histograms.
Two example SUSY models are superimposed for illustrative purposes.}
\label{fig:lgg_myy}
\end{figure}
 
\begin{figure}[tp]
\begin{subfigure}[b]{0.5\textwidth}
\centering\includegraphics[width=\textwidth]{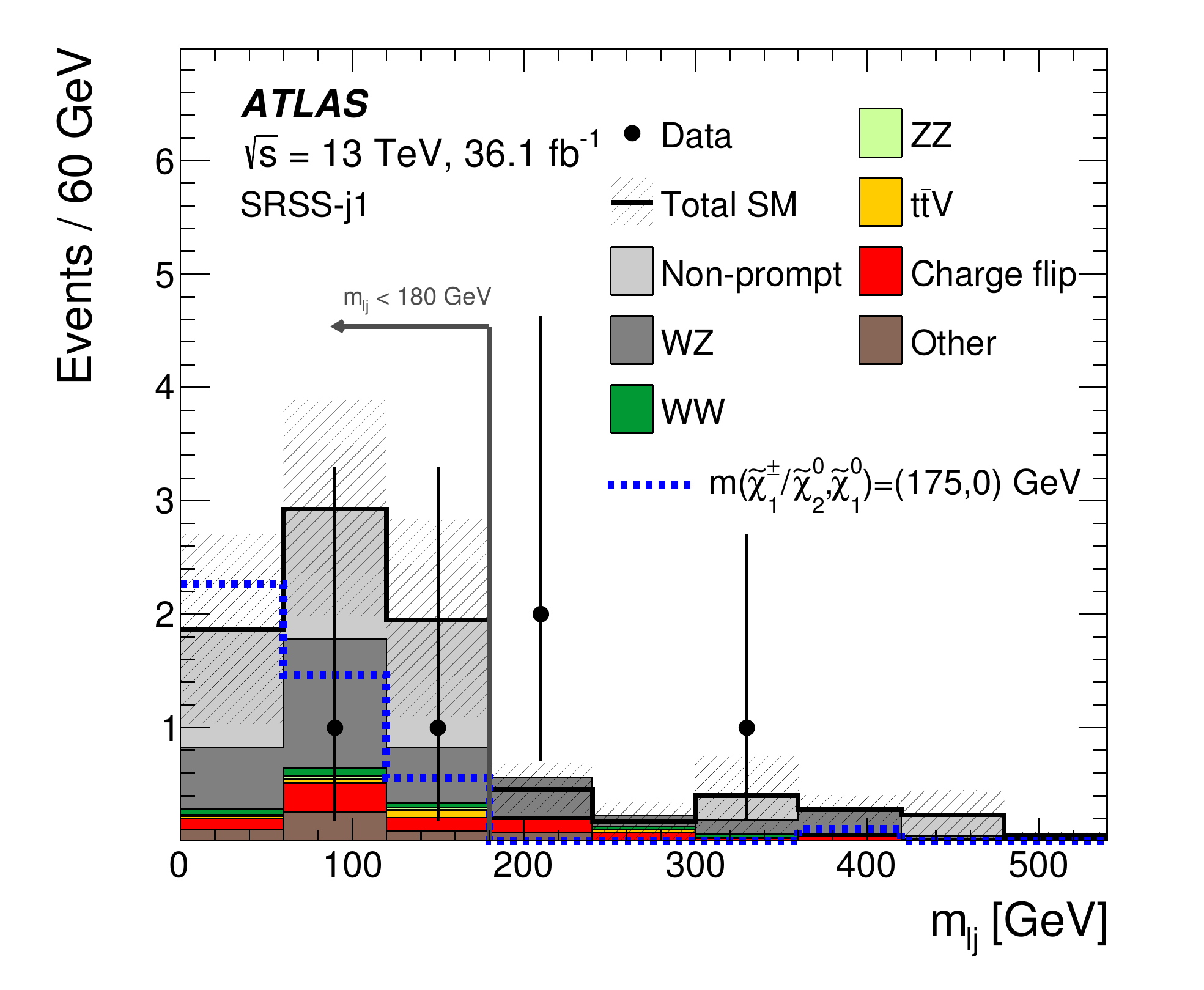}
\caption{\mljj in \SRSSJone}
\label{fig:twoL-results-mlj-SRjet1}
\end{subfigure}
\begin{subfigure}[b]{0.5\textwidth}
\centering\includegraphics[width=\textwidth]{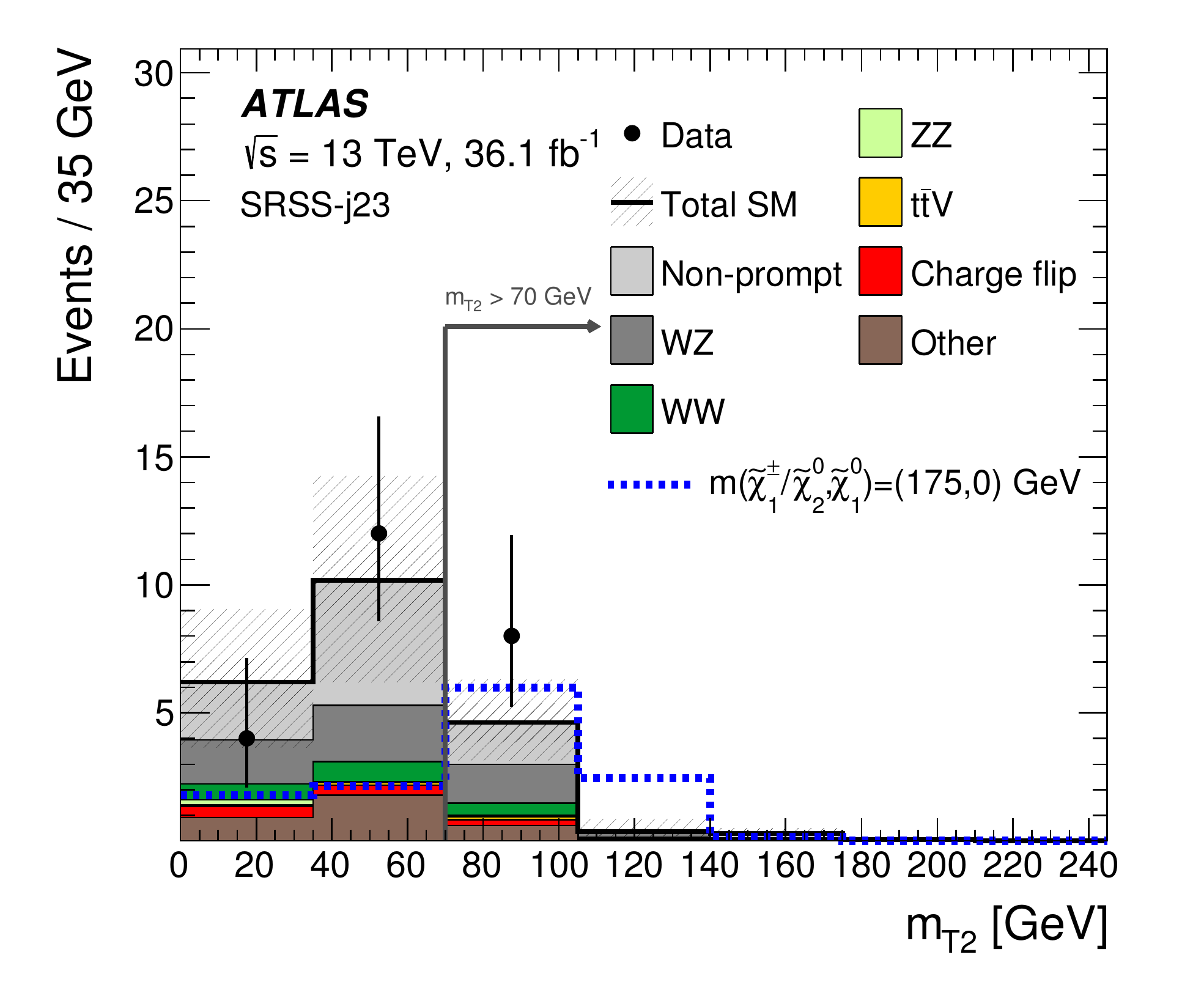}
\caption{\mttwo in \SRSSJtwothr}
\label{fig:twoL-results-mttwo-SRjet23}
\end{subfigure}
\caption{
Observed and predicted distributions for \subref{fig:twoL-results-mlj-SRjet1} \mljj
in \SRSSJone and \subref{fig:twoL-results-mttwo-SRjet23} \mttwo in \SRSSJtwothr.
All SRs selections but the one on the quantity shown are applied. All uncertainties are included in the uncertainty band. An example SUSY model is superimposed for illustrative purposes.}
\label{fig:twoL-results}
\end{figure}
 
\Fig{\ref{fig:results-0Lbb-exclusion}} shows the observed and expected exclusion contours for the simplified models shown in \Fig{\ref{fig:diagram-had}} for the \hadronic analysis channel.
The signal region (either \SRHMhad or \SRLMhad) used for each hypothesis for the $\chinoonepm/\ninotwo-\ninoone$ mass difference is chosen according to which has better expected sensitivity. Experimental and theoretical systematic uncertainties, as described in \Sect{\ref{sec:systematics}}, are applied to background and signal samples. \Fig{\ref{fig:results-1Lbb-exclusion}} shows the observed and expected exclusion contours obtained for the \oneLbb channel: in this case, a statistical combination of the results from the three signal regions is performed.
Due to the large branching ratio of the Higgs boson into $b$-quark pairs, the sensitivity of the \hadronic and \oneLbb channels is best at high masses of the chargino and next-to-lightest neutralinos, and exclusion limits up to 680 GeV are achieved for massless neutralinos.
 
\Fig{\ref{fig:results-1Lgammagamma-exclusion}} shows the expected limits obtained for the \oneLgamgam channel.
The excess of events observed in this signal region precludes an exclusion limit, even when combining the two SRs.
Exclusion limits for the \twoL analysis, obtained with a statistical combination of the two signal regions,
are shown in \Fig{\ref{fig:results-twoL-exclusion}}. This channel is primarily sensitive at low $\chinoonepm/\ninotwo$
mass values and slightly extends the observed exclusion for models with small mass difference between $\chinoonepm/\ninotwo$ and $\ninoone$.
Finally, the sensitivity of the \threeL channel is small compared to other analysis channels due in large part
to not considering hadronic $\tau$ decay modes. The observed and expected cross-section exclusion contours,
based on the statistical combination of the \threeL SRs, are compared with those of other
channels in \Fig{\ref{fig:results-alldiag-exclusion-1D}} and \Fig{\ref{fig:results-allmassless-exclusion-1D}}
as a function of the $\chinoonepm/\ninotwo$ masses for $m(\ninotwo) - m(\ninoone) = 130~\GeV$, and for
a fixed value of the $\ninoone$ mass, respectively.
 
A summary of the exclusion contours from the analyses presented here is shown in \Fig{\ref{fig:results-combo-exclusion}}. Observed and expected contours as obtained from each channel are shown, with the exception of the \threeL~analysis, which has no sensitivity. The overall expected sensitivity varies from $m(\chinoonepm/\ninotwo) = 150~\GeV$ to $m(\chinoonepm/\ninotwo) = 635~\GeV$, including significant improvements compared to previous results towards large $m(\ninoone)$ masses near the kinematic limit of the processes considered.
The gain in sensitivity is largely due to the increased center-of-mass energy and dataset size relative to Run~1,  the improvements in the optimization of the signal and control region definitions, as well as the addition of the \hadronic~analysis channel.
 
\begin{figure}[tp]
 
\begin{subfigure}[b]{0.5\textwidth}
\centering\includegraphics[width=\textwidth]{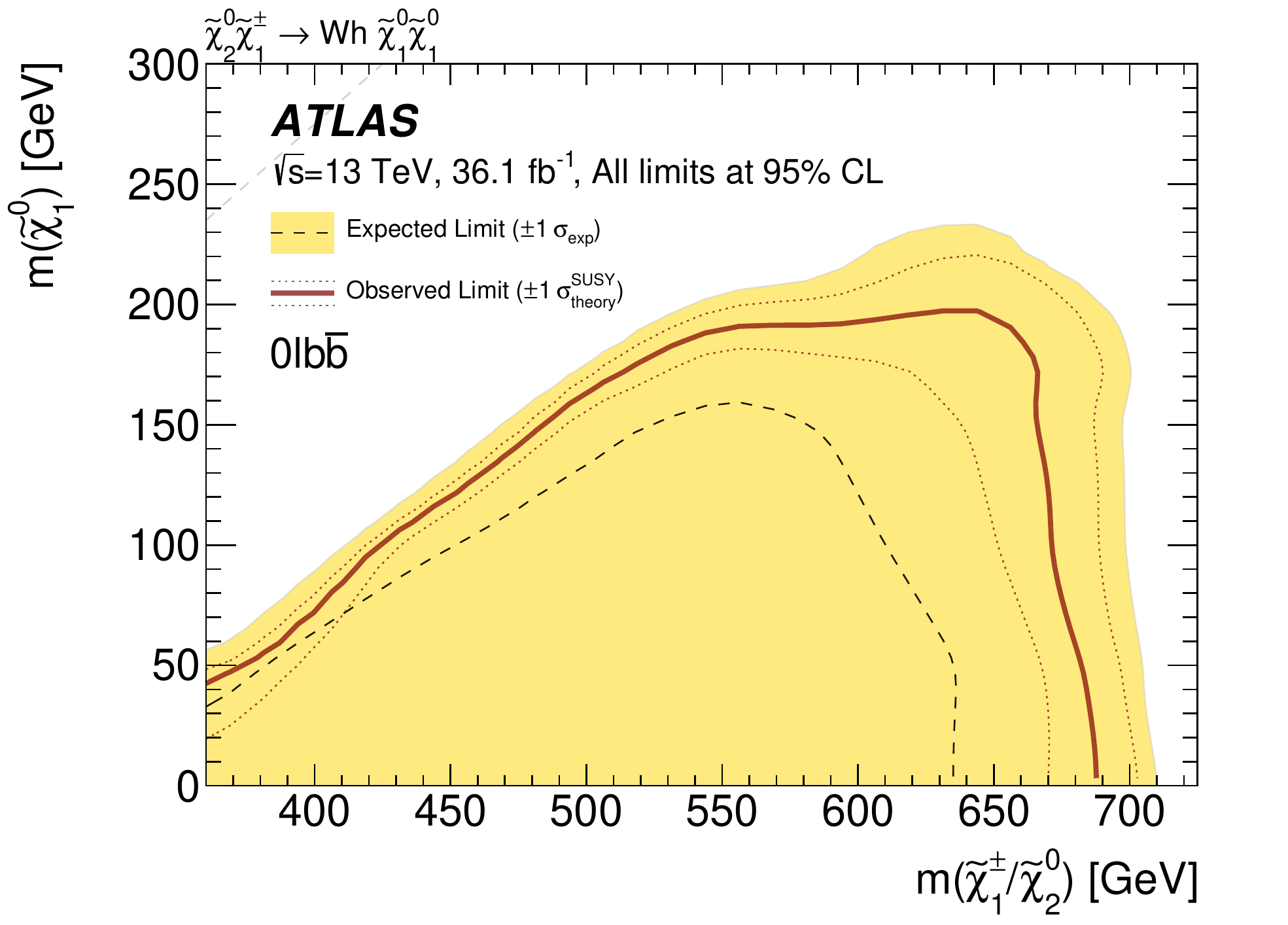}
\caption{\hadronic}
\label{fig:results-0Lbb-exclusion}
\end{subfigure}
\begin{subfigure}[b]{0.5\textwidth}
\centering\includegraphics[width=\textwidth]{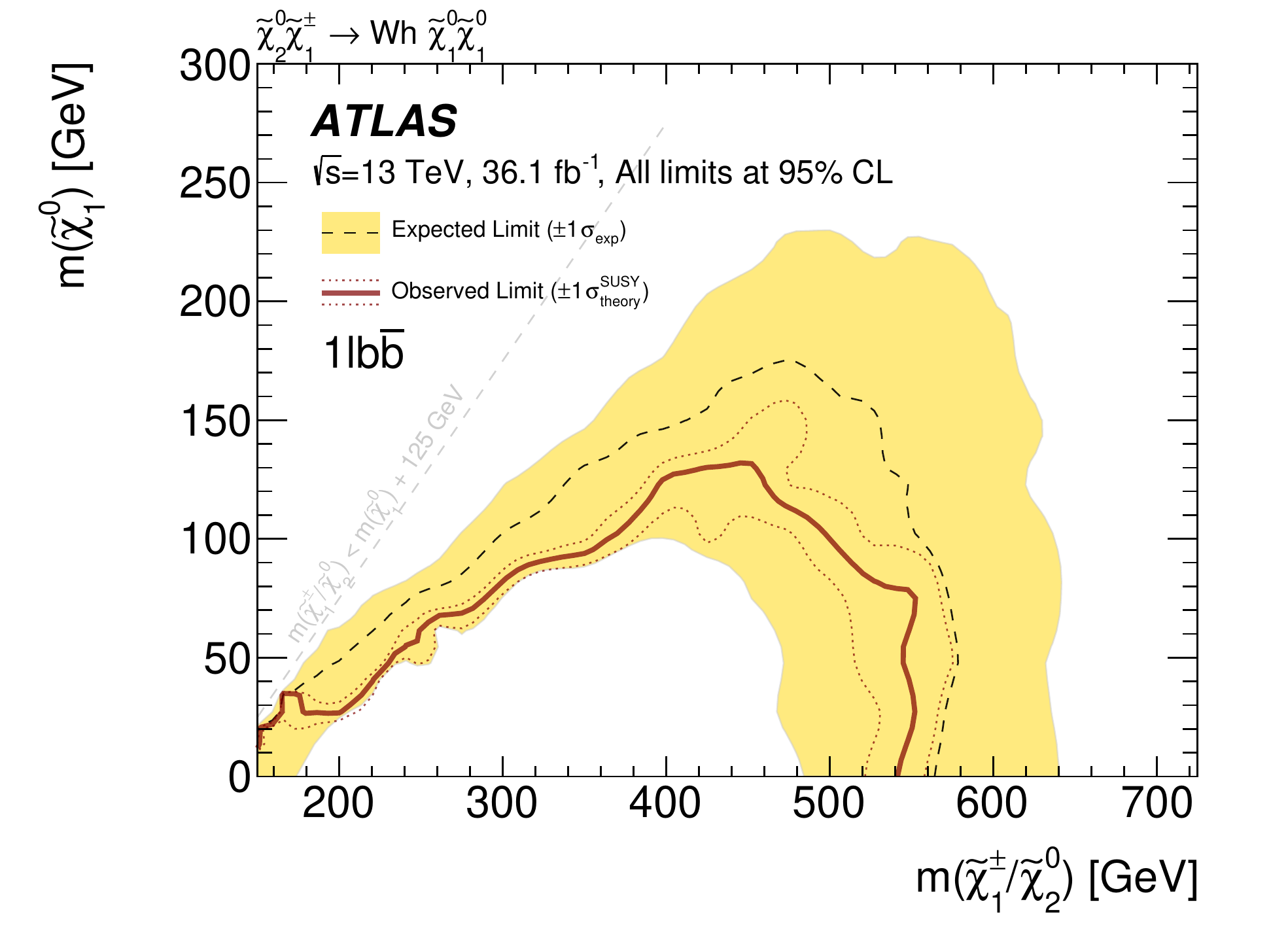}
\caption{\oneLbb}
\label{fig:results-1Lbb-exclusion}
\end{subfigure}
 
\begin{subfigure}[b]{0.5\textwidth}
\centering\includegraphics[width=\textwidth]{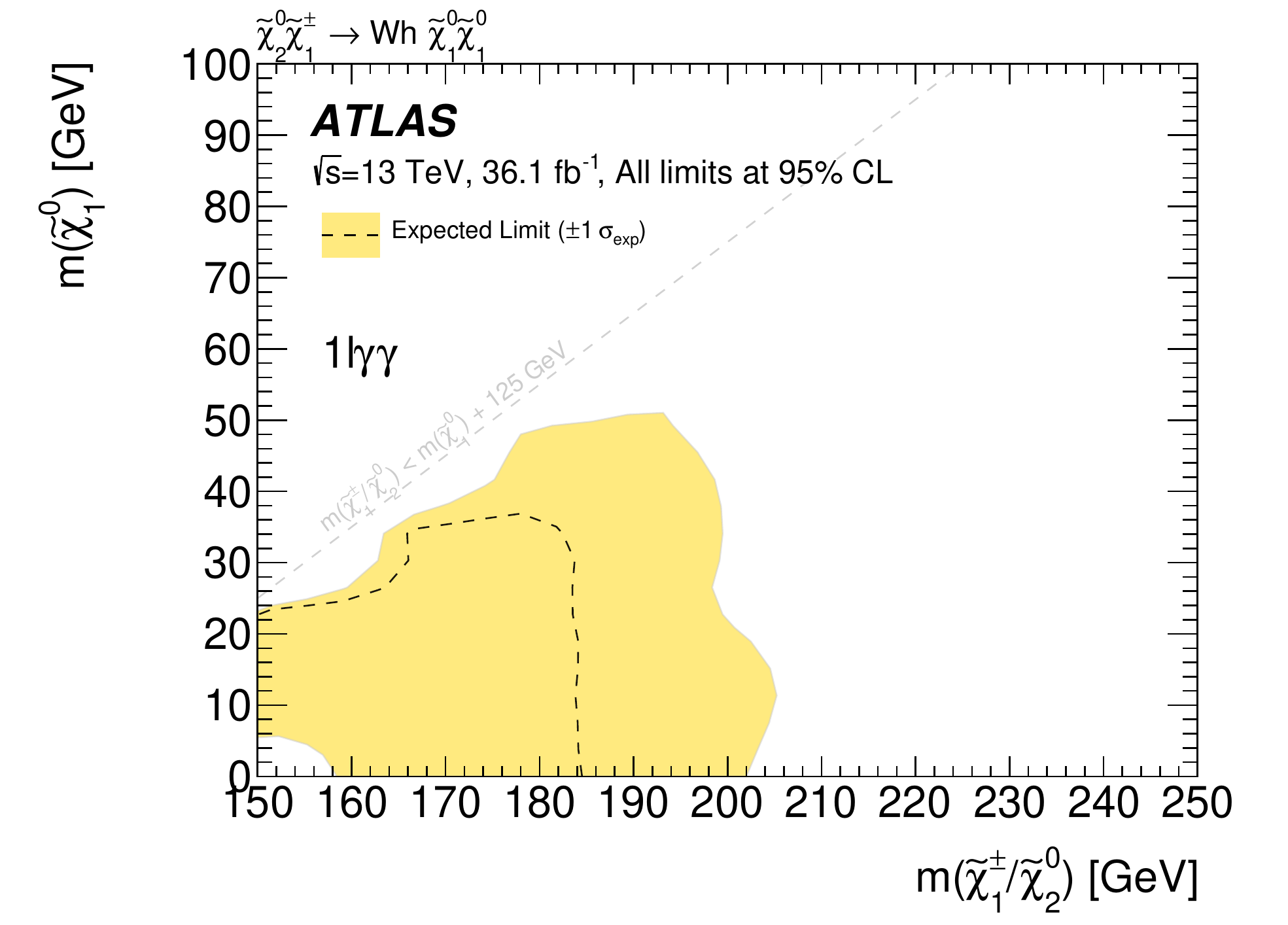}
\caption{\oneLgamgam}
\label{fig:results-1Lgammagamma-exclusion}
\end{subfigure}
\begin{subfigure}[b]{0.5\textwidth}
\centering\includegraphics[width=\textwidth]{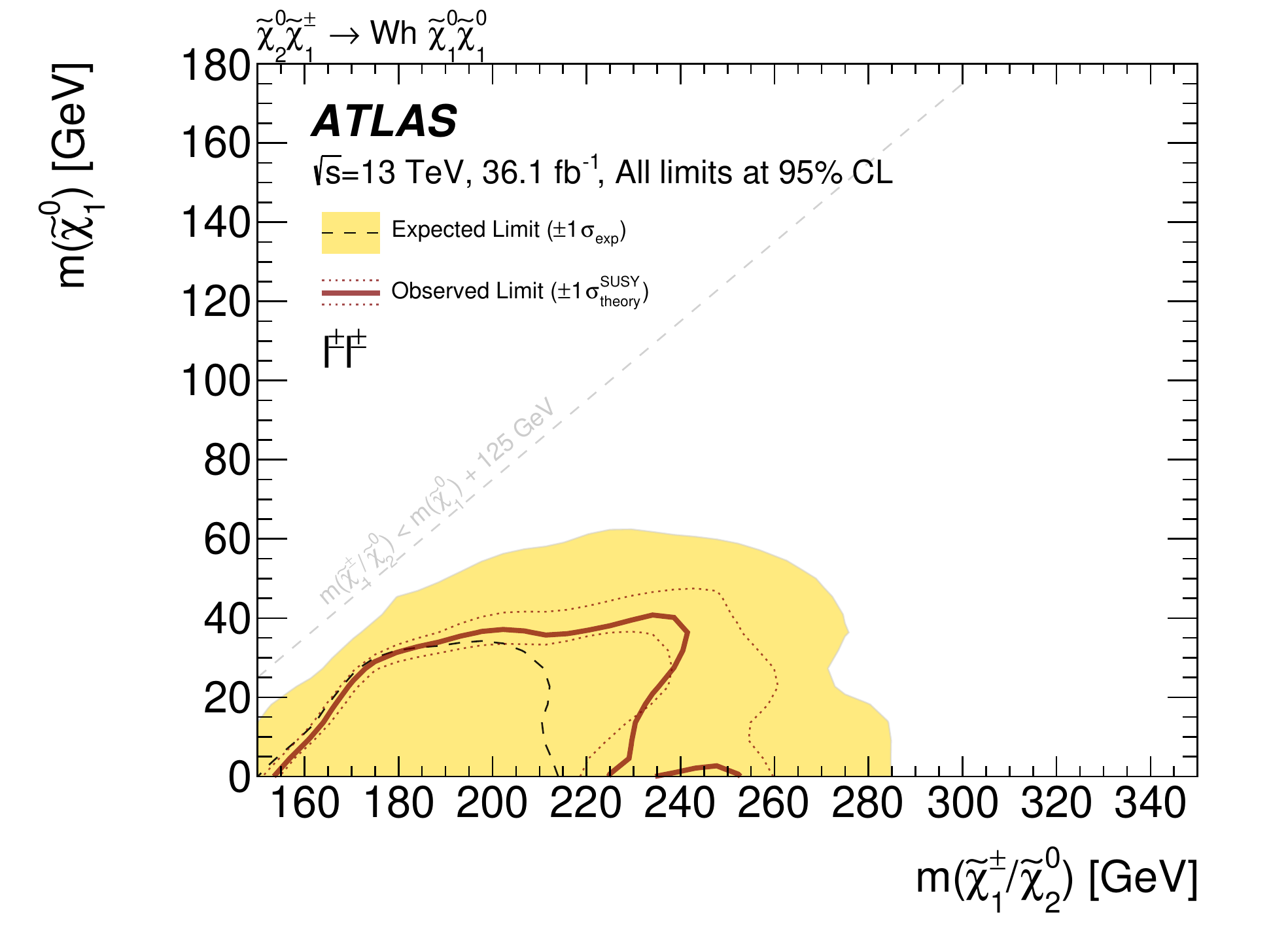}
\caption{\twoL}
\label{fig:results-twoL-exclusion}
\end{subfigure}
 
\caption{The expected and observed exclusion for the \hadronic, \oneLbb, \oneLgamgam, and \twoL channels. Experimental and theoretical systematic uncertainties, as described in \Sect{\ref{sec:systematics}}, are applied to background and signal samples and illustrated by the yellow band and the red dotted contour lines, respectively. The red dotted lines indicate the $\pm 1$ standard-deviation variation on the observed exclusion limit due to theoretical uncertainties in the signal cross-section.}
\label{fig:limits}
\end{figure}
\begin{figure}[!ht]
\centering
\begin{subfigure}[b]{0.5\textwidth}
\centering
\includegraphics[width=\textwidth]{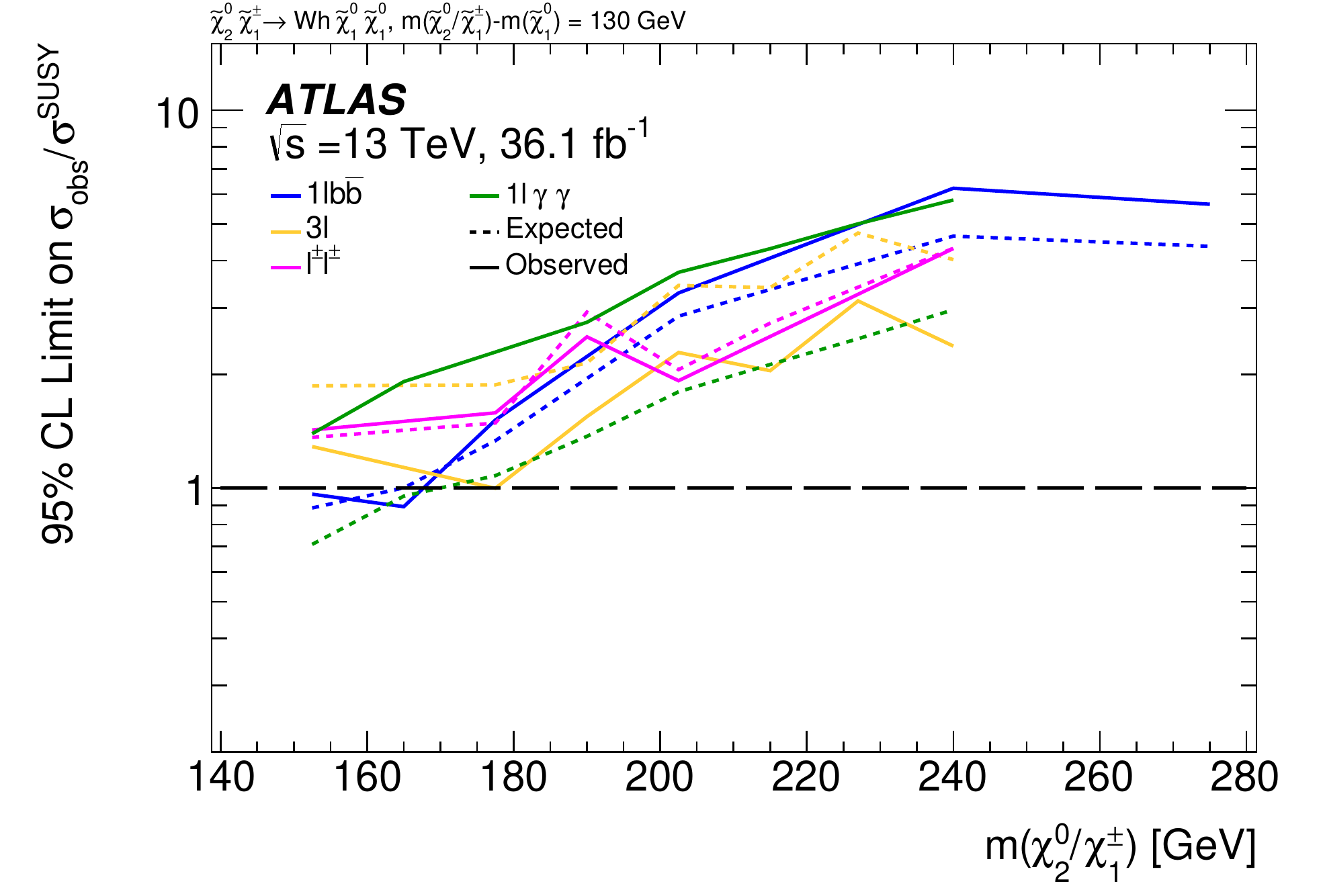}
\caption{1-D exclusion for $m(\ninotwo) - m(\ninoone) = 130~\GeV$.}
\label{fig:results-alldiag-exclusion-1D}
\end{subfigure}
\begin{subfigure}[b]{0.5\textwidth}
\centering
\includegraphics[width=\textwidth]{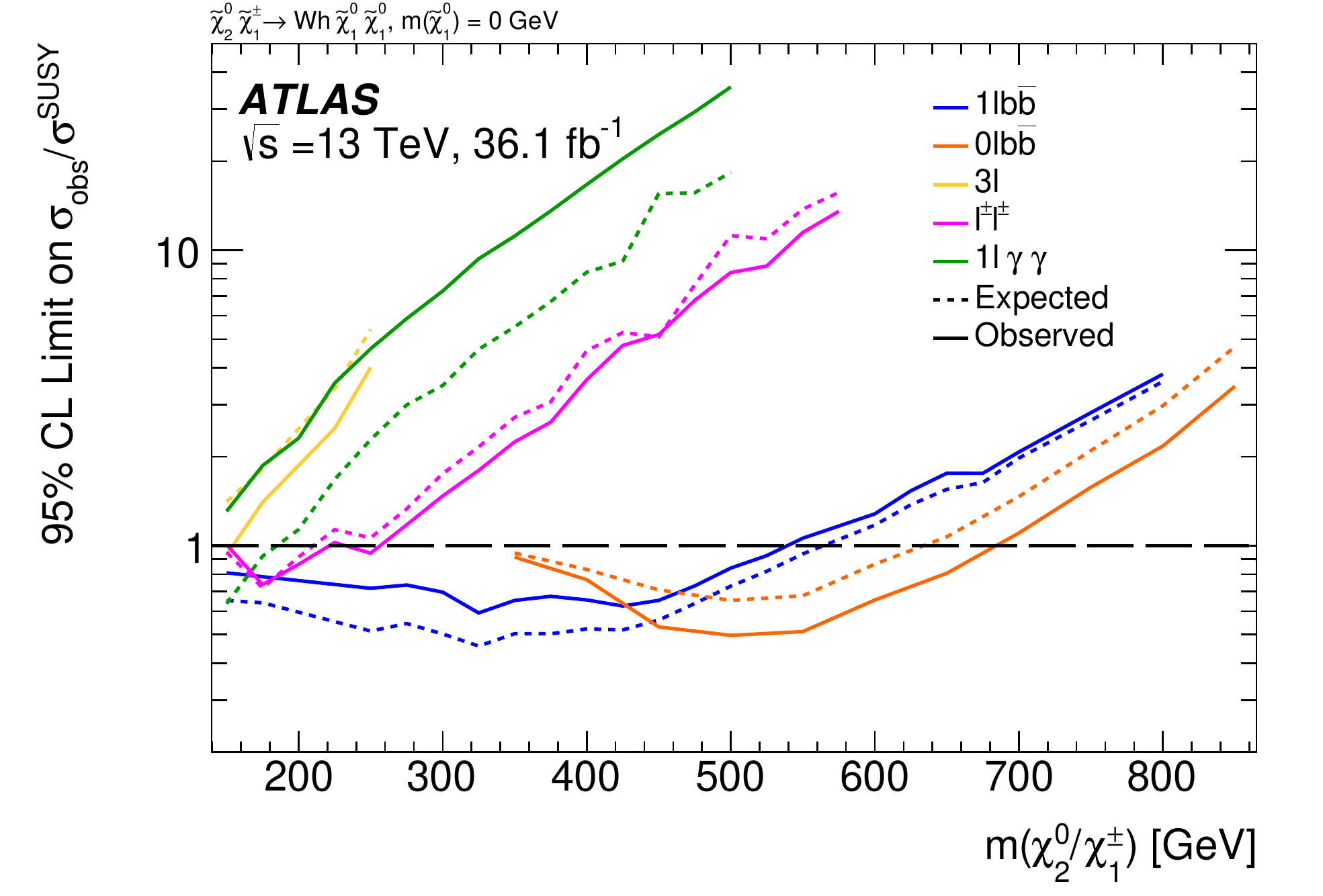}
\caption{1-D exclusion for $m(\ninoone)$ = 0 GeV.}
\label{fig:results-allmassless-exclusion-1D}
\end{subfigure}
\caption{The expected and observed cross-section exclusion as a function of the $\ninotwo/\ninoone$ masses for all channels \subref{fig:results-alldiag-exclusion-1D} for signal models with $m(\ninotwo) - m(\ninoone) = 130~\GeV$ and \subref{fig:results-allmassless-exclusion-1D} assuming $m(\ninoone)$ = 0 GeV.}
\label{fig:results-exclusion-all_1D}
\end{figure}
 
\begin{figure}[!ht]
\centering
\includegraphics[width=0.90\textwidth]{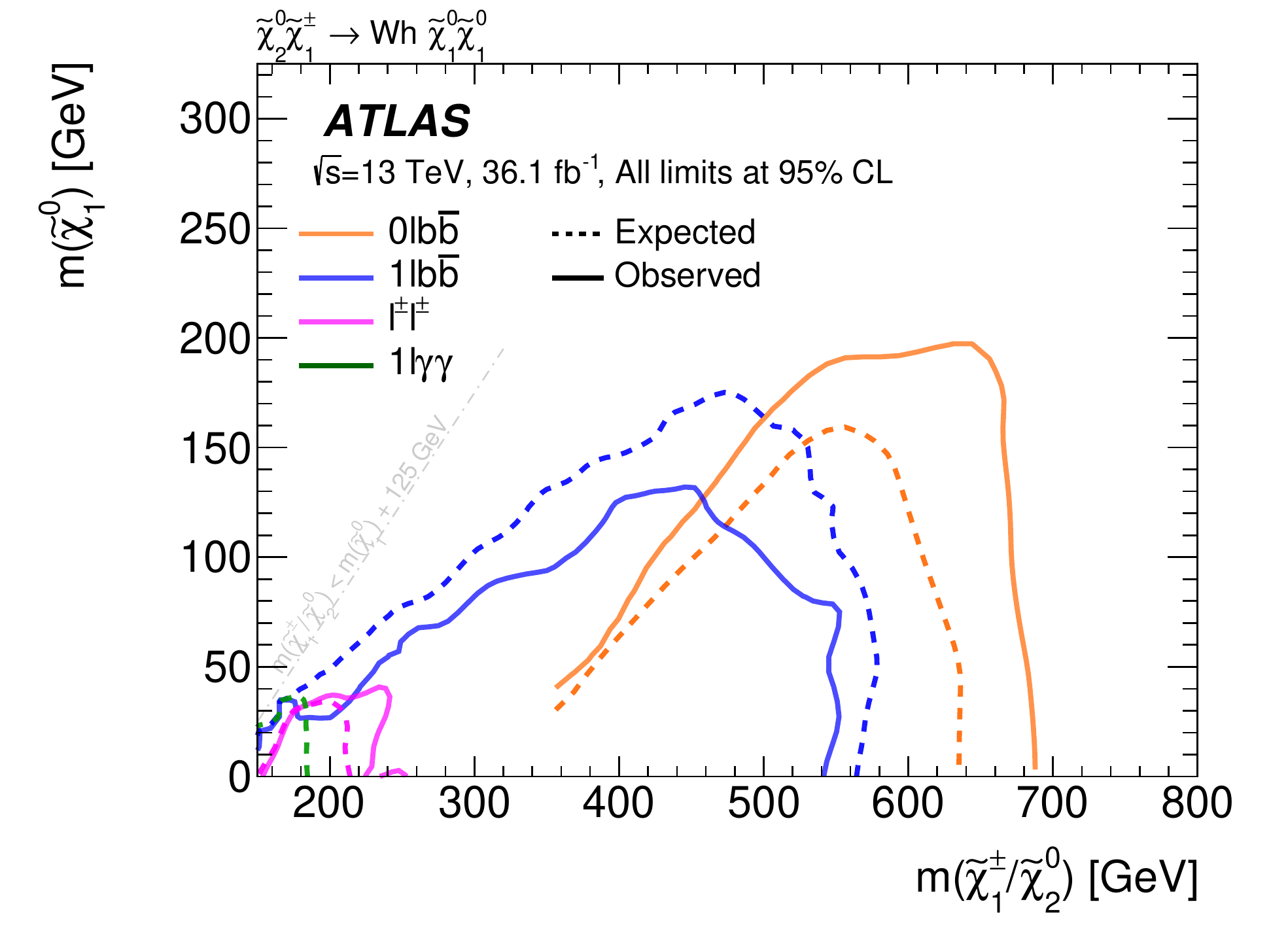}
\caption{Comparison of the expected and observed exclusions for each analysis channel studied. Only the expected exclusion is shown for the \oneLgamgam channel since the observed exclusion does not appear due to the excess observed. }
\label{fig:results-combo-exclusion}
\end{figure}

 
\FloatBarrier
 
\section{Conclusion}
\label{sec:conclusion}
 
Results of a comprehensive search for the electroweak pair production of a chargino and a neutralino $\pp\ra\chinoonepm\ninotwo$ are presented,
based on 36.1 fb$^{-1}$ of proton--proton collision data collected at $\sqrt{s} = 13$ TeV by the ATLAS experiment at the Large Hadron Collider.
Final states are considered where the chargino decays into the lightest neutralino and a \Wboson boson, $\chinoonepm\ra\ninoone\Wpm$, while the next-to-lightest neutralino decays into the lightest neutralino and a SM-like 125 GeV Higgs boson, $\ninotwo\ra\ninoone h$. The search includes \hadronic, \oneLbb, \lgg and multilepton final states with large missing transverse momentum in order to maximize sensitivity to signals of new physics processes involving \Wh and SUSY DM candidates. The searches based on final states containing \bjets (\hadronic and \oneLbb) provide unprecedented sensitivity to high-mass electroweak production for this benchmark scenario. The multilepton and \lgg~searches provide sensitivity in the region of low masses, which is more difficult to access.
Crucially, exploiting the various branching ratios of the Higgs boson into bottom quarks, photons, and multileptons, and designing an overall strategy that benefits from the complementarity of the various search channels is essential for the wide sensitivity of this analysis.
No evidence of new physics processes is observed and stringent limits are placed on the existence of electroweak production of SUSY particle pairs with significant improvements over previous searches for high $\chinoonepm\ninotwo$ masses.
In the context of the considered SUSY model, masses of $\chinoonepm$ and $\ninotwo$ smaller than 680 GeV are excluded at 95\% confidence level for a massless neutralino.
 
 
\section*{Acknowledgments}
 
 
We thank CERN for the very successful operation of the LHC, as well as the
support staff from our institutions without whom ATLAS could not be
operated efficiently.
 
We acknowledge the support of ANPCyT, Argentina; YerPhI, Armenia; ARC, Australia; BMWFW and FWF, Austria; ANAS, Azerbaijan; SSTC, Belarus; CNPq and FAPESP, Brazil; NSERC, NRC and CFI, Canada; CERN; CONICYT, Chile; CAS, MOST and NSFC, China; COLCIENCIAS, Colombia; MSMT CR, MPO CR and VSC CR, Czech Republic; DNRF and DNSRC, Denmark; IN2P3-CNRS, CEA-DRF/IRFU, France; SRNSFG, Georgia; BMBF, HGF, and MPG, Germany; GSRT, Greece; RGC, Hong Kong SAR, China; ISF and Benoziyo Center, Israel; INFN, Italy; MEXT and JSPS, Japan; CNRST, Morocco; NWO, Netherlands; RCN, Norway; MNiSW and NCN, Poland; FCT, Portugal; MNE/IFA, Romania; MES of Russia and NRC KI, Russian Federation; JINR; MESTD, Serbia; MSSR, Slovakia; ARRS and MIZ\v{S}, Slovenia; DST/NRF, South Africa; MINECO, Spain; SRC and Wallenberg Foundation, Sweden; SERI, SNSF and Cantons of Bern and Geneva, Switzerland; MOST, Taiwan; TAEK, Turkey; STFC, United Kingdom; DOE and NSF, United States of America. In addition, individual groups and members have received support from BCKDF, CANARIE, CRC and Compute Canada, Canada; COST, ERC, ERDF, Horizon 2020, and Marie Sk{\l}odowska-Curie Actions, European Union; Investissements d' Avenir Labex and Idex, ANR, France; DFG and AvH Foundation, Germany; Herakleitos, Thales and Aristeia programmes co-financed by EU-ESF and the Greek NSRF, Greece; BSF-NSF and GIF, Israel; CERCA Programme Generalitat de Catalunya, Spain; The Royal Society and Leverhulme Trust, United Kingdom.
 
The crucial computing support from all WLCG partners is acknowledged gratefully, in particular from CERN, the ATLAS Tier-1 facilities at TRIUMF (Canada), NDGF (Denmark, Norway, Sweden), CC-IN2P3 (France), KIT/GridKA (Germany), INFN-CNAF (Italy), NL-T1 (Netherlands), PIC (Spain), ASGC (Taiwan), RAL (UK) and BNL (USA), the Tier-2 facilities worldwide and large non-WLCG resource providers. Major contributors of computing resources are listed in Ref.~\cite{ATL-GEN-PUB-2016-002}.
 

\FloatBarrier
\clearpage
\newpage
 
\printbibliography

\clearpage 
 
\begin{flushleft}
{\Large The ATLAS Collaboration}

\bigskip

M.~Aaboud$^\textrm{\scriptsize 35d}$,    
G.~Aad$^\textrm{\scriptsize 100}$,    
B.~Abbott$^\textrm{\scriptsize 127}$,    
O.~Abdinov$^\textrm{\scriptsize 13,*}$,    
B.~Abeloos$^\textrm{\scriptsize 131}$,    
D.K.~Abhayasinghe$^\textrm{\scriptsize 92}$,    
S.H.~Abidi$^\textrm{\scriptsize 166}$,    
O.S.~AbouZeid$^\textrm{\scriptsize 40}$,    
N.L.~Abraham$^\textrm{\scriptsize 155}$,    
H.~Abramowicz$^\textrm{\scriptsize 160}$
% ...
\bigskip
\\

% ...
$^{13}$Institute of Physics, Azerbaijan Academy of Sciences, Baku; Azerbaijan.\\
% ...
$^{35}$$^{(a)}$Facult\'e des Sciences Ain Chock, R\'eseau Universitaire de Physique des Hautes Energies - Universit\'e Hassan II, Casablanca;$^{(b)}$Facult\'{e} des Sciences, Universit\'{e} Ibn-Tofail, K\'{e}nitra;$^{(c)}$Facult\'e des Sciences Semlalia, Universit\'e Cadi Ayyad, LPHEA-Marrakech;$^{(d)}$Facult\'e des Sciences, Universit\'e Mohamed Premier and LPTPM, Oujda;$^{(e)}$Facult\'e des sciences, Universit\'e Mohammed V, Rabat; Morocco.\\
% ...
$^{40}$Niels Bohr Institute, University of Copenhagen, Copenhagen; Denmark.\\
% ...
$^{92}$Department of Physics, Royal Holloway University of London, Egham; United Kingdom.\\
% ...
$^{100}$CPPM, Aix-Marseille Universit\'e, CNRS/IN2P3, Marseille; France.\\
% ...
$^{127}$Homer L. Dodge Department of Physics and Astronomy, University of Oklahoma, Norman OK; United States of America.\\
% ...
$^{131}$LAL, Universit\'e Paris-Sud, CNRS/IN2P3, Universit\'e Paris-Saclay, Orsay; France.\\
% ...
$^{155}$Department of Physics and Astronomy, University of Sussex, Brighton; United Kingdom.\\
% ...
$^{160}$Raymond and Beverly Sackler School of Physics and Astronomy, Tel Aviv University, Tel Aviv; Israel.\\
% ...
$^{166}$Department of Physics, University of Toronto, Toronto ON; Canada.\\
% ...
$^{a}$ Also at Borough of Manhattan Community College, City University of New York, New York NY; United States of America.\\
$^{b}$ Also at Centre for High Performance Computing, CSIR Campus, Rosebank, Cape Town; South Africa.\\
$^{c}$ Also at CERN, Geneva; Switzerland.\\
$^{d}$ Also at CPPM, Aix-Marseille Universit\'e, CNRS/IN2P3, Marseille; France.\\
$^{e}$ Also at D\'epartement de Physique Nucl\'eaire et Corpusculaire, Universit\'e de Gen\`eve, Gen\`eve; Switzerland.\\
% ...
$^{*}$ Deceased

\end{flushleft}


\end{document}